\documentclass[11pt]{article}

\usepackage{mathrsfs}
\usepackage[scr=esstix,cal=boondox]{mathalfa}

 \usepackage{ bm}
\usepackage{eucal}
\usepackage{color}

\usepackage{tikz}
\usepackage{amsmath,bbm}
\usepackage{amssymb}
\usepackage{epic}

\usepackage{graphicx}
\usepackage{pict2e}

\usepackage{pdfsync}

  \usepackage{soul}

\title{ Transversal Cusp-Airy versus Cusp-Airy for Lozenge Tilings }

\author{Mark Adler \thanks{Brandeis University, Waltham, MA
} 
  ~~~~ ~~~
  Pierre
 van Moerbeke \thanks{UCLouvain, Louvain-la-Neuve, Belgium \& Brandeis University, Waltham, MA 
 \newline Keywords: Lozenge tilings, nonconvex domains, tacnode kernels.}
 }

\date{}

\newcommand{\MAT}[1]{\left(\begin{array}{*#1c}}
\newcommand{\mat}{\end{array}\right)}
\newcommand{\qed}{\leavevmode\unskip\nobreak\penalty200\hskip2pt\null
\nobreak\hfill\rule{1.1ex}{1.1ex}
\medbreak }

\newcommand{\Hp}{H^{^{\tiny\mbox{p}}}}

\newcommand{\phat}{\widehat\pi}
\newcommand{\I}{{\rm i}}

\newcommand{\AR}{{\cal A}}
\newcommand{\CR}{{\cal C}}
\newcommand{\DR}{{\cal D}}
\newcommand{\ER}{{\mathcal E}}
\newcommand{\FR}{{\cal F}}

\newcommand{\HR}{{\cal H}}
\newcommand{\IR}{{\cal I}}

\newcommand{\KR}{{\cal K}}
\newcommand{\LR}{{\cal L}}
\newcommand{\RR}{{\cal R}}

\newcommand{\PR}{{\cal P}}
\newcommand{\QR}{{\mathcal Q}}
\newcommand{\SR}{{\cal S}}

\newcommand{\WR}{{\cal W}}

\newcommand{\BC}{{\mathbb C}}

\newcommand{\BH}{{\mathbb H}}

\newcommand{\BZ}{{\mathbb Z}}

\newcommand{\iy}{\infty}
\newcommand{\pl}{\partial}
\newcommand{\al}{\alpha}

\newcommand{\Id}{\mathbbm{1}}
\newcommand{\lcont}{\nwarrow \atop  \nearrow}
\newcommand{\rcont}{\nearrow \atop  \nwarrow}

\newenvironment{proof}{\medskip\noindent{\it Proof:\/} }{\qed}
\newenvironment
        {remark}{\medskip\noindent\underline{\it Remark:\/} }{\medbreak}

\newcommand{\om}{\omega}

\newcommand{\la}{\langle}
\newcommand{\ra}{\rangle}
\newcommand{\ga}{\gamma}
\newcommand{\ka}{\kappa}
\newcommand{\Ga}{\Gamma}
\newcommand{\dt}{\delta}
\newcommand{\Dt}{\Delta}
 \newcommand{\vr}{\varepsilon}
\newcommand{\sg}{\sigma}
\newcommand{\ze}{\zeta}
\newcommand{\BR}{{\mathbb R}}
\newcommand{\lb}{\lambda}

\newcommand{\dis}{\displaystyle}

\newcommand{\WF}{\widehat{\widehat \WR}_{k,\ell} }

\newcommand{\bl}{\begin{aligned}}
\newcommand{\el}{\end{aligned}}

\newcommand{\diag}{\operatorname{diag}}

\def\be#1\ee{\begin{equation}#1\end{equation}}
\def\bea#1\eea{\begin{eqnarray}#1\end{eqnarray}}
\def\bean#1\eean{\begin{eqnarray*}#1\end{eqnarray*}}

 \newtheorem{definition}{Definition}[section]
 \newtheorem{theorem}[definition]{Theorem}
 \newtheorem{lemma}[definition]{Lemma}
 \newtheorem{corollary}[definition]{Corollary}
 \newtheorem{proposition}[definition]{Proposition}

\catcode `!=11

\newdimen\squaresize
\newdimen\thickness
\newdimen\Thickness
\newdimen\ll! \newdimen \uu! \newdimen\dd! \newdimen \rr! \newdimen
\temp!

\def\sq!#1#2#3#4#5{%
\ll!=#1 \uu!=#2 \dd!=#3 \rr!=#4
\setbox0=\hbox{%
 \temp!=\squaresize\advance\temp! by .5\uu!
 \rlap{\kern -.5\ll!
 \vbox{\hrule height \temp! width#1 depth .5\dd!}}%
 \temp!=\squaresize\advance\temp! by -.5\uu!
 \rlap{\raise\temp!
 \vbox{\hrule height #2 width \squaresize}}%
 \rlap{\raise -.5\dd!
 \vbox{\hrule height #3 width \squaresize}}%
 \temp!=\squaresize\advance\temp! by .5\uu!
 \rlap{\kern \squaresize \kern-.5\rr!
 \vbox{\hrule height \temp! width#4 depth .5\dd!}}%
 \rlap{\kern .5\squaresize\raise .5\squaresize
 \vbox to 0pt{\vss\hbox to 0pt{\hss $#5$\hss}\vss}}%
}
 \ht0=0pt \dp0=0pt \box0
}

\def\vsq!#1#2#3#4#5\endvsq!{\vbox to \squaresize{\hrule
width\squaresize height 0pt%
\vss\sq!{#1}{#2}{#3}{#4}{#5}}}

\newdimen \LL! \newdimen \UU! \newdimen \DD! \newdimen \RR!

\def\vvsq!{\futurelet\next\vvvsq!}
\def\vvvsq!{\relax
  \ifx     \next l\LL!=\Thickness \let\continue=\skipnexttoken!
  \else\ifx\next u\UU!=\Thickness \let\continue=\skipnexttoken!
  \else\ifx\next d\DD!=\Thickness \let\continue=\skipnexttoken!
  \else\ifx\next r\RR!=\Thickness \let\continue=\skipnexttoken!
  \else\def\continue{\vsq!\LL!\UU!\DD!\RR!}%
  \fi\fi\fi\fi
  \continue}

\def\skipnexttoken!#1{\vvsq!}

\def\place#1#2#3{\vbox to 0pt{\vss
\rlap{\kern#1\squaresize
  \raise#2\squaresize\hbox{$#3$}}
\vss}}

\catcode `!=12

\begin{document}

 \sloppy
\maketitle

 
  \begin{abstract}
 The fluctuations of lozenge tilings of hexagons with one or several cuts (nonconvexities) along opposite sides are governed by the (discrete-continuous) tacnode kernel ${\mathbb L}^{\mbox{\tiny dTac}}$, upon letting  
  the hexagon become very large (or, in other terms, keeping the hexagon fixed, with the tiles becoming very small). This is a point process with a fixed finite number $r$ of (continuous) points along a discrete set of parallel lines within a specific strip of width $\rho$; outside the strip,  the number of points along the parallel lines increase. Both numbers $\rho$ and $r$ are specified by  the distance between the cuts and their sizes and are fixed parameters in the tacnode kernel (see \cite{AJvM1,AJvM2}). Letting $r\to\iy$ and keeping $\rho$ fixed, one finds a liquid phase inscribed in the polygon, whose boundary (arctic curve) has a  cusp near each cut, with two (different) solid phases descending into the cusp (split-cusp). Duse-Johansson-Metcalfe  \cite{DJM} show that near the cusp the tile-fluctuations   obey the cusp-Airy statistics. It would seem natural to   see this cusp-Airy phenomenon in the neighborhood of the cut, for the limit  $r\to \iy$  of the tacnode kernel ${\mathbb L}^{\mbox{\tiny dTac}}$. As it turns out, another statistic appears:  the {\em transversal cusp-Airy} statistics, which was    puzzling   to all of us. For $\rho=0$, these statistics are derived and fully explained  in this paper. Finally, we formulate a conjecture for large $r$,  when $\rho>0$;  to give evidence, we prove an $\ell$-spaced Christoffel-Darboux formula for Hermite polynomials and its double Fourier transform. 
   \end{abstract}

 \tableofcontents


\newpage

{\bf Acknowledgement}: The authors thank Kurt Johansson for getting us involved in this problem;  we had many  insightful discussions over the years. Kurt had  excellent suggestions and intuitions! Thanks also to the Mittag-Leffler Institute, Djursholm, for their hospitality during the fall semester 2024. Our thanks also go to Christophe Charlier who did many simulations, which were crucial for the   understanding of this problem. We also thank Christophe Charlier and Tom Claeys for interesting discussions. Mark Adler and Pierre van Moerbeke like to thank Tom Claeys and UCLouvain for their kind hospitality and their unwavering support over the years.
 
\section{General introduction }

Planar dimer models on bipartite graphs, or simply put, tiling models, yield a rich source of physical and mathematical phenomena, touching upon statistical physics and combinatorics. In 1911, MacMahon \cite{McMa} counted the number of lozenge tilings of a hexagon, while Fisher and Temperley \cite{Temp} counted the number of domino tilings on a rectangular chessboard. Kasteleyn \cite{Kast} expressed the number of dimer covers on a planar bipartite graph and the associated entropy in terms of a determinant of the ``Kasteleyn matrix'', a fundamental tool to obtain correlation kernels for tiling models. These ideas have been used to model spontaneous magnetization by Kaufman and Onsager\cite{Onsager}; recently Borodin and Berggren \cite{BB2} studied phase transitions going with crystallization. More recently insights and tools from Random Matrix Theory (RMT) have led to a better understanding of the fluctuations near phase transitions in tiling models, both macroscopically and microscopically. Probability distributions from RMT seem to describe the statistical fluctuations of the tiles near singular points, where different phases meet; they therefore qualify for the status of  ``universal distributions''. 

In this paper, we focus on tiling problems of polygonal regions; they have been linked to Gelfand-Zetlin cones (interlacing behavior) by Cohn-Kenyon-Propp \cite{CKP} and Petrov \cite{Petrov2}, to nonintersecting paths, to determinantal point processes, kernels and random matrices by Johansson\cite{Jo16}. In particular, lozenge tilings of hexagons were governed by a kernel consisting of Hahn polynomials; see the  2016 Harvard lectures of  Johansson\cite{Jo16} and Gorin\cite{Go}. Johansson \cite{Jo05c} showed that the tiles near generic points of the arctic curve fluctuate as the Airy-process, while Johansson-Nordenstam\cite{JN} show that, appropriately scaled, the tiles near tangency points fluctuate like the interlacing eigenvalues of the principal minors of a GUE-matrix. Okounkov-Reshetikhin \cite{OR1} study Schur processes and random matrix distributions using geometric-combinatorial tools.   


This paper focusses on nonconvex regions, having two phases, a solid and a liquid phase. A lot of recent work has also been done in models having three phases (solid, liquid and gas), which  will not be discussed here; see, among others  \cite{KOS,KO,ChJo1,ChJo2,Cha,Cha1,BoDu,BB1,BB2}. See also Charlier-Claeys \cite{CC}.

For  large two-phase  nonconvex regions, it is known that the arctic curve (boundary between the two phases) can have inward simple cusps, besides the tangency points of the arctic curve with the boundary of the region; see Astala-Duse-Prause-Zhong \cite{AstDuse}. The fluctuations of the tiles near these critical points are of interest because of their universal nature. We focus here  on the neighborhood of the tangential direction to a cusp on the arctic curve. One finds a particular phenomenon; namely two different solid phases are present near the cusp, with the boundary (between the two) being tangent to the cusp. A local analysis of this situation by Duse-Johansson-Metcalfe \cite{DJM} identified the fluctuations as satisfying the {\em cusp-Airy} statistics, to be explained in Theorem 2.3. 




In \cite{AJvM1,AJvM2}, we considered  global tiling  of a hexagon with lozenges, having two or several nonconvexities (for short: {\em cuts}) on opposite sides of the hexagon.  For the sake of this paper, we consider the case of two cuts, facing each other (possibly slightly shifted)  along two opposite  sides of a hexagon; see Figs.1. This leads to two near-hexagon-like shapes, separated by a strip determined by the two cuts. Upon tiling this shape with lozenges, one notices, roughly speaking, two near-arctic ellipses inscribed in the two near-hexagon-like shapes, separated by a strip, where the interaction takes place. The strip contains filaments of lozenges of two types in a sea of tiles of the third type; these filaments connect the two ellipses; their number is purely determined by the geometry of the figure. Letting  the tiles become very small in comparison to the figure, the global statistics is governed by the discrete-tacnode kernel ${\mathbb L}^{\mbox{\tiny dTac}}    $ of Adler-Johansson-van Moerbeke\cite{AJvM1,AJvM2}. This kernel is continuous in one direction and discrete in the other. That very same kernel appears as well for a non-convex situation; namely, domino-tilings of Aztec-rectangles\cite{AJvM3}, which are such that the pair of upper-most (left-most) squares have the same orientation as the pair of lower-most (right-most) squares, unlike in the usual Aztec diamond. In \cite{AJvM0}, we considered a doubly continuous version of this kernel, involving two groups of nonintersecting paths merely touching.


{\bf Motivation and open questions:} This paper came about in the following context. The discrete tacnode kernel ${\mathbb L}^{\mbox{\tiny dTac}}  _{r,\rho,\beta} $ for $\rho=\beta=0$ shared many   features with the Duse-Johansson-Metcalfe cusp-Airy kernel \cite{DJM}, except for being quite a bit more complicated. Therefore it seemed very natural  to expect to see the cusp-Airy kernel, as the limit ($r\to \iy$) of the discrete tacnode kernel ${\mathbb L}^{\mbox{\tiny dTac}}$. And yet, we found something different, which was {\em very puzzling} to all of us! That contradicts a conjecture at the end of \cite{AstDuse}, as pointed out to us by Eric Duse. This paper addresses  this mystery.



The simulation of \textcolor{red}{Figs. 4a and 4b}  shows clearly the emergence of two cusps in the arctic curve near the top  and the bottom  of the figure. The mathematics of it will be discussed informally in section 10, from a macroscopic point of view. For the sake of this discussion, consider the upper-cut combined with the upper-cusp. This cusp could be tangent either to the extension of the $\{\rho\}$-line (the extension of the left-hand side of the cut) or the $L_1$-line (the extension of the right-hand side of the cut); see \textcolor{red}{Fig. 4b}.   According to the simulation of \textcolor{red}{Fig. 4},  the cusp-Airy fluctuations (of the yellow tiles) take place along the set of parallel lines to $L_1$.

However, the discrete tacnode kernel describes a point process (of blue tiles) along parallel lines  to the  $\{\rho\}$-line.  
 This point process is thus  transversal to the $L_1$-line and therefore {\em transversal} to the cusp. And, indeed, the scaling limit of the discrete tacnode kernel ${\mathbb L}^{\mbox{\tiny dTac}}$ for large $r$ leads to a new kernel, the transversal cusp-Airy kernel $ {\mathbb L}^{\mbox{\tiny T-cuspAiry}} $, which is different  from the cusp-Airy kernel $ {\mathbb L}^{\mbox{\tiny cuspAiry}} $. This result leads to more open questions, like obtaining the cusp-Airy and transversal cusp-Airy kernels directly from the discrete kernel ${\mathbb L}^{\mbox{\tiny blue}}$, as in (\ref{limit}), letting $r\to \iy$, together with the scaling (\ref{geomscale}) and (\ref{coordscale}). Picking a strip  $\{\rho\}$ of size $\rho=0$ greatly simplifies the calculations. We will nevertheless formulate a conjecture for the case $\rho>0$, and give a convincing argument for it. 

   \section{The tiling model, the discrete-tacnode kernel and the   transversal cusp-Airy kernel} 
   
     {\em The model and main results.}  This paper deals with tilings  of nonconvex hexagons with lozenges and in particular it deals with the edge behavior near the nonconvexity when the hexagon becomes large or equivalently when the hexagon remains fixed and the tiles become very small. The tiles described in Figure 1 have the color blue, red and green (corresponding to right-leaning, straight and left-leaning). An affine transformation maps the tiles on the left above Fig. 1 into the ones on the right above Fig. 1, the latter being more convenient from the point of view of coordinates. The two cuts define {\em a strip} which is bounded by two parallel lines, ~the first one extending the
     

            \newpage
            
            \vspace*{-2.2cm}




 \vspace*{6.5cm}

  

  
 
  \setlength{\unitlength}{0.015in}\begin{picture}(0,0)

       {\makebox(500,-190) {\rotatebox{0} {\includegraphics[width=240mm,height=270mm] {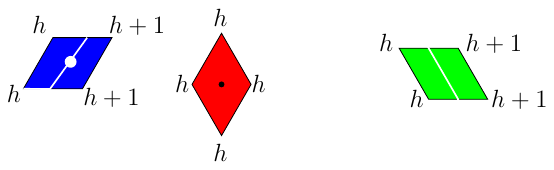}    }}}
 \end{picture}
 
 \vspace{1cm}
  
  \setlength{\unitlength}{0.015in}\begin{picture}(0,0)
  {\makebox(310,190) {\rotatebox{-90} {\includegraphics[width=135mm,height=191mm] {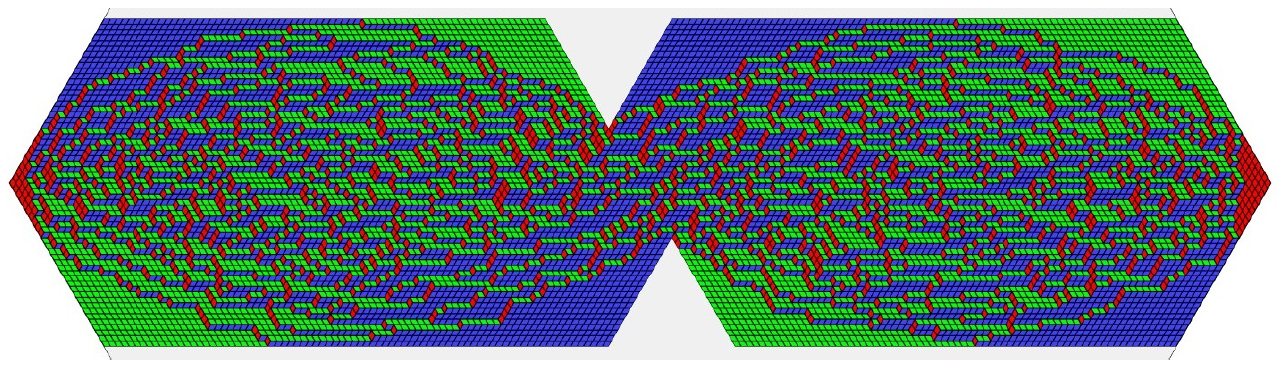}    }}}
  
   \put( -335, 120){$b$} \put( -335, 70){$c$}
    \put( 20, 120){$c$} \put( 20, 70){$b$}
    \put( -215, 150){$n_1$} \put( -120, 150){$n_2$}
     \put( -215, 150){$n_1$}
     \put( -182, 145){$\leftarrow d\rightarrow$}
      \put( -120, 150){$n_2$}
       \put( -215,  35){$m_1$}
     \put( -162,  40){$\leftarrow d\rightarrow$}
      \put( -100,  35){$m_2$}

  \end{picture}

    \vspace*{-1cm}
      
      Figure 1\label{fig1}: Tiling simulation of a hexagon of size $n_1=m_2=70, ~n_2=m_1=80, ~b=c=30$ with two opposite cuts (of size $d=20$ by means of blue, red and green tiles, as indicated above the picture, with $\rho=0$ and $r=10=\#$of filaments. (Courtesy of Christophe Charlier)
      
      \vspace*{1cm}
      
        \noindent right hand side of the lower-triangular cut and the second one extending the  left hand side of the upper-triangular cut. As it turns out from {Figs 1  and Figs. 4 and 5},   this strip $\{\rho\}$of width $\rho$ (called that way  throughout the paper), plays a prominent role in this work. 
            
               The array of parallel lines to this strip defines a point process, containing the points corresponding to the center of the blue tiles in Fig 1. A first observation is that this defines a {\em doubly interlacing point process}, viewed from the perpendicular direction to the strip; see \textcolor{red}{Fig 5}. To be precise, within the strip (including the boundary) the number of blue dots per line 
   remains the same (called $r$ throughout) and form an interlacing set, whereas by moving away from the strip, either to the left or to the right, the number increases, at least for a while, by one and still interlace; see the doubly interlacing  example in \textcolor{red}{Fig. 6}. For a general survey of doubly interlacing point processes, see \cite{AvM3}.


      Given a random tiling of the model, {\em the point process of blue dots is governed by a kernel ${\mathbb L}^{\mbox{\tiny blue}}(\eta,\xi;\eta',\xi')$} in discrete variables, where $\eta$ parametrizes the array of parallel lines mentioned above and where $\xi$ parametrizes the discrete locations of the blue dots on the lines. Another way of describing this point process is to assign {\em heights and  level lines} to each of the tiles; the height is specified by the indices $h, ~h+1$ on each tile, as done on top of Fig. 1 (red tiles are flat and the height of the blue and green ones are tilted upwards from left to right). The level lines on the tiles extend  to a set of successive level lines on the hexagon; they contain the same information as the tiling. Then the blue dots coincide with the intersection of the level lines with the parallel lines in the array, mentioned before; see \textcolor{red}{Figs. 1 and 5}. The kernel ${\mathbb L}^{\mbox{\tiny blue}}(\eta,\xi;\eta',\xi')$, as given in \cite{AJvM1,AJvM2}, is then given by $r+2$-fold multiple contour integrals of rational functions and Vandermonde determinants. 
      
      Letting the size of the polygon tend to $\iy$, properly rescaled (see (\ref{geomscale}) and (\ref{coordscale})), while maintaining $r$ and $\rho$ constant, leads to {\em the discrete tacnode kernel ${\mathbb L}^{\mbox{\tiny dTac}}  _{r,\rho,\beta}( \tau_1, \theta_1 ;\tau_2, \theta_2)$},  given by formula (\ref{Final0}), where the $\tau_i\in \BZ$'s parametrize the parallel lines in the array, whereas the $\theta_i\in \BR$ parametrize the blue dots moving {\em continuously} along those lines; $\beta$ is an extra free parameter. This is done in \cite{AJvM1,AJvM2}.

      The purpose of this paper is to understand the behavior of the blue dots near the cuts, when its number becomes large. This will be done in the case that the strip $\{\rho\}$ has width $\rho=0$ and that the free parameter $\beta=0$ in (\ref{Final0}). 
      When $\rho=0$, the formula for the discrete tacnode kernel ${\mathbb L}^{\mbox{\tiny dTac}}_{r,\rho,\beta} $ simplifies considerably. It will be given in a form convenient  for asymptotic analysis when $r\to \iy$. 
      
      Before stating Theorem \ref{Th:prodop''}, we need  to 
      %
 define  three operators, for $\tau\in \BZ$ and $\xi \in \BR$; they will be used throughout:
\newline {\em (i)} A $\tau$-fold {\em (anti)-derivative operator}  for $\xi\in \BR$ and $\tau\in \BZ$,\footnote{We have: $\DR_{\xi}^\tau(u)[g(u)]=\left(\frac{\pl}{\pl \xi}\right)^\tau \int_\BR \dt(u-\xi) g(u)du
  = \int_\BR \left(\frac{\pl}{\pl \xi}\right)^\tau \dt(u-\xi) g(u)du$. For $\tau<0$, the latter gives the second formula in (\ref{Doper}).}
 \be\bl\label{Doper}
   \DR_\xi^{ \tau}(u)[g(u)]&=\left(\frac{\pl}{\pl \xi}\right)^\tau g(\xi)
    \mbox{,  for $\tau\geq 0$}
 \\&=-\int_\xi^\iy du \frac{(\xi-u)^{-\tau-1}}{(-\tau-1)!}g(u)
 \mbox{,  for $\tau< 0$},
 \el\ee
having the property that  $\DR_\xi^\tau \DR_\xi^{\tau'}= \DR_\xi^{\tau+\tau'}$ for any $\tau,\tau'\in \BZ$.
\newline {\em (ii)} An {\em integral operator},   involving integration about the imaginary line $L_+$ to the right of a small circle $\Ga_0$ about $0$ and  : \be\label{Eoper''}\bl
   \ER^{   \tau }_{ \xi }(v)g(v)
  &:=\sqrt{2\pi}
\oint_{\Ga_0} \!\frac{dw ~ e^{- \frac{w ^2}{2}-  \xi  w }}{2\pi \I (-w )^{ - \tau}}  
\int_{L_+} \frac{dv ~e^{ \frac{  v ^2}2} }{2 \pi \I  (w \!-\!v )}g(v) \\&= (-1)^{\tau+1}
\int_{L_+} \frac{dv ~e^{ \frac{  v ^2}2} 
 F_{-\tau}^{  \xi}( {v} )}{\sqrt{2\pi}~ \I v^{-\tau } }g(v)~~~~(=0 \mbox{ for }\tau\geq 0)
.\el\ee
The second expression uses the generating function $  F^{\xi}_ \tau(v)$, defined in (\ref{tildeF}) below, and will follow from (\ref{EFtilde}).
\medbreak
%
\noindent {\em (iii)} We now consider the sum of these two operators, combined with a Fourier transform,
  \be
\label{FR0}
\FR^{(\tau)}_\xi(u):=\DR^{ \tau }_{ \xi }(u )
+   \ER_{  \xi }^{   \tau   }(v
 )
 \int_{\BR}\frac{d u }{\sqrt{2\pi}}
 e^{ u v 
  },~~v\in L_+
. \ee
 We also need the generating function in $v$ for the probabilistic Hermite polynomials $\Hp_j $, truncated at power $\tau-1$ in $v$: (see (\ref{HermP}))
  \be \label{tildeF}\bl   F^{\xi}_ \tau(v) &:=\bigl[e^{-\frac{v^2}2-\xi v   }\bigr]_{ {[0,\tau-1 ]}}   =\sum_{j=0}^{\tau-1}\frac{H^{\tiny\mbox{p}}_j(-\xi)}{j!}  v ^j 
, ~~~~~(=0 \mbox{,  for }\tau\leq 0)
\\   E^{\xi}_ \tau(v)&:=
e^{-\frac{v^2}2-\xi v   }- F^{\xi}_ \tau(v)
\el\ee

   \begin{theorem} \label{Th:prodop''}
  For $\rho=\beta=0$ and given $ \FR^{(\tau)}_\xi(u)$ as in (\ref{FR0}) and $  F^\xi_\tau(v)$ as in (\ref{tildeF}), the discrete tacnode kernel ${\mathbb L}^{\mbox{\tiny dTac}}_{r,\rho,\beta} $ takes on the following form for arbitrary choices of $\vr_i=\pm 1$: (using $\theta=\xi\sqrt2$ in the kernel  (\ref{Final0})): 
\be\label{prodop''}\bl
     & ( \sqrt 2)^{\tau_2-\tau_1}   {\mathbb L}^{\mbox{\tiny dTac}}_{r,\rho,\beta}  (\tau_1, \xi_1\sqrt2;\tau_2, \xi _2\sqrt2)\sqrt2 d\xi\Bigr|_{\rho=\beta=0}
     \\ &  =   -     \BH^{\tau_1-\tau_2}(\xi_2-\xi_1)  d\xi  
   \\&~ +   \oint_{ 
    L_{+}} 
   \frac{e^{\frac{v^2}2}dv}{2\pi \I v^{\tau_1-\tau_2}}   \left(e^{ \xi_2 v} F^{\xi_1}_{\tau_1}(v) +e^{-\xi_1v}   F^{-\xi_2}_{-\tau_2}(v) 
  -e^{  \frac{v^2}2}  F^{\xi_1}_{\tau_1}(v) F^{-\xi_2}_{-\tau_2}(v)   \right)  d\xi 
\\ &  ~
+ (-  1)^{\tau_1-\tau_2}  \vr_1 ^{\tau_1}\vr_2^{\tau_2}
 \FR^{(-\tau_1)}_{\vr_1\xi_1}(u_1)
 \FR^{(\tau_2)}_{-\vr_2\xi_2}(u_2 )
 K_r( \vr_1u_1,- \vr_2 u_2)  d\xi
%
%
 %
 %
\\&=:({\mathbb L}_0+{\mathbb L}_1 +{\mathbb L}_2)(\tau_1,\xi_1;\tau_2,\xi_2)d\xi =:({\mathbb L}_0+(\sum_{i=1}^3{\mathbb L}^+_{1i}) +{\mathbb L}_2)(\tau_1,\xi_1;\tau_2,\xi_2)d\xi, \el\ee
   where $\BH^{\tau }(\xi )$ and $K_r(u,v)$ denote the Heaviside function and the standard GUE-kernel, defined later in (\ref{Theta}) and (\ref{KGUE}).  The three parts on the right hand side of (\ref{prodop''}) will be denoted throughout by ${\mathbb L}_i:=
   {\mathbb L}_i(\tau_1,\xi_1;\tau_2,\xi_2)$, with the subkernel ${\mathbb L}_1$ decomposed into three parts ${\mathbb L}_{1i}$. 
   
   \end{theorem}
 
   \remark The ${\mathbb L}_i$ correspond to the first three lines on the right hand side of this formula, whereas ${\mathbb L}_1$ decomposes into a sum of three terms ${\mathbb L}_{1i}^\pm$ for the $+$ sign, the $\pm$ referring to the vertical line $L_\pm$ to the right or left of $0$. The ${\mathbb L}_{1i}^\pm$ will be discussed in subsection 7.4.

 
 
 %


  
    \medbreak
      
  We now consider the scaling limit of the kernel ${\mathbb L}^{\mbox{\tiny dTac}}  _{r,\rho,\beta}( \tau_1, \theta_1 ;\tau_2, \theta_2)|_{\rho=\beta=0}$ (as in (\ref{Final0})) at a distance $\theta=\pm 2\sqrt r$ along the line $\{\rho\}$ for $\rho=0$ (i.e., towards the upper- and lower-cut), measured from the middle point of the line. This describes the fluctuations of the blue dots near the cuts; so, this leads to a new kernel, {\em the Transversal Cusp-Airy kernel 
 ${\mathbb L}^{\mbox{\tiny T-cuspAiry}}   ( \tau_1,  \xi'_1 ;\tau_2,  \xi'_2)$  } for $\tau_i\in \BZ$ and $\xi'_i\in \BR$, which will be given in Theorem \ref{MainTheo}.    Given the 
  Airy cubic, 
\be\label{Airycub}
\AR_u(\xi)=\frac{u^3}3-u\xi      ,\ee  
we define the $\tau$-Airy function (see \cite{ADvM,DJM})), involving integration over the following contours, either ${\mathcal C}_L:=$ $\nwarrow \atop  \nearrow$ $0$, a contour from $e^{\frac{4\pi}{3}}\iy$ to $e^{\frac{2\pi}{3}}\iy$ to the left of $0$
 , or $\CR_R:=$$0$ $ {\nearrow}\atop { \nwarrow}$, a reflection of the first contour $\CR_L$ about the imaginary axis,   (setting $u\to -u$)
\be\bl\label{r-Airy}
A^{(0)}_{\tau}(\xi)  =\int_{\CR_L+\Ga_0
 }e^{-\AR_u(\xi)
  }\frac{du}{2\pi \I u^\tau}
 =(-1)^\tau\int_{\CR_R-\Ga_0
 }e^{\AR_u(\xi) 
 \ }\frac{du}{2\pi \I u^\tau}
 \el\ee
and an  ``{\em extended $r $-Airy function}", generalizing the former\footnote{Compared to $A^{(0)}_{\tau}(\xi)$ the function $A _{\tau}(\xi)$ has the cubic term  missing in the $\Ga_0$-integration !}: (see Fig. 5) 
\be\label{Atau}\bl A_\tau (\xi) &= \!\!\left[\int_{\CR_L
 }\!e^{-\AR_u(\xi)
  }\!+\!\oint_{\Ga_0} {e^{u\xi} } 
   \right]\frac{du}{2\pi \I u^\tau}\!=\!\!\left[\!\int_{\CR_R
  }\! e^{\AR_u(\xi)
    }\!+\!\oint_{-\Ga_0} {e^{-u\xi} } 
\right]\frac{(-1)^{\tau}du}{2\pi \I u^\tau}
 \el\ee
 %
 \vspace*{-1.2cm}
 
 \setlength{\unitlength}{0.017in}\begin{picture}(0,60)
 \put(120, -90){\makebox(0,0) {\rotatebox{0}{\includegraphics[width=91mm,height=105mm]
  {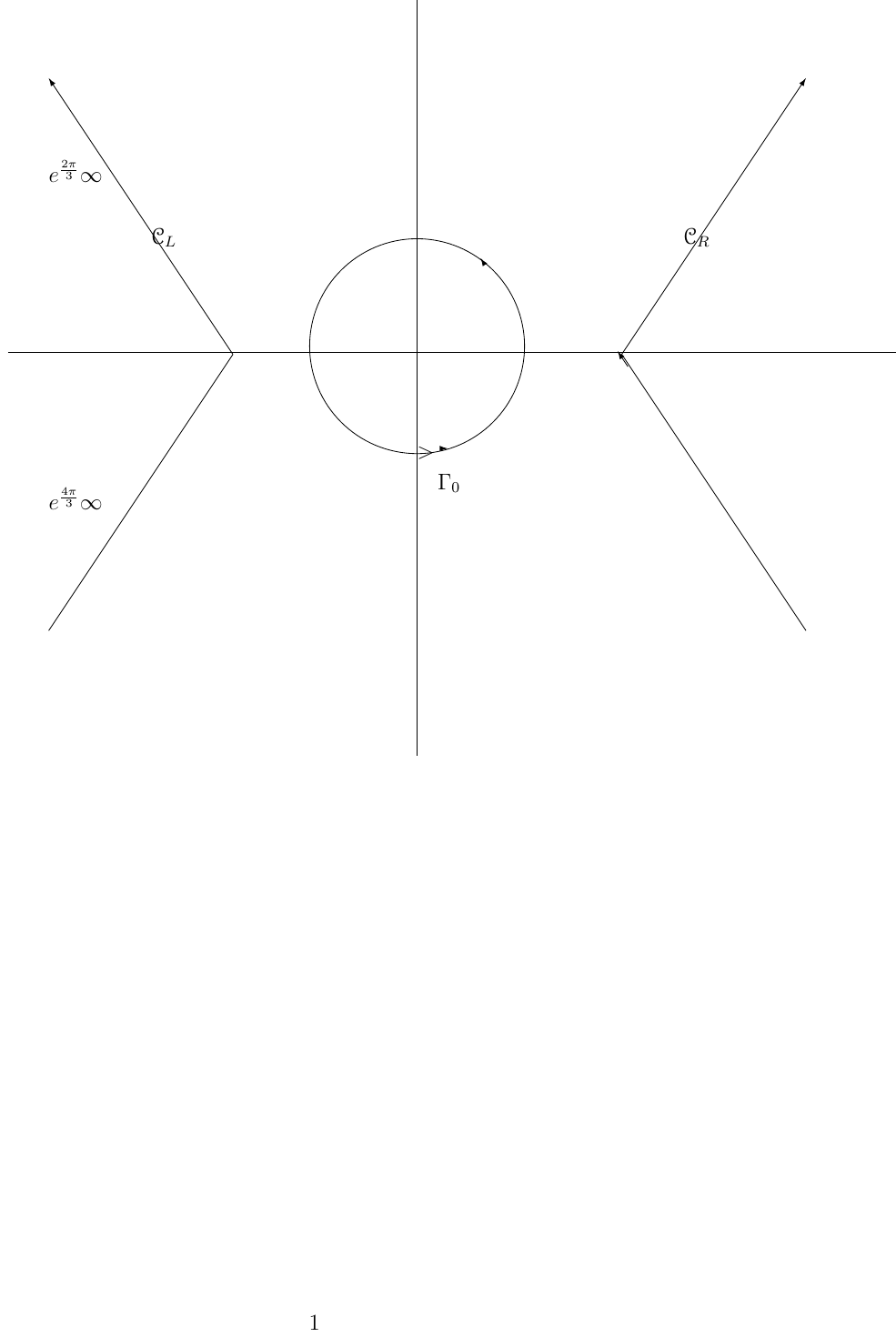} } } } 
\end{picture}

\vspace*{3.9cm}

Figure 2. Contours for the transversal cusp-Airy kernel ${\mathbb L}^{\mbox{\tiny T-cuspAiry}}$. 

\vspace{.5cm}
 
\begin{theorem}  \label{MainTheo}Then
 the following limit holds for all $\tau_i, \in \BZ$, near the upper-cut ($\vr=1$) or the lower-cut ($\vr=-1$), except in the sector $\tau_2<0<\tau_1$, 
 \be\label{limLtac-Ledge}\bl
 \lim_{r\to \infty}(-\vr r^{1/6})^{\tau_1-\tau_2}    &{\mathbb L}^{\mbox{\tiny dTac}} _{r,\rho,\beta} ( \tau_1, \sqrt2 \xi_1 ;\tau_2, \sqrt2\xi_2) \sqrt2d\xi \Bigr|
_{_{{\rho=\beta=0}\atop{\xi_i=\vr\bigl(\sqrt{2r}+\frac{\xi'_i}{\sqrt{ 2} r^{1/6}}\bigr)}}}
\\&~~~~~ =
 {\mathbb L}^{\mbox{\tiny T-cuspAiry}}
 ( \tau_1,  \xi'_1 ;\tau_2,  \xi'_2)d \xi' 
\el \ee
where the kernel of blue dots is given by (with $A_\tau(\xi)$ defined in (\ref{Atau})):
%
\be\bl
 \label{Ledge0}   {\mathbb L}^{\mbox{\tiny T-cuspAiry}}   ( \tau_1,  \xi'_1 ;\tau_2,  \xi'_2) 
 &=\left(-\Id_{0\leq\tau_2< \tau_1}\Id_{\xi'_1\geq \xi'_2 }+ \Id_{ \tau_2< \tau_1\leq 0}\Id_{\xi'_2\geq \xi'_1 } \right)
\frac{(  \xi'_1\!-\!  \xi'_2)^{\tau_1-\tau_2-1}}{(\tau_1-\tau_2-1)!} 
\\&~~~+(-1)^{\tau_2}\int_0^\iy
d\mu A_{\tau_1}(\mu+\xi'_1) A_{-\tau_2}(\mu+\xi'_2).~ \el\ee
The kernel ${\mathbb L}^{\mbox{\tiny T-cuspAiry}}$  has a natural involution $\IR$,
\be\IR:~(\xi'_1,\xi'_2, \tau_1,\tau_2)\to (\xi'_2,\xi'_1, -\tau_2,-\tau_1)
\label{Invol0}
\ee
acting on ${\mathbb L}^{\mbox{\tiny T-cuspAiry}}   ( \tau_1,  \xi'_1 ;\tau_2,  \xi'_2)$ by conjugation and having the virtue of exchanging $ \Id_{0\leq\tau_2< \tau_1}\Id_{\xi'_1\geq \xi'_2 } $ and $   \Id_{ \tau_2< \tau_1\leq 0}\Id_{\xi'_2\geq \xi'_1 }$ in (\ref{Ledge0}):
\be\label{InvolL}\IR\left[{\mathbb L}^{\mbox{\tiny T-cuspAiry}}   ( \tau_1,  \xi'_1 ;\tau_2,  \xi'_2)\right]=(-1)^{\tau_1-\tau_2}
{\mathbb L}^{\mbox{\tiny T-cuspAiry}}   ( \tau_1,  \xi'_1 ;\tau_2,  \xi'_2)
\ee
In the sector $\tau_2<0<\tau_1$, the kernel 
  ${\mathbb L}^{\mbox{\tiny dTac}} _{r,\rho,\beta}$ is oscillatory in the limit $r \to \iy$.
 \end{theorem}
 
 
 \remark
 The invariance under the involution implies that {\em the probabilities are the same}: whether one considers a configuration $(\tau_i,\xi_i)$ to the left of the line $\{\rho\}$ or its mirror image. In going from the upper-cut to the lower-cut, the $\xi$-axis changes direction.


\vspace{.4cm}

\noindent{\bf  This adds another critical behavior to the known critical points: the Transversal-cusp-Airy process :}
\medbreak
 
 \noindent\underline{\em Transversal-Cusp-Airy points}. The simulation in Fig. 4 depicts a hexagon of size $m_1+d+m_2,~b,c,n_1+d+n_2,b,c$ with an upper- cut and a lower-cut, both of size $d$.   
   The new kernel  ${\mathbb L}^{\mbox{\tiny T-cuspAiry}}( \tau_1,  \xi'_1 ;\tau_2,  \xi'_2)$ in  Theorem \ref{MainTheo} describes {\em the fluctuations of the blue tiles along  the lines (labeled $\tau\in \BZ$) parallel to the line $\{\rho\}$ near the arctic curve between the frozen and the stochastic region, in the neighborhood  of each of the cuts}; see the circle in Fig.4a. From the middle of the line $\{\rho\}$ (see black dot in Fig. 4a), we rescale in the neighborhood of $\xi=\sqrt{2r} $ (see tip of the arrow in Fig. 4a) and $\xi=-\sqrt{2r} $ for the scale $ \frac{1}{{\sqrt{ 2}} r^{ 1/6}}$. Figs. 4a and its zoomed version 4b show very clearly the parallel lines to the blue edge of the cut; they intersect the blue tiles, and never pierce the red or yellow ones. See section  10 for further clarifications.  The transversal-Cusp-Airy point  process deserves the {\em ``universality"} label, because it also appears in the context of  tilings of skew-Aztec rectangles, mentioned before; see the forthcoming paper \cite{AvM4}.

  \noindent \newline \underline{\em Turning points}. The arctic boundary in this configuration touches the boundary of the figure at a number of so-called  turning points. At these points, the natural limit process is given by the {\em GUE-minor process}. This leads to a single interlacing set of dots; see \cite{JN,OR1}. 
 
\noindent \newline  \underline{\em Cuspidal points  and split tacnode}.  Instead of considering the strip $\{\rho\}$ formed by  extending the right edge of the lower-cut and the left-edge of the upper-cut, one considers here the strip $(L_1,L_2)$ formed by extending the left-edge of the lower cut and the right edge of the upper-cut. The pictures  Fig.4a and 4b show   two solid phases (blue and red tiles) in the neighborhood of the tip of the cuts, which descend into a cusp in the arctic boundary; here one has a doubly interlacing set of yellow tiles along parallel lines to $ L_1$ (or 
 $L_2$). It seems very reasonable to state  that here {\em the fluctuations of the yellow tiles} are given by the Cusp-Airy process, as discovered by Duse-Johansson-Metcalfe in  \cite{DJM}. Notice here that the parallel lines intersect the yellow tiles, but never the blue or red ones.

\begin{theorem}(Duse-Johansson-Metcalfe \cite{DJM}): The limiting cusp-Airy kernel of yellow dots ${\mathbb L}^{\mbox{\tiny cuspAiry}} $  is given by the formula below for all $\tau_i\in \BZ$ and $\xi'_i\in \BR$: (see Contours in Fig. 3)
\be\label{cusp-Airy'}
\bl
&{\mathbb L}^{\mbox{\tiny cuspAiry}}    ( \tau_1,  \xi'_1 ;\tau_2,  \xi'_2)
\\&~~~~=-\Id_{ \tau_2< \tau_1} \Id_{\xi'_1\geq \xi'_2 }\frac{(  \xi'_1\!-\!  \xi'_2)^{\tau_1-\tau_2-1}}{(\tau_1-\tau_2-1)!} 
+ \int_{\CR_L+\Ga_0^+}dz \int_{\CR_R-\Ga_0} \frac{dv}{v-z}\left(\frac{v^{\tau_2}e^{\frac{v^3}3- v\xi'_2
 }}{ z^{\tau_1} e^{\frac{z^3}3-z\xi'_1  }}\right)\el\ee
Except for the sector $\tau_2<0<\tau_1$, the cusp-Airy kernel ${\mathbb L}^{\mbox{\tiny cuspAiry}}$ above has an expression  reminiscent of (\ref{Ledge0}) in terms of $A^{(0)}_{\tau}(\xi)$ (see (\ref{r-Airy})), rather than $A^{ }_{\tau}(\xi)$  (see (\ref{Atau})) :%
\be\bl
 \label{Ledge}   {\mathbb L}^{\mbox{\tiny cuspAiry}}   ( \tau_1,  \xi'_1 ;\tau_2,  \xi'_2) 
 &=  {{-\Id_{ \tau_2< \tau_1} \Id_{\xi'_1\geq \xi'_2 }}  } 
\frac{(  \xi'_1\!-\!  \xi'_2)^{\tau_1-\tau_2-1}}{(\tau_1-\tau_2-1)!} 
\\&~~~+(-1)^{\tau_2}\int_0^\iy
d\mu A^{(0)}_{\tau_1}(\mu+\xi'_1) A^{(0)}_{-\tau_2}(\mu+\xi'_2).~ \el\ee

  
\end{theorem}

\vspace*{-4cm}

 \setlength{\unitlength}{0.017in}\begin{picture}(0,60)
 \put(120, -150){\makebox(0,0) {\rotatebox{0}{\includegraphics[width=90mm,height=90mm]
  {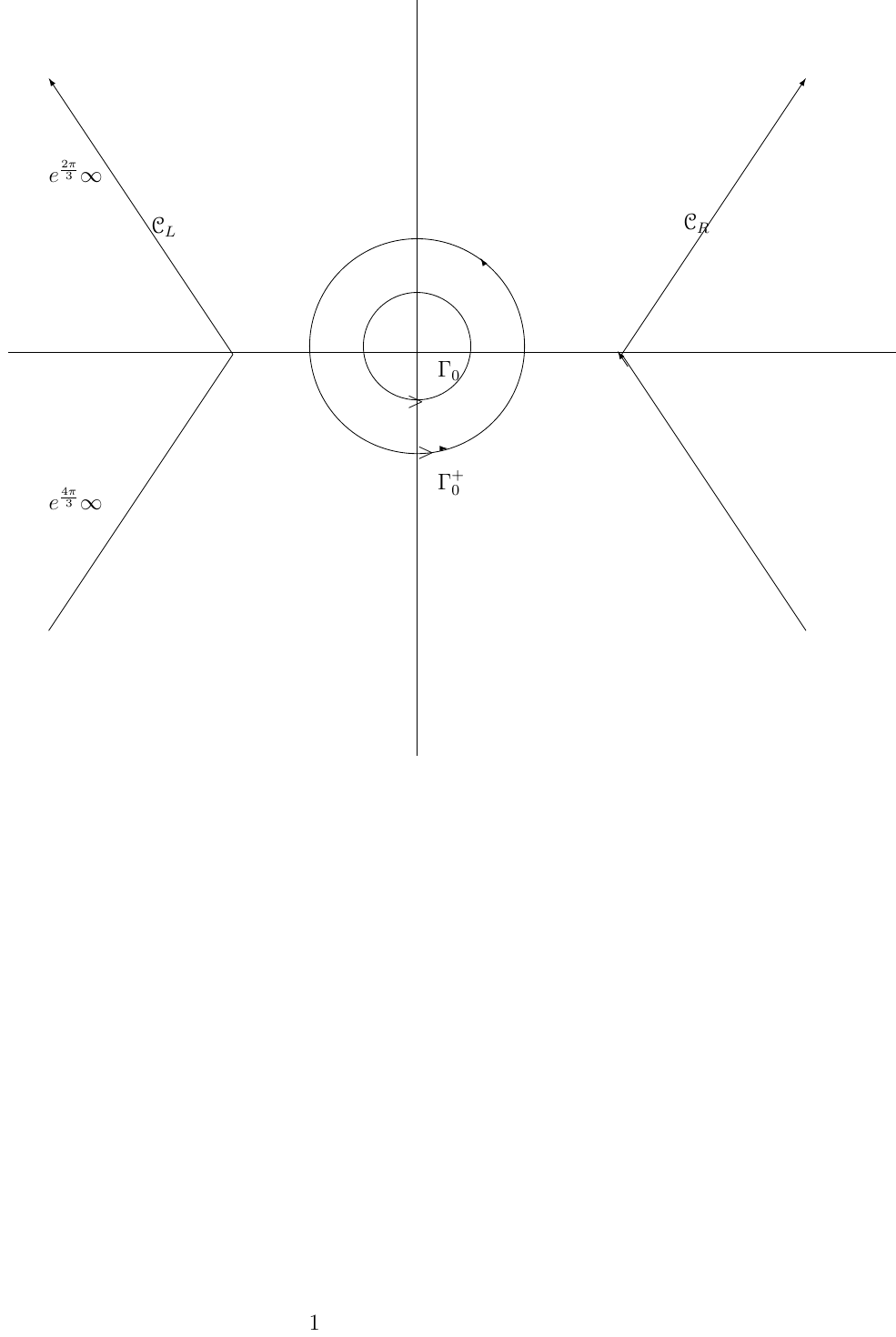}}}} 
\end{picture}

\vspace*{6cm}

 

   Figure 3. Contours for the cusp-Airy kernel.
   
   \vspace*{1cm}
 \remark The cusp-Airy kernel ${\mathbb L}^{\mbox{\tiny cuspAiry}} $, as in (\ref{cusp-Airy'}), satisfies the same involution  identity (\ref{InvolL}) as $ {\mathbb L}^{\mbox{\tiny T-cuspAiry}}$.
 
 Finally, we formulate a conjecture about the case where the strip $\{\rho\}$ of Fig. 5 has width $\rho\geq 0$. In the simulation of Fig.1 this would be obtained by moving the lower cut to the left by $\rho$ steps. In  Fig. 5, the strip $\{\rho\}$ has width $\rho=2$.
 \medbreak
 
  \noindent {\bf Conjecture 2.4}
%
{\em Given the strip $\{\rho\}$ of width $\rho\geq 0$, with $\tau_1\leq 0$ and $\tau_2\geq \rho$, 
the following limit holds, with $A_{\tau}$ as in (\ref{Atau}):
\be\label{mainlimit}\bl\lim_{r\to \iy}
(-r^{1/6})^{\tau_1-\tau_2}&
 {\mathbb L}_{r,\rho,\beta}^{\mbox{\tiny dTac}}  ( \tau_1, \theta_1 ;\tau_2, \theta_2) \sqrt{2} d\xi \Bigr|
_{_{{\rho\geq 0,~~\beta=0}\atop{\xi_i=  \sqrt{2r}+\frac{\xi'_i}{\sqrt{ 2} r^{1/6}} }}}
\\&= (-1)^{\tau_2-\rho}   
\int_0^\iy d\mu A_{\tau_1-\rho}(\mu+\xi'_1)A_{-(\tau_2-\rho)}(\mu+\xi'_2)
;\el\ee
i.e., it is the same formula as for $\rho=0$, except that the levels $\tau_i$ are shifted backwards by $\rho$.}


 A 'convincing proof' of this conjecture will appear in  section 11.

  \newpage


  
    


      
      
 
      
      


          \newpage
          
           \setlength{\unitlength}{0.017in}\begin{picture}(0,60)
\put(120, -70){\makebox(0,0) {\rotatebox{0}{\includegraphics[width=200mm,height=250mm]
 {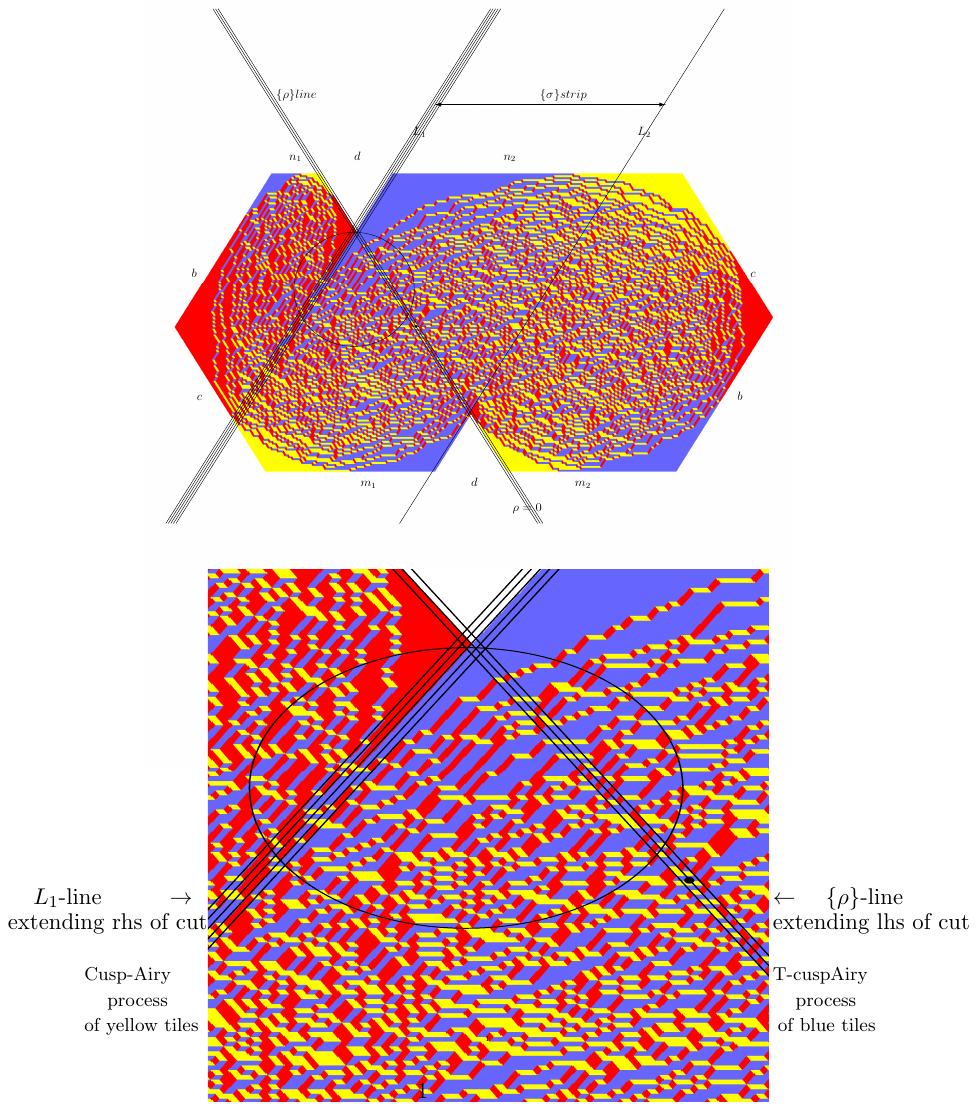}}}}
 \end{picture}
 
 \vspace*{13cm}
 
 Figure 4a. Simulation of Lozenge tilings with $m_1=m_2=70,~~n_1=20,~n_2=120, ~d=30,~b=80,~c=75, \mbox{and so}~r=b-d=50,~\rho=   n_1-m_1+b-d=0$ and $\sigma=m_1-n_1+c-d=95$. For the sake of clarity we have replaced in this simulation the green tiles by yellow ones.       

      Figure 4b. Zooming Fig. 4a near the upper-cut. \\(Courtesy of Christophe Charlier)

          \newpage

\newpage

          \newpage

      \newpage


     \newpage

 
      %
   
  \newpage
   \vspace*{-.8cm}

    \section{Some background on tilings of nonconvex hexagons and the discrete tacnode kernel}

{\em The geometry of non-convex hexagons with cuts and lozenge tilings}
 : This is a brief outline of the results in \cite{AJvM1,AJvM2}. As mentioned, consider the tiling  of the non-convex hexagon ${\bf P}$ (with two opposite cuts of the same size $d$), as in Figs. 1 and 5. The lower cut is at distances $m_1$ and $m_2$ from the lower-left and lower-right corner of the hexagon and the upper cut is at a distance $n_1$ and $n_2$ from the upper-left and upper-right corner of the hexagon.

 \noindent That $m_i$ and $n_i$ satisfy $m_1+m_2=n_1+n_2$ is needed for tilability. The two remaining parallel edges have sizes $b $ and $c\sqrt{2}$, with $N:=b+c$, with $b,c\geq d$ to guarantee tilability. 
 
 The hexagon ${\bf P}$ is embedded into the discrete rectangular  grid of coordinates $(x,n)$, such that the left-most and right-most corners of the grid are given by $x=-d-b-c-1/2$ and $x=m_1+m_2-1/2$; the horizontal dotted lines are parametrized by $n$, with $n=0$ and $n=N$ being the bottom and top-line of ${\bf P}$. Moreover the $x$-origin  $x=0$ is the black dot on the lower-line $n=0$.
%
Filling up the two cuts with red tiles is a way to deal with the cuts.

 Besides the $(m,x)$ coordinates, we have {\em another system of coordinates} $(\eta, \xi)$, and its inverse
    \be\eta=m+x+\tfrac 12,~~~\xi=m-x-\tfrac 12 ~~~~~ \Leftrightarrow ~~~m(\eta,\xi): =\tfrac 12(\eta+\xi),~~x(\eta,\xi):=\tfrac 12 (\eta-\xi-1) .\label{Lcoord}\ee
    with $x$ being the running (integer) variable along the horizontal (dotted) lines in Fig. 5. 
     The origin $(\eta,\xi)=(0,0)$ is the little circle lying $1/2$ to the left of $(n,x)=(0,0)$. 
      
         To this configuration Fig. 5, we associate {\em the strip} of parallel oblique lines $$\{\rho\}=\{\eta \in \BZ \mbox{, such that }m_1\leq \eta\leq m_1+\rho\}$$ 
      of width  \be\label{rho'} \rho=  n_1-m_1+b-d  =m_2-n_2+b-d,\ee
      determined by  extending the oblique line $\eta=m_1$ of the lower cut and the oblique line $\eta=m_1+\rho=  n_1 +b-d$ belonging to the upper cut. It is also of interest to consider the vertical strip  $\{\sigma\}$ formed by the extension of the vertical segments of the cuts (see Fig. 5); this strip has width
      \be\label{sigma}\bl\sigma&:=m_1-n_1+c-d
      \\&=b+c-2d \mbox{, for $\rho=0$}.\el\ee
      
      \newpage
      
 %
%

 
 \newpage

\newpage
\vspace*{-6.4cm}

\setlength{\unitlength}{0.017in}\begin{picture}(0,170)
\put(175,-70){\makebox(0,0) {\rotatebox{0}{\includegraphics[width=140mm,height=200mm]
 {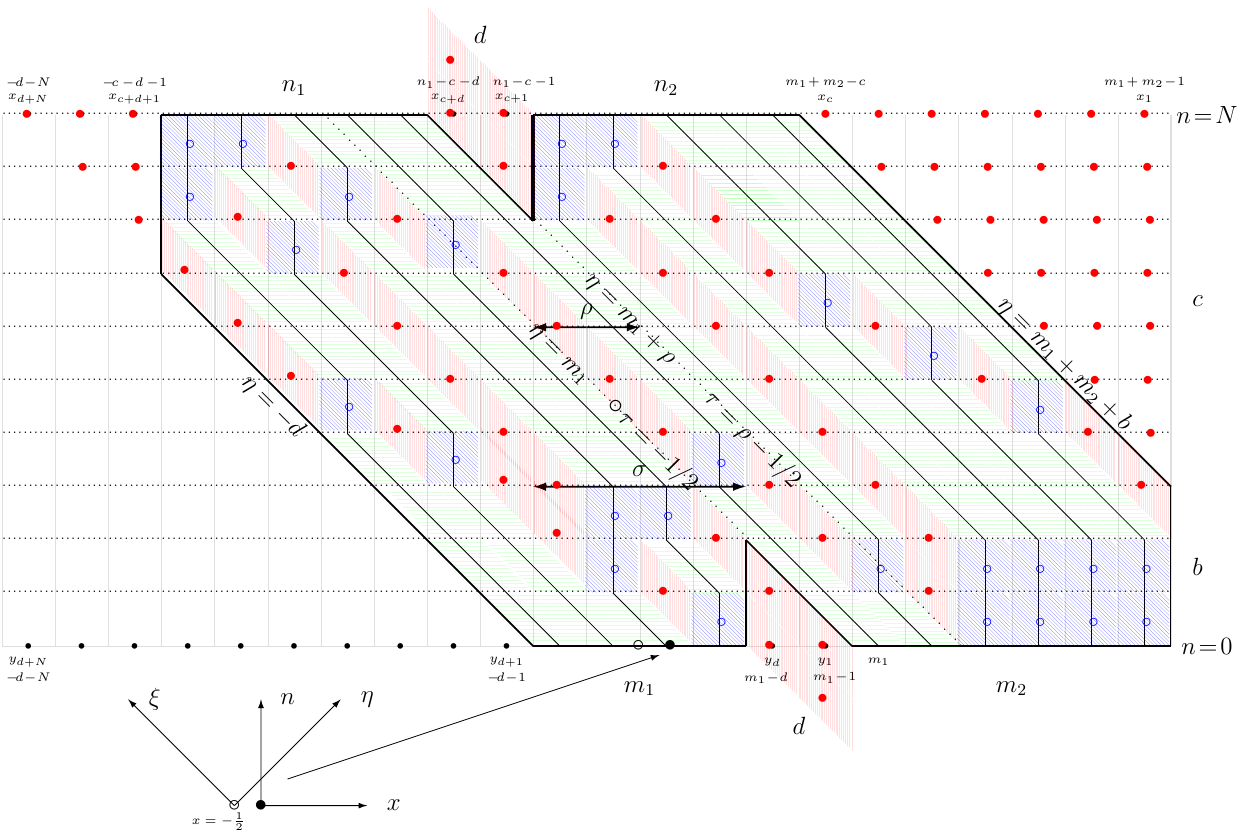} }}}
 
 \begin{picture}(0,170)\put(175,-100){\makebox(0,0) {\rotatebox{0}{\includegraphics[width=140mm,height=200mm]{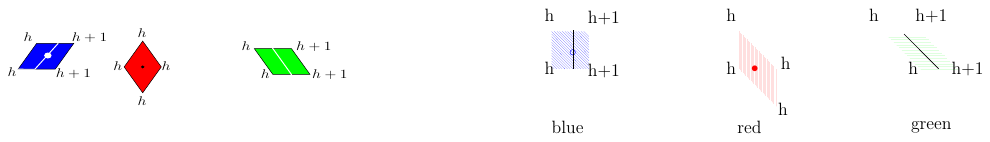} }}}\end{picture}

 

\end{picture}

\vspace*{8.2cm}

Figure 5. The upper figure gives the affine transformation of the tiles to a new set of tiles, with level lines for the appropriate height function. The lower figure gives the tiling of a hexagon, with  two opposite cuts of equal size; a tiling  with red, blue and green tiles, carrying the level lines. Here $d=2$, $n_1=n_2=5,~m_1=4,~m_2=6,~ b=3,~c=7$, and thus $~ r=1,~ \rho=2,~  \sigma =4 .$ The $(x,n)$-coordinates have their origin at the black dot and the $(\xi, \eta)$-coordinates at the small circle given by $(x,n)=(-\tfrac 12,0)$.
%
Blue tiles carry blue dots, belonging to oblique lines $\eta=k$ for $-d+1\leq k\leq m_1+m_2+b-1$. The left and right boundaries of the strip $\{\rho\}$ are given by the dotted oblique lines $\eta=m_1$ and $\eta=m_1+\rho$. 

 \vspace*{1.3cm}

\setlength{\unitlength}{0.015in}
 \begin{picture}(0,130)
 
 \put(188,0){\makebox(0,0){\includegraphics[width=145mm,height=180mm]{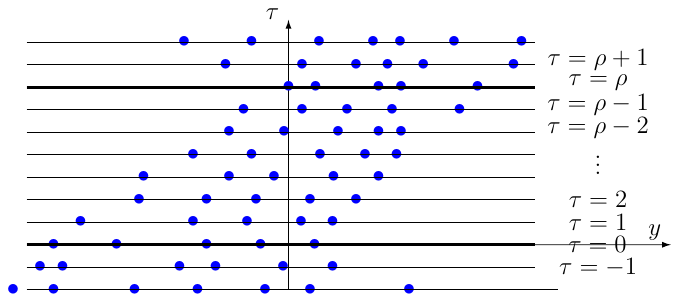}}}
   
 \end{picture}
 
 \vspace*{-2.3cm}
 
 Figure 6. The strip $\{\rho\}$ and the doubly interlacing set of blue dots.

 \newpage

 
   
 
 

 
  \vspace{.5cm}


 %
%
  
  \newpage
  
  \vspace*{1cm}
  
  Putting a blue dot at the center of each blue tile, we now consider the point process  {\bf ${\mathbb L}^{\mbox{\tiny blue}}$ of
 {\bf blue dots}, belonging to the set of oblique lines $\eta=k\in \BZ$}:
 $$ {\mathbb L}^{\mbox{\tiny blue}}(\eta_1,\xi_1;\eta_2,\xi_2),\mbox{ with } (\eta,\xi)=(k,2\ell-k-1),~~(k,\ell)\in \BZ^2,$$
 so that $\eta+\xi\in \BZ_{\tiny\mbox{odd}}.$
%
  The blue dots on the oblique lines doubly interlace in the $\xi$-variable as in Fig. 6, going from left to right. 
   Some height function arguments show that the number $r$ of blue dots along each of the $\rho+1$ parallel oblique lines $\eta=k$ within the strip $\{\rho\}$ (i.e., $m_1\leq \eta\leq m+\rho$) is always the same and equals:
  \be \label{r}r:=b-d.\ee

     The number of blue dots grows by $1$ each time you go one step to the left of the strip, namely along the lines $\{\eta \in \BZ\mbox{, such that }
m_1 -d \leq \eta\leq m_1-1\}$ and this number grows by $1$  when going one step to the right from the strip in the range $\{\eta\in \BZ \mbox{, such that } m_1+\rho+1\leq \eta\leq m_1+\rho+d\}
$. 
Beyond this range the number of blue dots along the oblique lines will go down one by one upon going left and right. As we will see, all the action will take place in the range $\{\eta \in \BZ\mbox{, such that }m_1-d\leq \eta\leq m_1+d+\rho  \}$.

 %
   
.
  
The discrete point process   ${\mathbb L}^{\mbox{\tiny blue}}(\eta_1,\xi_1;\eta_2,\xi_2)$ of blue dots
will in the scaling limit  tend to the 
 discrete-tacnode kernel $ {\mathbb L}_{r,\rho,\beta}^{\mbox{\tiny dTac}}  ( \tau_1, \theta_1 ;\tau_2, \theta_2)$ in the variables $(\tau_i,\theta_i)\in \BZ\times \BR$; see \cite{AJvM1,AJvM2}. It is defined by  the following expression, where the integrations along $  L _{\pm }$ refer to upwards oriented vertical lines   to the right or left of a (counterclock) contour $\Gamma_0$ about the origin and where we assume $r\geq 0$. Later in the paper, we will use the new variable
  $\theta_i=\sqrt{2}\xi_i$ (not to be confused with the $\xi_i$ introduced earlier).
  \be \label{Final0}\begin{aligned}
 {\mathbb L}^{\mbox{\tiny dTac}}  _{r,\rho,\beta}(&\tau_1, \theta_1 ;\tau_2, \theta_2)  
  :=  - 
{\mathbb H}^{\tau_1-\tau_2}(  \theta_2-  \theta_1) 
\\& +\oint_{\Ga_0}\frac{dv} {(2\pi\I)^2}\oint_{ L_{ +}}  \frac{ dz}{z-v}\frac{v^{\rho-\tau_1}}{z^{\rho-\tau_2}}
\frac{e^{-v^2  -  \theta_1    v  }}
{e^{-z^2 - \theta_2   z  }}
      \frac{ \Theta_r( v, z )} { \Theta_r(0,0)}  
      ~~~~~\left\{\bl& { \mbox{$\neq 0$ only} }\\&{\mbox{when $\rho< \tau_1$} }\el\right.
\\& +\oint_{\Ga_0}\frac{dv} {(2\pi\I)^2}\oint_{  L_{ +}}\frac{ dz}{z-v} \frac{v^{  \tau_ 2}}{z^{  \tau_1}}
\frac{e^{ -v^2 +  (\theta_2- \beta  )v  }}{e^{- z^2  + (   \theta_1- \beta    )z  }}
     \frac{ \Theta_r( v  , z )} { \Theta_r(0,0)} ~~~\left\{\bl& { \mbox{$\neq 0$ only} }\\& {\mbox{when $\tau_2< 0$} }\el\right.
\\& +r\oint_{ L_{ +}  }    \frac{dv} {(2\pi\I)^2} \oint_{ L_{ +}}dz \frac{v^{ -\tau_1}}{z^{\rho-\tau_2}}
\frac{e^{ v^2 -    (\theta_1- \beta     )v  }}
{e^{-z^2 - \theta_2    z  }}
    \frac{ \Theta^+_{r-1}(  v, z )} {  \Theta_r(0,0)} ~~~~\mbox{always $\neq 0$}
 \\
 & -\tfrac1{r+1}\oint_{\Ga_0}\frac{dv} {(2\pi\I)^2} \oint_{\Ga_0}dz 
 \frac{v^{ \rho-\tau_1}}{z^{ -\tau_2}}
\frac{e^{-v^2 - \theta_1 v  }}
{e^{ z^2 - (\theta_2- \beta    )z  }}
    \frac{ \Theta^-_{r+1}( v, z  )} {  \Theta_r(0,0)}~\left\{
    \bl
     &  \neq 0 \mbox{ only when}
     \\&\mbox{ $\tau_2<0<\rho<\tau_1$}
  \el
  \right.  
    \\& =:
    \sum_0^4 {\mathbb L}^{\mbox{\tiny dTac}}_i  ,
\end{aligned}\ee
where
  \be   \begin{aligned}
    \BH^{m}(z)&:=\frac{z^{m-1}}{(m-1)!}\Id _{z\geq 0}\Id_{m\geq 1},~~~~(\mbox{Heaviside function})
    \\
 \Theta_r( v, z)&:=\left[     
\prod_1^r\oint_{ L_{ +}}\frac{e^{    2w_\al^2+\beta   w_\al}}{    w_\al  ^{\rho }}
 ~\left(\frac{z\!-\!w_\al}{v\!-\!w_\al}\right) \frac{dw_\al}{2\pi \I}\right]\Dt_r^2(w_1,\dots,w_r)
 \\  \Theta^{\pm}_{r\mp1}(  v, z)     &:=\left[     
\prod_1^{r\mp 1}\oint_{  L_{ +}}\frac{e^{    2w_\al^2+\beta  w_\al}}{    w_\al  ^{\rho }}
 ~\left( ({z\!-\!w_\al} )\ ({v\!-\!w_\al})\right)^{\pm 1} \frac{dw_\al}{2\pi \I}\right]
 \\
 &~~ ~~\qquad~ ~\qquad \qquad~~~~~~~~~~~~~~~~~~~~\times~\Dt_{r\mp1}^2(w_1,\dots,w_{r\mp1}).
 \label{Theta}
\end{aligned}\ee

   

\medbreak

 {\bf The scaling}. Let both, the sides  $b,c,n_1,n_2,m_1,m_2$ and the size $d$ of the two cuts of the polygon $\bf P$ tend  to $ \infty$, {\em keeping $r,\rho$ finite and fixed}. Consider fixed (new) real parameters $1<\gamma<3$,   $a:=2\sqrt{\frac{\ga}{\ga-1}}$, $  \beta_1<0$, ~$  \beta_2, ~ \gamma_1,~ \gamma_2\in \BR$; so $2<\tfrac{\ga+1}{\ga-1}<\iy$. So, let $d,b,c,n_1,n_2,m_1,m_2$ go to $\iy$, such that , : 
\be\begin{array}{lllll}
b=d+r&& c=\ga d
\\
n_i = m_i-(-1)^i(\rho-r) &&  m_i=\tfrac{\ga+1}{\ga-1}
 ( d+\tfrac a2  \beta_i\sqrt{d}+ \ga_i)\mbox{  for  }i=1,2.  
\end{array} \label{geomscale}\ee
For $i=1,2$, we have the following rescaling $$(\eta_i,\xi_i)\in \BZ^2 \to (\tau_i,\theta_i)\in \BZ\times\BR, \mbox{ with $\xi_i-\eta_i\in 2\BZ+1,$}$$ 
   about  the halfway point $(\eta_0,\xi_0)$ along the left boundary of the strip $\{\rho\} $, shifted by $(-\tfrac 12, \tfrac 12)$    (see the circle along the line $\eta=m_1$ in Fig.5):  
\be\label{coordscale}\begin{aligned}
 (\eta_i,\xi_i) &=(\eta_0,\xi_0)+(\tau_i,~\tfrac {\ga+1}a (\theta_i +  \beta_2)\sqrt{d} )\mbox{    with   }
  (\eta_0,\xi_0)  =(m_1,N-m_1 -1).\end{aligned}
\ee
The variables $\tau\mbox{ and }\eta \in \BZ$  are related by $\tau=\eta-m_1$; in particular, $\tau=0$ and $\tau=\rho$ refer to the left most and right most boundary of the strip $\{\rho\}$. With this scaling and after a conjugation, the discrete-discrete kernel ${\mathbb L}^{\mbox{\tiny blue}}$ 
  tends to the discrete-continuous kernel ${\mathbb L}^{\mbox{\tiny dTac}} $, as in (\ref{Final0}), depending only on three parameters, the width $\rho$ of the strip $\{\rho\}$, the number $r=b-d$ of blue dots on the oblique lines in the strip $\{\rho\}$ and the parameter 
$ \beta :=- \beta_1- \beta_2$.

\begin{proposition}{  (Adler-Johansson-van Moerbeke~\cite{AJvM1,AJvM2})}.\label{prop:Final0} Given the geometrical and coordinate scalings (\ref{geomscale}) and (\ref{coordscale})\footnote{We need $d<b$, to guarantee $r>0$.
Moreover, we assume $x_i\geq y_i$ and in particular (see \cite{AJvM1}, p.290 above formula (1) and in  \cite{AJvM1}p.9 formula (1))
$y_d\notin \{ x\mbox{-coordinates of an upper-cut}\} 
$; i.e.,  
  $x_{c+1}<y_d<x_c$. 
%
%
 These conditions are met in the asymptotics (see \cite{AJvM2})
using the expansion of $x_c$, $x_{c+1}$ and $y_d$ in formulas (65) and (64) in [4].} 
, the point process of blue dots, governed by the kernel ${\mathbb L}^{\mbox{\tiny blue}}$ is given in the limit when $d\to \iy$ by the tacnode kernel\footnote{The $1/2$ going with the increment $\Dt \xi_2$ is present because the $\xi_i$'s go up by $2$ for fixed $\eta$.} ${\mathbb L}_{r,\rho,\beta}^{\mbox{\tiny dTac}}$, where $\gamma,a$ are defined in (\ref{geomscale}):
 \be\begin{aligned}
 \lim_{d\to \infty} (-1)^{\tfrac 12 
   (\eta_1+\xi_1-\eta_2-\xi_2)}
&\left(\sqrt{d}\frac{\ga\!+\!1}{2a}\right)^{\eta_2-\eta_1 
 }
{\mathbb L}^{\mbox{\tiny blue}}( \eta_1 ,\xi_1;\eta_2,\xi_2)\frac 12 \Dt\xi_2\Bigr|_{(\eta,\xi)\to (\tau,\theta)}
\\
&=  
{\mathbb L}^{\mbox{\tiny dTac}} _{r,\rho,\beta}(\tau_1, \theta_1;\tau_2, \theta_2)d\theta_2.\end{aligned}
 \label{limit}\ee 
 %
%
This kernel is invariant under the involution :
\be \label{involution}
\tau_1\leftrightarrow \rho-\tau_2\mbox{  and  }\theta_1\leftrightarrow \beta-\theta_2,\ee
with ${\mathbb L}^{\mbox{\tiny dTac}}_1\leftrightarrow {\mathbb L}^{\mbox{\tiny dTac}}_2$ and with ${\mathbb L}^{\mbox{\tiny dTac}}_i$ being self-involutive for $i=0,3,4$.
As indicated in (\ref{Final0}), this kernel ${\mathbb L}_1^{\mbox{\tiny dTac}}$ has support on $\{\tau_1>\rho\}$, ${\mathbb L}_2^{\mbox{\tiny dTac}}$ has support on $\{\tau_2<0\}$ and ${\mathbb L}_4^{\mbox{\tiny dTac}}$ on $\{\tau_1>\rho\}\cap \{\tau_2<0\}$.
 
\end{proposition}

\remark Notice that, from the scaling  (\ref{geomscale}), the lower cut sits asymptotically in the middle of the lower side of $\bf P$ ($m_1\simeq m_2$) and the upper cut at $n_1\simeq m_1-r$ for $\rho=0$. This is so in Fig.4.

 \section{Hermite polynomials, the GUE-kernel and a useful identity}
\subsection{Hermite polynomials}
 The Hermite polynomials $H_j(x)$, the orthonormal Hermite polynomials $h_j(x)$, the monic polynomials $\pi_j(x)$ and the (monic) probabilistic Hermite  polynomials $\Hp_j(x)=\widetilde H_j(x)$ are defined by
 \be\label{Herm}\bl
  H_j(x) &=
j!\oint _{\Ga_0}e^{-z^2+2xz}\frac{dz}{2\pi \I z^{j+1}}\\&=
 2\sqrt{\pi} 2^{j }  \int_{L_+} e^{(w-x)^2 }\frac{w^jdw}{2\pi \I}\stackrel{ }{=}
   (-2\I)^{j }  \int_\BR e^{-(\xi-\I x)^2 }\frac{  \xi ^jd\xi}{\sqrt{\pi}  }\\
&=e^{x^2}\left(-\frac{d}{dx}\right)^j e^{-x^2}=(-1)^jH_j(-x)= 
   2^j x^j+\dots
\\
h_j(x)&=\frac1{c_j} \pi_j(x)=\frac{1}{2^j c_j}H_j(x),\mbox{   with  } c_j=\frac{\sqrt{j!}\pi^{1/4}}{2^{j/2}}
\mbox{   and  } \pi_j(x)=x^j+\dots\el \ee
and\be\label{HermP}\bl
  \Hp_j(x):=\widetilde H_j(x) &:=
j!\oint _{\Ga_0}e^{-\frac{z^2}2+ xz}\frac{dz}{2\pi \I z^{j+1}}
\\& =
2^{-j/2}H_j(\frac{x}{\sqrt{2}})
=e^{x^2/2}\left(-\frac{d}{dx}\right)^j e^{-x^2/2}=x^j+\dots
\el\ee

All these Hermite polynomials   are orthogonal with regard to $e^{-x^2}$ or $e^{-x^2/2}$,
$$\bl
&\int_{\BR}H_j(x)H_k(x)e^{-x^2}dx =\sqrt{\pi}2^j j! \dt_{jk}\mbox{ ,~~ }  \int_{\BR}h_j(x)h_k(x)e^{-x^2}dx =\dt_{jk}\\& \int_{\BR}\pi_j(x)\pi_k(x)e^{-x^2}dx=c_jc_k\dt_{jk},~~~
\int_{\BR}\Hp_j(x) \Hp_k(x) e^{-\frac{x^2}2}dx=\sqrt{2\pi}  j! \dt_{jk}  
\el$$
The Hermite polynomials have the following generating function and a Rodrigues formula (setting for the last formula $ z=-\frac{\I}{\sqrt{2}} y$)
\be \bl\label{genfun}
&e^{-z^2+2xz}=\sum_{j=0}^\infty\frac{H_j(x)}{j!} z^j
,~~~ e^{-\frac{z^2}2+ xz}=\sum_{j=0}^\infty\frac{\Hp_j(x) }{j!} z^j,
\\& e^{\frac {y^2}2-\I \sqrt{2} xy}=\sum_{j=0}^\infty\frac{H_j(x) }{j!} \left(\frac{y}{\I\sqrt{2}}\right)^j. \el\ee
Consider now the integral $\Phi_k^{ \pm}(\eta)$:
  
  \be \label{Phi}\begin{aligned}
 \Phi_k^\pm (\eta) &:=\frac{1}{2\pi \I} \int_{L_{\pm}} \frac{e^{v^2+ 2\eta v}}{v^{k+1}} dv  
  \\&= \frac{\pm1}{  \sqrt{\pi}} 
    \left\{\begin{aligned}
  &   
  \int^\infty_0 \frac {(\pm 2\xi)^k} {k!} e^{- (\xi \mp \eta)^2} d\xi \left( =\int_{\mp\eta}^\iy\frac{(2(\eta\pm \xi))^k}{k!}
  e^{-   \xi  ^2} d\xi\right) \quad ,  k\geq 0
  \\   & (\pm 2)^k e^{-\eta^2}H_{-k  -1}(\mp \eta)  \quad ,\quad k \leq -1
 \end{aligned}\right. \\ 
  \end{aligned}. \ee
  with $\Phi_k^-(\eta)=(-1)^{k+1}\Phi_k^+(-\eta)
  $
  and, in terms of probabilistic Hermite polynomials (\ref{HermP}),
  $$\widetilde  \Phi_k^+(\eta):=\frac{1}{2\pi \I} \int_{  L_{ +}} \frac{e^{\frac{z^2}2+ \eta z}}{z^{k+1}} dv  
=\left\{\bl
&\int^\infty_0 \frac {y^k} {k!} e^{- \frac12 (y - \eta)^2} \frac{dy}{\sqrt{2\pi} }\mbox{~,~ for~}k\geq0
\\&\frac1{\sqrt{2\pi}}e^{-\frac{\eta^2}{2}}\Hp_{-k-1} (-\eta)\mbox{~,~ for~}k\leq -1
\el\right.$$
  %
%
%
%
The following functions will be used throughout: 
\be\label{f}  f(x,u):=\sqrt{2}e^{-u^2+2xu-\frac{x^2}2}\mbox{ and } f_\lb(x,u):=f(x,u)e^{\lb u}\ee
 including the identities:
 \be\label{fder}
\begin{aligned} 
 f(\xi,z )^{-1}  \left(-\frac{\pl}{\pl\xi}\right)^n f( \xi,z)
&=   \Hp_n \left( {\xi-2z} \right) 
\\ 
f(\xi,w )  \left( \frac{\pl}{\pl\xi}\right)^nf(\xi,w )^{-1}
 & =   (-\I)  ^n  \Hp_n
 \left(  { \I(\xi-2w )}  \right) 
 .\el\ee



\noindent Throughout the paper, we use the   Vandermonde determinants:
\be\label{vdm}\bl\Dt_n(v)&:=
  \det(v_{n-1-j}^{ i})_{0\leq i,j\leq n-1}
 =\prod_{0\leq i<j\leq n-1}(v_i-v_j)
\\ \widetilde\Dt_n(v)&:=  \det(v^i_j)_{ {0\leq i,j\leq n-1}}
 =(-1)^{\frac{n(n-1)}{2}}\Dt_n(v)
.\el\ee

%

\subsection{GUE and its Fourier transform revisited}

Define the extended GUE-operators, acting on analytic functions $g(u,w)$ and $g(u,w;\lb)$:\be\bl\KR^{(u,w,\lb)}_{n }&:=2\!\!\int_0^\iy\!\! d\lb\oint_{\Ga_0}\frac{du}{2\pi \I u^n }
\oint_{L_+} \frac{w^ndw }{2\pi \I 
 }
 \mbox{ and }\KR^{(u,w  )}_{n} :=2 \!\!\oint_{\Ga_0}\frac{du}{2\pi \I u^n}
\oint_{L_+}\frac{w^ndw 
 }{2\pi \I  (w\!\!-\!\!u)
 }
  \label{K-oper}\el\ee
%

\noindent Recall the {\em GUE-kernel $K_n(x,y):=K^{  GUE }_n(x,y)$}, the Christoffel-Darboux formula and the notation (\ref{f}) for the $f$ and $f_\lb$:
\be\label{KGUE}\bl
K_n&( x,y) :=
\frac{1}{c^2_{n-1}}
\frac{ \pi_n(x) \pi_{n-1}(y)-\pi_n(y) \pi_{n-1}(x)}{x-y}
  e^{-\frac{x^2+y^2}{2}}
\\&=\sum_{j=0}^{n-1}\frac{1}{c_j^2}\pi_j(x)\pi_j(y) e^{-\frac{x^2+y^2}{2}} =2\sum_{j=0}^{n-1}e^{-\frac{x^2}2}\frac{H_j(x)}{j!}e^{-\frac{ y^2}2}\frac{H_j(y)}{2\sqrt{\pi}2^j} 
\\&\stackrel {*}{=}2\oint_{\Ga_0}\frac{du}{2\pi \I}
\oint_{L+}\frac{dw}{2\pi \I (w-u)}\frac{w^n}{u^n}
\frac{e^{-u^2+2xu-\frac{x^2}2}}{e^{-w^2+2yw-\frac{y^2}2}}
=\KR_n^{u,w}\frac{f(x,u ) }{f(y,w )  }
%
\\&= 
2\int_0^\iy d\lb
\oint_{\Ga_0}\frac{du}{2\pi \I}
\oint_{L+}\frac{dw}{2\pi \I  }\left(\frac{w }{u }\right)^n
\frac{e^{-u^2+2(x+\frac{\lb} 2 )z-\frac{x^2}2}}{e^{-w^2+2(y+\frac{\lb} 2 )w-\frac{y^2}2}}
  =
 \KR^{(u,w,\lb)}_n\frac{f_\lb(x,u)}{f_\lb(y,w)}
\el \ee
The 3rd and 4th equalities follow from expanding $(w-z)^{-1}=\frac{1}{w(1-\frac zw)}=\int_0^\iy e^{-\lb(w-z)}d\lb$ for $|z|<|w|$ and using the first formula (\ref{Herm}) for the first $H_j$ and the second formula  (\ref{Phi}) for $H_j$. From(\ref{KGUE}), the GUE-kernel has the following symmetries:
\be\label{Ksymm}K_n(x,y)=K_n(y,x)=K_n(-x,-y).\ee
 For future use, define the GUE-kernel $\widetilde K_n(  w,  u)$ :
 \be\widetilde K_n(  w,  u):=K_n(-\I \sqrt{2} w,-\I \sqrt{2} u). \label{Ktilde}\ee
Also notice the eigenfunction identity:
\be\label{IntGUE}\int_{\BR}K_N(s,t)  e^{-\frac{   t^2 }2}dt=e^{-\frac{s^2}2}\sum_{k=0}^{N-1}\int_{\BR}h_k(s)h_k(t)e^{-t^2}dt=e^{-\frac{s^2}2}\int_{\BR}h_0(s)h_0(t)e^{-t^2}dt=e^{-\frac{s^2}2},
\ee
or in other coordinates ($t=-\I\sqrt{2}w,~s=-\I\sqrt{2}u$), using (\ref{Ktilde}),
\be 2\pi  \sqrt{2 } \int_{L \pm}
 K_n(-\I\sqrt{2}u,-\I\sqrt{2}w)  
 e^{w^2}\frac{dw}{2\pi \I}=e^{u^2}.
\label{eigenf}\ee
Also, from the fourth equality in (\ref{KGUE}) and (\ref{fder}), it follows readily that the following partials satisfy
\be\bl\label{partialGUE}
&\left(-\frac{\pl }{\pl x}\right)^{k }
 \left(\frac{\pl }{\pl y}\right)^{\ell }
K^{  }_n(x,y)=
\KR_n^{(u,w,\lb)}\frac{f_\lb(x,u)}{f_\lb(y,w)}\Hp_k \left( {x-2u} \right) 
\I^{\ell}\Hp_\ell
 \left(\frac{y-2w}{\I}\right).
 \el\ee
 Define the single and double Fourier transform of  functions $g(x)$ and $g(x,y)$ as follows:
 \be\label{FourierDef}\bl
 \widehat g(\xi,y):=
\int_\BR\frac{dx~e^{\I\xi x}}{\sqrt{2\pi}}
g(x,y)  ,~~
 \widehat{\widehat g}(\xi_1,\xi_2) 
&=\int_{\BR}\frac{dx ~e^{ \I \xi_1 x}}{\sqrt{2\pi}}
\int_{\BR}\frac{dy ~e^{ \I \xi_2 y}}{\sqrt{2\pi}}g(x,y) .
\el\ee

 \begin{lemma}\label{lemma:Fourier}
 The Fourier transform of Hermite polynomials $H_j$ and the function $f(x,u)$, as in (\ref{f}), reads as follows:
 \be \label{Hhat}
 \int_{\BR} \frac{dx}{\sqrt{2\pi}}e^{ \I \xi  x} H_j(x)e^{-\frac{x^2}2}=  \I ^j
 H_j(\xi)e^{-\frac{\xi^2}2},
 \ee
\be \label{Fourierf}\widehat f(\xi,u) =\int_\BR\frac{dx}{\sqrt{2\pi}}e^{\I \xi x}
f(x,u)  =f(\xi,\I u ).\ee
 In the notation (\ref{FourierDef}) and (\ref{KGUE}), the Fourier transform in $x$   of the GUE-kernel  $K_r(x,y)$ reads:
\be\bl \label{GUEhat}
\widehat K_r(\xi,y):=
\int_\BR\frac{dx}{\sqrt{2\pi}}e^{\I\xi x}
K_r(x,y)  
    =\KR_r^{u,w}\frac{f(\xi,\I u ) 
 }{f(y,w )}
,\el\ee
with
\be\label{Khatsym}\widehat K_r(\xi,y)=\widehat K_r( -\xi,-y)=\widehat K_r(y,\xi ),\ee
and the double Fourier transform $\widehat{\widehat K}_r$ of $K_r(x,y)$ equals (see (\ref{FourierDef})):
\be\label{Fourierdouble}\bl\widehat{\widehat K}_r (\xi_1,\xi_2) 
&
  =K_r(  \xi_1,-\xi_2)=K_r( - \xi_1, \xi_2)
.\el\ee


\end{lemma}
\proof Using the fact that the Gaussian integrates to $1$, the Fourier transform (\ref{GUEhat})  follows from
$$\bl  \int_\BR\frac{dx}{\sqrt{2\pi}}e^{\I \xi x}e^{2xu-\frac{x^2}2}=e^{\frac12 (2u+\I \xi)^2}\mbox{ and so }\int_\BR\frac{dx}{\sqrt{2\pi}}e^{\I \xi x}
f(x,u)&=\sqrt2 e^{u^2+2u\I\xi-\frac{\xi^2}{2}}
 \\&=f(\xi,\I u ), 
\el$$establishing (\ref{Fourierf}); this also proves (\ref{GUEhat}), while (\ref{Hhat}) follows from (\ref{genfun}).
 Finally, we have for the double Fourier transform:  
 $$
 \bl 
 \widehat{\widehat K}_r( -\xi_1,-\xi_2) 
  &=  {\int_{\BR} \frac{dx ~e^{-\I \xi_1 x }}{\sqrt{2\pi}} }  \int_{\BR}\frac{dy~e^{-\I \xi_2y}}{\sqrt{2\pi}}
 2\sum_{j=0}^{r-1}e^{-\frac{x^2}2}\frac{H_j(x)}{j!}e^{-\frac{ y^2}2}\frac{H_j(y)}{2\sqrt{\pi}2^j}
 \\&=2\sum_{j=0}^{r-1}(-1)^je^{-\frac{\xi_1^2}2}\frac{H_j(\xi_1)}{j!}e^{-\frac{ \xi_2^2}2}\frac{H_j(\xi_2)}{2\sqrt{\pi}2^j}=K_r(-\xi_1,\xi_2)
. \el$$
Finally we prove that $\widehat K_r(\xi,y)$ is a symmetric function; indeed
$$\bl
\widehat K_r(\xi,y) ={\int_{\BR} \frac{dx ~e^{  \I \xi  x }}{\sqrt{2\pi}} }K_r(x,y)
&=2\sum_{j=0}^{r-1}{\int_{\BR} \frac{dx ~e^{  \I \xi  x }}{\sqrt{2\pi}} }e^{-\frac{x^2}2}\frac{H_j(x)}{j!}e^{-\frac{ y^2}2}\frac{H_j(y)}{2\sqrt{\pi}2^j}\\&=2\sum_{j=0}^{r-1} e^{-\frac{\xi^2}2}
 \frac{( \I)^j H_j(\xi)}{j!}e^{-\frac{ y^2}2}\frac{H_j(y)}{2\sqrt{\pi}2^j}=\widehat K_r(y,\xi )\\&=\widehat K_r(-\xi,-y), \el$$
    ending the proof of Lemma \ref{lemma:Fourier}.\qed
 

%

\subsection{A useful  identity involving the integral operator $\FR^{ \tau}_{  \xi}$ acting on $f(u,z)$}

The reader is reminded of the definitions (\ref{Doper}), (\ref{Eoper''}) and (\ref{FR0}) of $\DR_\xi^{ \tau}$, $ \ER^{  -\tau }_{ \xi }$ and $ \FR^{-\tau}_{\vr \xi} =\DR^{- \tau }_{ \xi }(u )
+   \ER_{  \xi }^{  - \tau   }(v
 )
 \int_{\BR}\frac{d u }{\sqrt{2\pi}}
 e^{ u v 
  }$ and (\ref{f}) of $f(u,z)$.
%
 Notice that defining $g^{-}(u)=g(-u)$, we have that for $\tau<0$
 \be\label{Doper-}\bl\DR_{-\xi}^{- \tau}[g (-u)]=\DR_{-\xi}^{- \tau}(u)[g^-(u)]&=(-1)^\tau\int_{-\xi}^\iy du\frac{(\xi+u)^{\tau-1}}{(\tau-1)!}g(-u)\mbox{,  for $\tau\geq 1$}
 \\&=(-1)^\tau g^{(-\tau)}(\xi)
 \mbox{,  for $\tau\leq 0$}.
 \el\ee
 %
 %
  We now state: ($\BR^{\pm}$ denotes the real line slightly above (below) the origin)
 
 \begin{lemma}\label{lemma:bilinHerm}   The following $y$-integral over $\BR\mp$, depending on $\tau\in \BZ$ , $\xi \in \BR$, $z\in {\mathbb C}$, involving the infinite tail $ E_\tau^{ \xi}(u)$ as in (\ref{tildeF}) and $f(y,z)$ as in (\ref{f}), can be expressed as operator  $\FR^{-\tau}_{  \xi}$ defined in (\ref{FR0}), acting on  the function $f (-\vr u,\I z)$, valid for any  $\vr =\pm 1$:
\be\bl \label{IntcurlyF} 
  \int_{\BR^{-\vr} }\frac{e^{-\frac{y^2}{2}}  E_\tau^{  \xi}( {\I y} )
  dy}{\sqrt{2\pi}(-\I   y)^{\tau }}f(y,  z)
&  = \vr  ^\tau \FR^{-\tau}_{\vr \xi}(u)[f(   -\vr u,  \I   z)] .
\el\ee
For $\tau\leq 0$, we have that the second part $\ER_{\vr\xi}^{-\tau }$ of $\FR^{-\tau}_{\vr \xi}$, acting on $f$, vanishes:
 $$ \ER_{\vr\xi}^{-\tau }(u)[f(-\vr\I u,  z)]=0.$$

\end{lemma}

\proof  The left hand side of (\ref{IntcurlyF}) contains the infinite tail $ E_\tau^{  \xi}( {\I y} )$ of the Taylor series of $e^{\frac{y^2}{2}-\I \xi y  }$; therefore expressible as 
 $
   E_\tau^{  \xi}( {\I y} )
  =   e^{\frac{y^2}{2}-\I \xi y  } -
      F_\tau^{  \xi}( {\I y} )
 ,$ from (\ref{tildeF}). 
 So we need to compute the integral in (\ref{IntcurlyF}) above, using each of the pieces $e^{\frac{y^2}{2}-\I \xi y  }$ (in step 1) and $-
     F_\tau^{  \xi}( {\I y} )$ (in step 2):

\noindent {\em Step 1.} The following identity holds for $\vr=\pm 1$:
 \be\label{step1}\bl
 \int_{\BR^{-\vr} }&\frac{e^{-\frac{y^2}{2}} \left(e^{\frac{y^2}{2}-\I \xi y  }\right) dy}{\sqrt{2\pi}(-\I   y)^{\tau }} f(y,  \I z)= \vr ^\tau {\DR_{\vr  {\xi}} ^{-\tau}}(u)f(\vr u, z)
 .\el\ee
 Indeed, we have for $\tau\geq 1$,  
setting $y=-\I\sqrt{2}v$ in the integral 
$$\bl
 \int_{\BR^{\mp} }&\frac{e^{-\frac{y^2}{2}} \left(e^{\frac{y^2}{2}-\I \xi y  }\right) dy}{\sqrt{2\pi}(-\I   y)^{\tau }} f(y,  \I z) =\int_{\BR\mp}\frac{dy}{\sqrt{\pi}} 
\frac{ e^{\frac{y^2}{2}-\I \xi y } }{(-\I   y)^{\tau }e^{(y-\I z)^2}} 
\\&= \frac{   \sqrt{2}e^{z^2}}{\I(-\sqrt2)^\tau  }
\!\int_{L\pm}\!\frac{dv}{\sqrt{\pi}}\frac{ e^{ v^2+2(  \sqrt{2}z-\frac{\xi}{\sqrt2})v} }{v^\tau  }~~~(*)
\\&=   \mp (\sqrt2)^{\tau+1} e^{z^2}
\int_{\pm \frac{\xi}{\sqrt2}}^\iy du \frac{(\frac{\xi}{\sqrt2}\mp u)^{\tau-1}}{(\tau-1)!}
e^{-(u \mp  \sqrt{2}z)^2}\mbox{,~using (\ref{Phi})}
\\&= \mp (\sqrt2)^\tau\int_{\pm \frac{\xi}{\sqrt2}}^\iy du \frac{(\frac{\xi}{\sqrt2}\mp u)^{\tau-1}}{(\tau-1)!}f(\sqrt2 z,\pm u)\mbox{, for $f$ defined in (\ref{f})}
 \\&\stackrel{**}{=}(\pm1)^\tau (\sqrt2)^\tau\DR_{\pm \frac{\xi}{\sqrt2}}^{-\tau}(u)f(\pm\sqrt2 u, z),~~\mbox{using (\ref{Doper}),~   (\ref{Doper-}),   }
 \\&=(\pm1)^\tau {\DR_{\pm  {\xi}} ^{-\tau}}(u)f(\pm  u, z)
.\el$$
In $\stackrel{**}{=}$, one uses $f(\sqrt2 z, \pm u)=f(\pm \sqrt2 u,z)$ and (\ref{Doper-}). This proves the identity of {\em Step 1} for $\tau\geq 1$.
For $\tau\leq 0$, the integral (*) above equals $(-1)^\tau \Hp_ {-\tau} (\xi-2z)f(\xi,z)$, upon using (\ref{Phi}), which then can be expressed as as a derivative $(\frac{\pl}{\pl \xi})^{-\tau}=\DR_\xi^{-\tau}$ of $f(\xi,z)$, using (\ref{fder},\ref{Doper}), thus proving {\em Step 1}.


%

\medbreak

\noindent
 {\em Step 2.}  We now prove the identity: 
 %
 \be\label{step2} -\int_{\BR^{-\vr} }  \frac{e^{-\frac{y^2}{2}}  F_\tau^{  \xi}( {\I y} )
  dy}{\sqrt{2\pi}(-\I   y)^{\tau }} f(y,  \I z)
 =\vr^\tau  \ER_{\vr \xi}^{-\tau}(u)
f(-\vr\I u,\I z)
.\ee
 Indeed, one checks, using the expression (\ref{tildeF}) of $ F_\tau^{  \xi}( {\I y} )$ in terms of a series of Hermite polynomials $\Hp(- {\xi} )$, combined with their integral representation (\ref{HermP}):
 \be\label{step2'}\bl
%
 - &\int_{\BR^{\mp} }  \frac{e^{-\frac{y^2}{2}}   F_\tau^{  \xi}( {\I y} )dy}{\sqrt{2\pi}(-\I   y)^{\tau }} f(y,  \I z)  = -\int_{\BR_\mp} 
\frac{  F_\tau^{  \xi}( {\I y} )dy}{\sqrt{\pi}(-\I  y)^{\tau }e^{(y-\I z)^2}}\mbox{,~then set $\I y=v$},
\\&=   \frac{      \sqrt{4\pi}  }{ (-1)^{\tau-1}}   
\!\int_{L\pm}\!\frac{dv}{2 \pi \I v^\tau }e^{ (v+z)^2 } 
\sum_{\ell=0}^{\tau-1}v ^\ell
\oint_{\Ga_0}e^{-\frac{w^2}{2}-  \xi w}\frac{dw}{2\pi \I w^{\ell+1}}
\\&=  \frac{     \sqrt{4\pi}  }{ (-1)^{\tau-1}}
\!\int_{L\pm}\!\frac{dv}{2 \pi \I v^\tau }e^{ (v+z)^2 } \oint_{\Ga_0}e^{-\frac{w^2}{2}-\xi w}\frac{dw}{2\pi \I  w^\tau}\frac{w^\tau-v^\tau}{w-v}
\mbox{, (geom. series)}
 \\&=   \frac{      \sqrt{4\pi}  }{ (-1)^{\tau-1}}
\oint_{\Ga_0}\frac{dw}{2\pi \I }  \int_{L\pm}\!\frac{dv}{2 \pi \I  (w-v)}
\frac{e^{ (v+z)^2 } }{e^{ \frac{w^2}{2}+\xi w}}(v^{-\tau}-w^{-\tau})\mbox{, exchanging integrations,}%
\\&=     \sqrt2   \sqrt{2\pi}   \oint_{\Ga_0}\frac{dw~ e^{- \frac{w^2}{2}\mp\xi w}}{2\pi \I (\mp w)^{ \tau}}  \int_{L+}\!\frac{dv~ e^{ (v\pm z)^2}}{2 \pi \I  (w-v)} 
~~~~~~(\dagger)
 \\&=  (\pm 1)^\tau\sqrt{2\pi}\oint_{\Ga_0}\frac{dw~ e^{- \frac{w^2}{2}\mp\xi w}}{2\pi \I (- w)^{ \tau}}  \int_{L+}\!\frac{dv~ e^{ \frac{v^2}2}f(\mp\I v,\I z)}{2 \pi \I  (w-v)}= (\pm 1)^\tau  \ER_{\pm \xi}^{-\tau}
f(\mp\I v,\I z).
\el\ee
The identity ($\dagger$) above is obtained by  setting $w\to -w$ and $v\to -v$, which changes $L_-$ to $L_+$, proving step 2.  

Then one adds the identities  (\ref{step1}) and (\ref{step2}) from {\em step 1} and {\em Step 2}, which (from (\ref{tildeF})) proves the first  identity below, after changing $z\to -\I z$; then,  the second identity follows from 
 $f( \vr x, y)=f( x, \vr y)$:
\be\bl \label{bilinHerm2'} 
  \int_{\BR^{-\vr} }\frac{e^{-\frac{y^2}{2}}  E_\tau^{  \xi}( {\I y} )
  dy}{\sqrt{2\pi}(-\I   y)^{\tau }}f(y,  z)
&  = \vr  ^\tau\left(\DR^{-\tau}_{\vr \xi}(u)[f( \vr  u, -\I z)]+   \ER_{\vr\xi}^{-\tau }(v)[f(-\vr\I v,  z)]\right).
 \\&=\vr  ^\tau \left(\DR^{-\tau}_{\vr \xi}(u)[f(u ,  -\vr\I z )]+ \ER_{\vr\xi}^{-\tau }(v)[f( -\I v,   \vr z)
 \right)
 \\&\stackrel{*}{=}\vr  ^\tau \left(\DR^{-\tau}_{\vr \xi}(u) + \ER_{\vr\xi}^{-\tau }(v)\int_{\BR}\frac{du}{\sqrt{2\pi}} e^{uv}\right) f( u,  -\I \vr z)
. \el\ee
 This establishes (\ref{IntcurlyF}).
 The last equality $\stackrel{*}{=}$ above uses the following Fourier transform (see  (\ref{Fourierf})) : 
$$\bl
\int_{\BR}\frac{du}{\sqrt{2\pi}} e^{uv}f( u,-\I \vr z)
&=\int_{\BR}\frac{du}{\sqrt{2\pi}} e^{\I u(-\I v)}f(  u,-\I \vr z)
 =f(-\I v, \vr z)
.\el$$
This ends the proof of Lemma   \ref{lemma:bilinHerm}. \qed

\begin{corollary}\label{cor:DRf} For all $\tau\in \BZ$,
\be\label{DRf}\bl \DR_x^{ \tau } (\mbox{\tiny$\bullet$})
f(\mbox{\tiny$\bullet$},u)&=
\oint_{L^{-\vr}}\frac{e^{\vr x\al}d\al}{\I \sqrt{2\pi}(\vr \al)^{-\tau}}f(\vr\I \al,\I u)
.\el\ee
\end{corollary}

\proof This follows at once from the identity (\ref{step1}) in {\em step 1} by rotating $\BR^{\mp}\to L^\mp$.\qed

 %

\remark The integrals in the sequence of formulas (\ref{step1}) and (\ref{step2}) have natural bilinear expressions in terms of Hermite polynomials; namely
for $\tau\in \BZ,x \in \BR,$ and also for $\tau=0$:
$$ \bl   \int_{\BR^{-\vr} }\!\!\!  \frac{e^{-\frac{y^2}{2}}   E_\tau^{x\sqrt2}(\I y)
  dy}{\sqrt{2\pi}(-\I   y)^{\tau }} f(y,  \I z)
 &=\!\!
 \int_{\BR\mp}\frac{dy}{\sqrt{\pi}}
\!\frac{  E_\tau^{x\sqrt2}(\I y)
 }{(-\I   y)^{\tau }e^{(y-\I z)^2}}  
 =
(\sqrt{2})^{-\tau} \!\!\!\!\!\!\!\! \sum_{j= \max(0,-\tau)} ^\iy 
\frac{H_{j+\tau}( {x} )H_{j }(z)}{(\tau+j)!(\sqrt{2})^{3j}}
\\&\stackrel{\tau=0}{=}\sum_{j=0}^\iy 
  \frac{H_{j}(x)H_j(z)}{j! ( \sqrt{2})^{3j} } =f(\sqrt2 x,z) , \el$$
 where the last equality for $\tau=0$ follows from formula (\ref{bilinHerm2'}).
 

%









\section{Ratios of functions $\Theta^\pm_r$ in terms of the GUE-kernel for $\rho=0$}

 
%

Define the $j$-fold integral $Z_j$, expressible in terms of the $c_\al^2$, as defined in (\ref{Herm}):
\be \label{Z}Z_j:=\left[\prod_{\al=1}^j\int_{\BR} 
 e^{-x_\al^2}
 \frac{dx_\al}{2\pi \I}\right] \Dt_j(x)^2
= j! \prod_{\al=0}^{j-1}c_\al^2,
 \ee 
 We now show that the ratios of the functions $\Theta$ as in (\ref{Theta}), appearing in the kernel (\ref{Final0}), can be expressed for $\rho=0$ and $\beta=0$ in terms of the GUE-kernel (\ref{KGUE}). Indeed, remembering the short hand notation $\widetilde K_r(  u,  v):=K_r(-\I \sqrt{2} u,-\I \sqrt{2} v)$, as in (\ref{Ktilde}), we have the following Lemma\footnote{with formula II due to Kurt Johansson.}:
\begin{lemma}\label{Theta-KGUE} The following holds for $\rho=\beta=0$ and $r>0$:
$$\bl 
&(I):~~~~
\frac{ \Theta^{ }_{r }(  u,v) }{\Theta _{r }(  0,0)}=1-2 \pi \sqrt{2}(v-u)\oint_{  L_{ +}  } \frac{dz}{2\pi \I}\widetilde K_r(  v,  z)
\frac{e^{z^2-v^2 }}{z-u}
 \\&(II):~~~ 
r\frac{ \Theta^{+}_{r-1}(  u,v) }{\Theta _{r }(  0,0)}=2\pi\sqrt{2} e^{-u^2-v^2}
\widetilde K_r(  u,  v)
%
\el$$
 \be\label{Theta-ratio}\bl
\\&
 (III):~~  
 \frac{-1}{r\!+\!1}  \frac{\Theta_{r+1}^-(u,v )}{\Theta_r(0,0)} \\&~~~~~~~~~= 
2\pi\sqrt{2}
 \left[
\bl 
 &   -\frac{ 1}{2\pi\sqrt{2}} \int_L \frac{dw}{ 2\pi \I }
\frac{e^{ 2w^2}}{(w-u )(w-v )}
\\& +\int_{L_+} \frac{e^{w^2}dw}{ 2\pi \I (w\!-\!v ) }\int_{L_+}  \frac{e^{ z^2}dz}{ 2\pi \I  (z\!-\!u )}
\widetilde K_r(  w, z)
\el
\right] 
.\el\ee

\end{lemma}

%


\proof The proof of this Lemma uses the results of Baik-Deift-Strahov \cite{BDS}, (Theorem 2.10), on expectations of products and ratios of characteristic polynomials in terms of orthogonal polynomials $\pi_k(x)$ and their Cauchy transform (with regard to a measure $d\al(t)$), which applied to a complex function $ f (t)$ is defined by:
\be\label{Cauchy}\widetilde f(z):=\frac{1}{2\pi\I}\int_\BR \frac{f(t)}{t-z }d\al(t) 
,~~~z\in \BC\backslash \BR.\ee
Namely, consider the characteristic  polynomial $$p_n(x, \lb)=\prod_1^n(x-\lb_i)$$ of a random Hermitian matrix (of size $n$) with probability measure
$$\bl d{\mathbb P}_n(\lb)=\frac1{Z_n} \Dt_n^2(\lb_1,\dots \lb_n)\prod_{i=1}^n d\al(\lb_i)
\mbox{, for } d\al(\lb)=e^{-\lb^2}d\lb\el .$$
Then the expectation is expressed in terms of (monic) orthogonal polynomials $\pi_k(x)$ and its Cauchy transform $\widetilde\pi_k(x)$ (see (\ref{vdm}) for the Vandermonde $\widetilde\Dt_k$)
\be\label{BDS}{\mathbb E}_n\left(\frac{\prod_{j=1}^kp(x_j,\lb) }{\prod_{j=1}^\ell p(y_j,\lb) }
\right)=(-1)^{\frac{\ell(\ell-1)}{2}}\frac{\prod_{j=n-\ell}^{n-1}\ga_j}{\widetilde\Dt_k(x)\widetilde\Dt_\ell(y)}\det\left(\bl 
&(\widetilde \pi_{n-\ell+j}(y_i))_{{1\leq i\leq \ell}\atop{0\leq j\leq \ell+k-1}}
\\&
 (\pi_{n-\ell+j}(x_i))_{{1\leq i\leq k}\atop{0\leq j\leq \ell+k-1}}
\el\right)
\ee
  in terms of (see (\ref{Z})):
\be\label{BDS'}\ga_{j-1}=-\frac{2\pi \I}{c_{j-1}^2}=-2\pi \I j \frac{Z_{j-1}}{Z_{j }}\mbox{ and } Z_j=j! \prod_{\al=0}^{j-1}c_\al^2\mbox{ and so } j\frac {Z_{j-1}}{Z_j}=\frac1{c_{j-1}^{ 2}}.\ee

 \noindent{\em Proof of identity (I) in (\ref{Theta-ratio})} : At first set $u=\frac{\I x}{\sqrt2}$ and $v=\frac{\I y}{\sqrt2}$ in the left hand side of (\ref{Theta-ratio})(I). 
 Then apply (\ref{BDS}) to that formula, with $n=r$, $k=\ell=1$. Then we use $\ga_{r-1}=-\frac{2\pi \I}{c_{r-1}^2}$ as in (\ref{BDS'}), the Cauchy transform (\ref{Cauchy}) and the Christoffel-Darboux representation (\ref{KGUE}) of the GUE-kernel and, in the last equality, we use the eigenfunction expression (\ref{IntGUE}),
$$\bl
 \frac{ \Theta^{ }_{r }( \frac{ \I x}{\sqrt2},  \frac{ \I y}{\sqrt2}) }{\Theta _{r }(  0,0)}
 &={\mathbb E}_r\left(\frac{ p(y,\lb) }{  p(x,\lb) }
\right)
   =-\frac{2\pi \I}{c_{r-1}^2}
\det\left(\begin{array}{cccc}
\widetilde \pi_{r-1}(x)&\widetilde \pi_{r}(x)
\\ \pi_{r-1}(y)& \pi_{r}(y)
\end{array}\right)
\\&\hspace*{-1cm}=  \frac{1}{c^2_{r-1}}\int_\BR 
dz\frac{\pi_r(y)\pi_{r-1}(z)-\pi_{r-1}(y)\pi_{r }(z)}{y-z}  
 e^{-\frac{(y^2+z^2)}2}\frac{z\!-\!y}{z\!-\!x}e^{\frac{(y^2-z^2)}2}
\\&\hspace*{-1cm}
= \int_{\BR}K_r(y,z)dz\left(1-\frac{y\!-\!x }{z\!-\!x}\right)e^{\frac{ y^2-z^2 }2}
=1-(y\!-\!x)\int_{\BR}K_r(y,z)\frac{e^{\frac{ y^2-z^2 }2}}{z-x}dz
.\el$$
Then setting $x=-\I \sqrt{2}u,~y=-\I \sqrt{2}v,$ again and $ z=-\I \sqrt{2}z'$ yields the first identity (\ref{Theta-ratio}) for 
$\frac{ \Theta^{ }_{r }(  u, v) }{\Theta _{r }(  0,0)}$, after inserting $1/2\pi \I$ in the integral and dropping the $'$ in $z'$.

\medbreak

\noindent{\em Proof of identity (II) in (\ref{Theta-ratio})}: 
Using the same substitution for $(u,v) $ as above, noticing that 
  $\Theta^+_{r-1}$ and $\Theta_r(0,0)$ contain $1/(2\pi \I)^{r-1}$ and $1/(2\pi \I)^{r }$ and applying (\ref{BDS}) to the expectation for $n=r-1$ and $k=2$,~ $\ell=0$, with $rZ_{r-1}/Z_r$ as in (\ref{BDS'}), we have:  
$$\bl
\frac {r\Theta^+_{r-1}(\frac{\I x}{\sqrt 2},\frac{\I y}{\sqrt 2})}{\Theta_r(0,0)}
 &  =   2\pi \I   
\frac{ (-\I \sqrt2)^{r^2}}{(-\I \sqrt2)^{r^2-1}}\frac{rZ_{r-1}}{Z_r }
{\mathbb E}_{r-1} \left(p_{r-1}(x,\lb)p_{r-1}(y,\lb)
\right)
\\&\hspace*{-1cm}=2\pi \sqrt2 c^{-2}_{r-1}
{\mathbb E}_{r-1} \left(p_{r-1}(x,\lb)p_{r-1}(y,\lb)
\right)
\\&\hspace*{-1cm}=2\pi \sqrt2 c^{-2}_{r-1}
\frac{1}{ y-x }
\det\left(
\begin{array}{cccc}
\pi_{r-1}(x)&\pi_{r}(x)
\\\pi_{r-1}(y)&\pi_{r}(y)\end{array}
\right)
  =2\pi \sqrt2e^{\frac{x^2+y^2}{2}}K_{r }(x,y),
\el$$
  ending the proof of identity II, upon resubstituting $x=-\I \sqrt{2}u,~y=-\I \sqrt{2}v $.
 
 \medbreak
 

\noindent{\em Proof of identity (III) in (\ref{Theta-ratio})}: 
As before,
using $\frac{Z_{r+1}}{Z_r(r+1)}=c_r^2$, applying (\ref{BDS}) for $n=r+1,~k=0,~\ell=2$, noticing that in that expression we have $\ga_{r-1}\ga_r=\frac{(2\pi \I )^2}{c_{r-1}^2c_r^2}$, 

$$\bl
\frac { \Theta^-_{r+1}(\frac{\I x}{\sqrt 2},\frac{\I y}{\sqrt 2})}{(r+1)\Theta_r(0,0)}
 &  = \frac1{  2\pi \I  } 
\frac{ (-\I \sqrt2)^{r^2}}{(-\I \sqrt2)^{r^2-1}}\frac{ Z_{r+1}}{(r+1)Z_r }
{\mathbb E}_{r+1} \left(\frac{1}{p_{r+1}(x,\lb)p_{r+1}(y,\lb)}
\right)
\\&=-\frac{c_r^{ 2} }{\pi\sqrt2  }   
{\mathbb E}_{r+1} \left(\frac1{p_{r+1}(x,\lb)p_{r+1}(y,\lb)}
\right)\\&=\frac{c_r^{ 2}}{\pi\sqrt2 } 
\frac{(2\pi \I)^2}{c_r^2c_{r-1}^2 (y-x)  }
 \det\left(\begin{array}{cccc}
\widetilde \pi_{r-1}(x)&\widetilde \pi_{r}(x)
\\\widetilde \pi_{r-1}(y)& \widetilde \pi_{r}(y)
\end{array}\right)
.\el$$
Then, using (55) and again the Christoffel-Darboux representation (\ref{KGUE}) of the GUE-kernel,  the equalities above continue as follows. In the last equality one may replace $\frac{1}{t-x}$ by $\frac{1}{s-x}$, because the integrand is symmetric in $s$ and $t$. So,
$$\bl
 =&\frac{ 1}{\pi\sqrt2( y-x) } 
\int_\BR dt\int_\BR ds \frac{\pi_{r-1}(t)\pi_{r  }(s)-\pi_{r}(t)\pi_{r-1 }(s)}{ c_{r-1}^2(t-x)(s-y)}e^{-t^2-s^2}
\\ = &\frac{1}{\pi\sqrt2( y-x )} 
\int_\BR dt\int_\BR ds~ e^{-\frac{t^2+s^2}2}
K_{r}(t,s)\frac{s-t}{(t-x)(s-y)}
 \\=& 
  \frac{1}{ \pi\sqrt2(y-x)  } 
\int_\BR dt\int_\BR ds~ e^{-\frac{t^2+s^2}2}
K_{r}(t,s)\left(\frac{y-x}{(t-x)(s-y)}-\frac{1}{s-y}+\underbrace{\frac{1}{t-x}}_{\to \frac{1}{s-x} }\right)
\\&
= \frac{ 1}{\pi\sqrt2}
 \int_\BR dt\int_\BR  ds~ e^{-\frac{t^2+s^2}2}
 K_{r}(t,s)\left(\frac{1}{(t-x)(s-y)}- \frac{1}{(s-x)(s-y)}    \right)
.\el$$
 Then one uses (\ref{IntGUE}) to integrate out the $t$-integral in the second term in the brackets. This gives:
$$\bl   &
 \frac { \Theta^-_{r+1}(\frac{\I x}{\sqrt 2},\frac{\I y}{\sqrt 2})}{(r+1)\Theta_r(0,0)}=  \frac{ 1}{\pi\sqrt2}
 \left(\int_\BR dt\int_\BR  ds~ \frac{e^{-\frac{t^2+s^2}2}K_{r}(t,s)}{(t-x)(s-y)}
 {}   -\int_\BR ds\frac{e^{-s^2}}{(s-x)(s-y)}     
 \right).
\el$$
Finally, one uses the change of variables $x=-\I\sqrt2 u,~y=-\I\sqrt2 v,~t=-\I\sqrt2 z,~s=-\I\sqrt2 w$; this leads to identity (III) in (\ref{Theta-ratio}), thus ending the proof of Lemma \ref{Theta-KGUE}.\qed



\section{The discrete-tacnode kernel for $\rho=0$}

This will be done in two stages:

\subsection{The expression for  the  kernel  ${\mathbb L}^{\mbox{\tiny dTac}}_{r,\rho,\beta}$ for $\rho=\beta=0$}

%

This section deals with specializing and reorganizing -for the sake of asymptotics- the kernel  ${\mathbb L}^{\mbox{\tiny dTac}}_{r,\rho,\beta} (\tau_1,\theta_1;\tau_2,\theta_2)$, as in (\ref{Final0}), for $\rho=\beta=0$. 
 We state the following  Proposition, using $E^{ \xi }_{ \tau }(u),~F^{ \xi }_{ \tau }(u)$ as in (\ref{tildeF}), and the GUE kernel $K_r$ as in (\ref{KGUE}):

 \begin{proposition} \label{prop:L-kernel}Setting $\rho=\beta=0$, the discrete tacnode kernel  ${\mathbb L}^{\mbox{\tiny dTac}}_{r,\rho,\beta} $ becomes \footnote{The $\pm$ in ${\mathbb L}^{\pm}_{1,i} $ refers to the $v$-integration about the complex line $L_\pm$, as will be important later.}
 \be\label{L-kernel}\bl
     & ( \sqrt 2)^{\tau_2-\tau_1}   {\mathbb L}^{\mbox{\tiny dTac}}_{r,\rho,\beta}  (\tau_1, \xi_1\sqrt2;\tau_2, \xi _2\sqrt2)\sqrt2 d\xi\Bigr|_{\rho=\beta=0}  \stackrel{*}{=}  -     \BH^{\tau_1-\tau_2}(\xi_2-\xi_1)  d\xi 
   \\&+   \oint_{ 
    L_{+}} 
   \frac{e^{\frac{v^2}2}dv}{2\pi \I v^{\tau_1-\tau_2}}   \left(e^{ \xi_2 v}  F^{\xi_1}_{\tau_1}(v) +e^{-\xi_1v}   F^{-\xi_2}_{-\tau_2}(v) 
  -e^{  \frac{v^2}2} F^{\xi_1}_{\tau_1}(v)  F^{-\xi_2}_{-\tau_2}(v)
   \right)  d\xi 
 \\&  + \int_{L+}  
  \frac{  e^{ \frac{u^2}2}du}{ \sqrt{2\pi} \I  u^{\tau_1}} \int_{L+} \frac{ e^{\frac{v^2}2}dv}{ \sqrt{2\pi} \I  v^{-\tau_2}}  K_r( -\I u,-\I v )  E^{ \xi_1}_{ \tau_1}(u)   E^{-\xi_2}_{-\tau_2}(v)d\xi
%
%
 %
 %
 \\&=
 \sum_{i=0}^2 {\mathbb L}_i
 (\tau_1,\xi_1;\tau_2,\xi_2)  d\xi=
\left( {\mathbb L}_{0} +( \sum_{i=1}^3{\mathbb L}^+_{1i})+ {\mathbb L}_{ 2}
\right) (\tau_1,\xi_1;\tau_2,\xi_2) d\xi. \el\ee
This uses the notation ${\mathbb L} _{i}$ of the kernel, as announced in (\ref{prodop''}) of Theorem \ref{Th:prodop''}. This kernel is invariant under the involution below, with ${\mathbb L}^+_{11}\leftrightarrow{\mathbb L}^+_{12}$ and with ${\mathbb L}^+_{13}$  and ${\mathbb L} _{2}$ self-involutive,
  \be\label{invol'}
  \xi_1\leftrightarrow -\xi_2\mbox{ and }
  \tau_1\leftrightarrow -\tau_2 
 . \ee
 In the sector $\tau_1<0<\tau_2$ the kernel reduces to ${\mathbb L} _{2}$.
\end{proposition}%

\remark The expression for the kernel (\ref{L-kernel}) (just after $\stackrel{*}{=}$)   can be rewritten  concisely, in terms of $E^{ \xi }_{ \tau } ,~F^{ \xi }_{ \tau } $ (as in (\ref{tildeF})), $\HR$ and the GUE-kernel $K_r$: $$( \sqrt 2)^{\tau_2-\tau_1}   {\mathbb L}^{\mbox{\tiny dTac}}_{r,\rho,\beta}  
 \sqrt2~\big |_{\rho=\beta=0}
\!=\!-     \BH^{\tau_1-\tau_2}(\xi_2-\xi_1)  
  + \oint_{L_+}\!\frac{e^{y^2/2}}{2\pi\I y^{\tau_1-\tau_2}}
 \left(E^{\xi_1}_{\tau_1}  \HR 
   \!+\!
 F^{\xi_1}_{\tau_1}e^{\xi_2 y}\right),$$
 where
 \be\label{H'tilde}\bl  \HR (y) 
  &=   e^{\frac{y^2}2}{  F_{-\tau_2}^{-\xi_2}(y)} 
  + 2\pi\oint_{L^+}\left(\tfrac{v}{y}\right)^{\tau_2}  \frac{dv}{2\pi \I} {e^{\frac{v^2}{2}    }    E_{-\tau_2}^{-\xi_2}(v)  }   K_r( -\I y,-\I v).
 \el \ee

It will be more convenient to carry out this computation for the discrete tacnode kernel in the $\theta_i$-coordinates. For doing so, we need expressions $\widetilde F^{\theta}_ \tau(u)$ and $\widetilde E^{\theta}_ \tau(u)$, in a different scale, such that $\widetilde F^{\sqrt2\xi}_{\tau}(\frac u{\sqrt2})= F^{ \xi}_{\tau}(u)$ and $\widetilde E^{\sqrt2\xi}_{\tau}(\frac u{\sqrt2})=   E^{ \xi}_{\tau}(u)$, in terms of  
 (\ref{tildeF}); namely
\be\bl \label{EF}
 \widetilde F^{\theta}_ \tau(u)&:=\bigl[e^{-u^2-\theta  u}\bigr]_{ {[0,\tau-1 ]}}= \sum_{\ell=0}^{\tau-1}\frac{u^\ell}{\ell!}H_\ell(-\tfrac{\theta}2)=
 {e^{-u^2-\theta  u}} - {\widetilde E^{\theta}_ \tau(u)} 
,\el\ee 
 with $F^{\theta}_ \tau(u)=0$, when $\tau\leq 0$. Then, for an entire $f(v)$ we have the following identity, using the contour $\Ga_\iy\simeq \Ga_0+\Ga_{\{u\}}$ and so $\Ga_0\simeq -\Ga_{\{u\}}+\Ga_\iy$, \be\label{Gamma0}
\oint_{\left\{{\Ga_0}\atop{\Ga_\infty}\right\}}\frac{f(v)
 dv} { 2\pi\I v^{\tau }( u-v )}=\{\pm\} u^{-\tau}\left[f(u)
  \right]_{\left\{{[0,\tau-1]}\atop{[\tau,\infty]}\right\}}
\mbox{ for }\left\{{u\mbox{ outside }\Ga_0}\atop{u\mbox{ inside }\Ga_\iy}\right\},\ee
   the following integral representation follows, given that $u\mbox{ outside }\Ga_0$ for the first formula and $u\mbox{ inside }\Ga_\iy$ for the second formula:
\be\label{EF'}
\frac{\widetilde F^{\theta}_ \tau(u)}{u^\tau}  = 
\oint_{ \Ga_0 }\frac{dv} { 2\pi\I }\frac{e^{ -v^2  -  \theta     v  }}{v^{\tau }(u-v )} 
\mbox{ and }
\frac{\widetilde E^{\theta}_ \tau(u)}{u^\tau}  = 
\oint_{ \Ga_\iy }\frac{dv} { 2\pi\I }\frac{e^{ -v^2  -  \theta     v  }}{v^{\tau }(v-u)} ,
\ee
and similarly in the $ F^{\xi}_ \tau$-world,
\be\label{EFtilde}
\frac{  F^{\xi}_ \tau(u)}{u^\tau}  = 
\oint_{ \Ga_0 }\frac{dv} { 2\pi\I }\frac{e^{ -\frac{v^2}2  -  \xi    v  }}{v^{\tau }(u-v )} .
\ee

%

\begin{lemma} \label{lemma:Lkernel'}
For any $\tau_i\in\BZ$, and $\rho=\beta=0$, the kernel ${\mathbb L}^{\mbox{\tiny dTac}}_{r,\rho,\beta}$ becomes, using the notation for the terms ${\mathbb L}^{\mbox{\tiny  }}_{i}$ in\footnote{The terms ${\mathbb L}^{\mbox{\tiny dTac }}_{i}$ in (\ref{Final0}) do not correspond to the ${\mathbb L}_i$ in (\ref{L-kernel}).} (\ref{Final0}) : 
  \be\label{L-kernel'}\bl
    & {\mathbb L}^{\mbox{\tiny dTac}}_{r,0,0} d\theta :=    {\mathbb L}^{\mbox{\tiny dTac}}_{r,\rho,\beta} (\tau_1,\theta_1;\tau_2,\theta_2)d\theta\Bigr|_{\rho=\beta=0}\\  &=   - \BH^{\tau_1-\tau_2}(\theta_2-\theta_1)d\theta\hspace*{6.9cm}
 ( =\LR_0 d\theta)
\\ &~~+\oint_{ 
    L_{+}} \! \! \frac{e^{u^2 }du}{2\pi \I u^{\tau_1-\tau_2}}\left(  e^{\theta_2u  }\widetilde F^{\theta_1}_{\tau_1}(u) \! + \! e^{-\theta_1u  }\widetilde F^{-\theta_2}_{-\tau_2}(u)-e^{u^2}\widetilde F^{\theta_1}_ {\tau_1}(u)\widetilde F^{-\theta_2}_ {-\tau_ 2}(u)
   \right)d\theta  
   \hspace*{.2cm} ( =\LR_1d\theta)
    \\&  
~~  +2\pi\sqrt{2}\int_{L+}  
 \frac{  e^{ u^2}du}{ 2\pi \I u^{\tau_1}} \int_{L+} \frac{ e^{v^2}dv}{ 2\pi \I v^{-\tau_2}}\widetilde K_r( u,v ) \widetilde E^{ \theta_1}_{ \tau_1}(u)\widetilde E^{-\theta_2}_{-\tau_2}(v)d\theta
  \hspace*{.7cm} (=\LR _{2} d\theta)
 \hspace*{5cm}~~ (=\LR _{3}d\theta )
%
%
 %
 %
 \\&=:\sum_{i=0}^2 \LR_i 
 (\tau_1,\tfrac{\theta_1}{\sqrt2};\tau_2,\tfrac{\theta_2}{\sqrt2})
d\theta
 =:\sum_{i=0}^2 \LR_i 
 (\tau_1,\xi_1;\tau_2,\xi_2)\sqrt2 d\xi,
  \el\ee
  with (the $\pm$ sign in $\LR^\pm_{1i}$ refers to the integration over the complex $L_{\pm}$)
  $$\LR_1=\LR^+_{11}+\LR^+_{12}+\LR^+_{13}.$$
   This kernel is invariant under the involution, 
  \be\label{invol'}
  \theta_1\leftrightarrow -\theta_2\mbox{ and }
  \tau_1\leftrightarrow -\tau_2,
  \ee
  with $\LR^+_{11} \leftrightarrow\LR^+_{12} $ and the other $\LR_i$'s being self-involutive.
  %

\end{lemma}

\proof
  Here all $ {\mathbb L}^{\mbox{\tiny dTac}}_i$ are $\neq 0$.
So we have the following:

\medbreak

\noindent\underline{\em Step 1:}  So, using the above and substituting in (\ref{Final0}) the formulas (\ref{Theta-ratio}) for the ratios of the $\Theta$'s  in (\ref{Final0}), we have the following expressions for ${\mathbb L}^{\mbox{\tiny dTac}}_i$; in some of these integrals below one can move the line $L_+\to L_-+\Ga_0^+$, which will produce a residue. So, we have, using the notation $\LR_i$ given in (\ref{L-kernel'}),
 $$\bl {\mathbb L}^{\mbox{\tiny dTac}}_0    =&- \BH^{\tau_1-\tau_2}(\theta_2-\theta_1) =\LR_0  
  \\
 {\mathbb L}^{\mbox{\tiny dTac}}_1
 =& \oint_{\Ga_0}\frac{dv} {(2\pi\I)^2}\oint_{  L_{ +}}  \frac{ dz}{z-v}\frac{v^{ -\tau_1}}{z^{ -\tau_2}}
\frac{e^{-v^2  -  \theta_1    v  }}
{e^{-z^2 - \theta_2   z  }}
      \frac{ \Theta_r( v, z )} { \Theta_r(0,0)} 
\\=&\oint_{\Ga_0}\frac{dv} {(2\pi\I)^2}\oint_{  L_{ + }}  
   \frac{ dz}{z-v}\frac{v^{ -\tau_1}}{z^{ -\tau_2}}
\frac{e^{-v^2  -  \theta_1    v  }}
{e^{-z^2 - \theta_2   z  }}
\\
&-2\pi\sqrt{2}
\oint_{\Ga_0}\frac{dv} { 2\pi\I }\oint_{  L_{ +}}\frac{  dz}{2\pi \I}\frac{v^{ -\tau_1}e^{   -  \theta_1    v  }}{z^{ -\tau_2}e^{  - \theta_2   z  }}
\oint_{  L_{ +}=  L_{ - }+\underbrace{  \mbox{\footnotesize$\Ga_0^+$}}_{* }  } \frac{du}{2\pi \I}\widetilde K_r( u, z)\frac{e^{u^2-v^2}}{u-v}
\\
{\mathbb L}^{\mbox{\tiny dTac}}_3   =&
 \oint_{  L_{ +}  }    \frac{dv} {2\pi\I} \oint_{  L_{ +}} \frac{dz} {2\pi\I} \frac{v^{ -\tau_1}}{z^{ -\tau_2}}
\frac{e^{ v^2 -     \theta_1 v  }}
{e^{-z^2 - \theta_2    z  }}
    \frac{r \Theta^+_{r-1}(  v, z )} {  \Theta_r(0,0)} 
\\
&= 2\pi\sqrt{2} \oint_{  L_{ +} = \underbrace{ \mbox{\footnotesize $\Ga_0^+$}}_{*} +  L_{  -} }    \frac{dv} {2\pi\I} \oint_{ L_{ +}} \frac{dz} {2\pi\I} \frac{v^{ -\tau_1}e^{   -  \theta_1    v  }}{z^{ -\tau_2}e^{  - \theta_2   z  }}
\widetilde K_r(  v, z).%
\el$$
  Summing up these three contributions, cancels the  two (*)-terms, as is seen from computing each of the residues in ${\mathbb L}^{\mbox{\tiny dTac}}_1$. Then the first $dv dz $-integrals over $\Ga_0$ and $L_+$ in ${\mathbb L}^{\mbox{\tiny dTac}}_1$ give $\LR_{11}^+$ in (\ref{L-kernel'}), upon using the integral representation (\ref{EF'}) of $\widetilde  F^{\theta_1}_{\tau_1}(u)$. This gives the following expression for:
  %
   %
 \be \label{L013}\bl
 &{\mathbb L}^{\mbox{\tiny dTac}}_0+ 
 {\mathbb L}^{\mbox{\tiny dTac}}_1+
 {\mathbb L}^{\mbox{\tiny dTac}}_3
  \\&=
\LR_0+\LR^+_{11}
  -2\pi\sqrt{2}
\oint_{\Ga_0=\Ga_\iy-\underbrace{\Ga_{_{\!\{u\}}}}_{(\dagger)}}\frac{dv} { 2\pi\I }\oint_{  L_{ +}}\frac{  dz}{2\pi \I}\frac{v^{ -\tau_1}e^{   -  \theta_1    v  }}{z^{ -\tau_2}e^{  - \theta_2   z  }}
\oint_{   L_{ - }   } \frac{du}{2\pi \I}\widetilde K_r(  u,  z)\frac{e^{u^2-v^2}}{u-v}
\\&
~~~ +2\pi\sqrt{2} \oint_{   L_{  -} }    \frac{dv} {2\pi\I} \oint_{  L_{ +}} \frac{dz} {2\pi\I} \frac{v^{ -\tau_1}e^{   -  \theta_1    v  }}{z^{ -\tau_2}e^{  - \theta_2   z  }}
\widetilde K_r(  v,  z) ~~~~~(\dagger)
 \\&=\LR_0+\LR^+_{11}+  2   \pi\sqrt{2}\int_{L_{ +}}\frac{  e^{ \theta_2 z} dz}{2\pi \I z^{-\tau_2}}\int_{L_{ +}} \frac{e^{ u^2}du}{2\pi \I} 
\widetilde K_r(  z,  u) \frac{\widetilde E^{\theta_1}_{\tau_1}(u)}{u^{\tau_1}} 
. ( **)
 \el\ee
 In the triple integral above, the contour $\Ga_0$ can be replaced by $\Ga_0\to \Ga_\iy-\Ga_{\{u\}}$ (see (\ref{EF'})), giving a cancellation of the terms $(\dagger)$. This leaves us with the last double integral in (\ref{L013}), labeled $(**)$. 
   \medbreak
   
   \noindent\underline{\em Step 2:} Compute ${\mathbb L}^{\mbox{\tiny dTac}}_2
 +{\mathbb L}^{\mbox{\tiny dTac}}_4$, using the formulas (\ref{Theta-ratio}) for the ratio  $\Theta_r(v,z)/\Theta_r(0,0)=1+...$ and using $\widetilde F^{-\theta_2}_{-\tau_2}(z)$ as in (\ref{EF}); notice the $1$ immediately produces $\LR^+_{12}$, as in (\ref{L-kernel'})
\be\label{L2'}\bl
{\mathbb L}^{\mbox{\tiny dTac}}_2&\stackrel{ }{=}
\oint_{\Ga_0}\frac{dv} {(2\pi\I)^2}\oint_{ 
     L_{ + }}  \frac{ dz}{z-v}\frac{v^{ \tau_2}}{z^{  \tau_1}}
\frac{e^{-v^2  +  \theta_2    v  }}
{e^{-z^2 + \theta_1   z  }}\frac{\Theta_r(v,z)}{\Theta_r(0,0)}
\\& 
=\LR^+_{12}-2\pi\sqrt{2}
\oint_{ \Ga_0 }\frac{dv} { 2\pi\I }\oint_{  L_{ +}}\frac{  dz}{2\pi \I}\frac{v^{  \tau_2}e^{      \theta_2    v  }}{z^{ \tau_1}e^{   \theta_1   z  }}
\oint_{   L_{ + }   } \frac{du}{2\pi  \I}\widetilde K_r(  z,  u)\frac{e^{u^2-v^2}}{u-v}  
.\el\ee
We now turn to expression ${\mathbb L}^{\mbox{\tiny dTac}}_4$ in (\ref{Final0}) containing $-\frac{1}{r+1}\frac{\Theta^-_{r+1}(v,z)}{\Theta _{r }(0,0)}$, which due to (\ref{Theta-ratio})(III), is given by the following expression after replacing $u\to v,~v\to z,~w\to u$. Again, referring to (\ref{Gamma0}), one  replaces $\Ga_0\to \Ga_\iy - \Ga_{\{u\}}$. Then: 
%
%
$$\bl
{\mathbb L}^{\mbox{\tiny dTac}}_4
&=
  \oint_{\Ga_0}\frac{dv} {(2\pi\I)^2} \oint_{\Ga_0}dz 
 \frac{v^{  -\tau_1}}{z^{ -\tau_2}}
\frac{e^{-v^2 - \theta_1 v  }}
{e^{ z^2 -  \theta_2 z  }} \left(-\frac{1}{r+1}\frac{\Theta^-_{r+1}(v,z)}{\Theta _{r }(0,0)}\right)
%
\\
&= 
   \oint_{\Ga_0=\Ga_\iy- \Ga_{\{u\}}}\frac{dv} {(2\pi\I)^2} \oint_{\Ga_0}dz 
 \frac{v^{  -\tau_1}}{z^{ -\tau_2}}
\frac{e^{-v^2 - \theta_1 v  }}
{e^{ z^2 -  \theta_2 z  }}   
 \\&~~~~\times 
\left[
\bl & - \int_{L_{ +}} \frac{du}{2\pi \I}
\frac{e^{ 2u^2}}{(u-z )(u-v )}
\\&\stackrel{ }{+} 2\pi\sqrt{2}\int_{L_{ +}} \frac{dw}{2\pi \I}\int_{L_{ +}} \frac{du}{2\pi \I}
 \frac{e^{w^2+u^2}}{(w\!-\!z )(u\!-\!v )}
\widetilde K_r(  w,  u)
  \el
\right]
 \el$$
$$\bl&= 
   \oint_{ \Ga_\iy }\frac{dv} {(2\pi\I)^2} \oint_{\Ga_0}dz 
 \frac{v^{  -\tau_1}}{z^{ -\tau_2}}
\frac{e^{-v^2 - \theta_1 v  }}
{e^{ z^2 -  \theta_2 z  }}   
 \\&~~~~\times 
\left[
\bl & - \int_{L_{ +}} \frac{du}{2\pi \I}
\frac{e^{ 2u^2}}{(u-z )(u-v )}~~~~~~~~~~~~~~~~~~~~~~~~~~~~~~~~~(A)
\\&\stackrel{ }{+} 2\pi\sqrt{2}\int_{L_{ +}} \frac{dw}{2\pi \I}\int_{L_{ +}} \frac{du}{2\pi \I}
 \frac{e^{w^2+u^2}}{(w\!-\!z )(u\!-\!v )}
\widetilde K_r(  w,  u)~~~~(B)
  \el
\right]
\\&\underbrace{ -\oint_{\Ga_0}\frac{dz} {(2\pi\I)^2} 
\left[\bl 
&  \int_{L_{0+}} \frac{du}{u-z } \frac{u^{  -\tau_1}}{z^{ -\tau_2}}
\frac{e^{u^2 - \theta_1 u   }}
{e^{ z^2 -  \theta_2 z  }}  ~~~~(=-\LR_{12})
\\&- 2\pi \sqrt{2} 
 \oint_{{L_{0+}}}\!\!dv
 \frac{v^{  -\tau_1}}{z^{ -\tau_2}}
 \frac{e^{- \theta_1 v  }}
{e^{  -  \theta_2 z  }}
\int_{L_{0+}} \frac{dw }{2\pi \I}
 \widetilde K_r( w,  v)\frac{e^{ w^2  -z^2}}{ w-z    }
\el \right] }_{-{\mathbb L}^{\mbox{\tiny dTac}}_2}
%
 , \el$$
  %
 where one notices that the last underbraced expression is precisely $=-{\mathbb L}^{\mbox{\tiny dTac}}_2$, as in (\ref{L2'}).  So, we have that ${\mathbb L}^{\mbox{\tiny dTac}}_2 +{\mathbb L}^{\mbox{\tiny dTac}}_4$ reads as follows, upon noticing that term $(A)$ can be rewritten by reshuffling the integrations and upon replacing $\Ga_0\to \Ga_\iy-\{w\}$  in term (B) below (again by the standard argument in (\ref{EF'})), and working out the residue at $z=w $ in (B) below:
  \be\label{L2L4}\bl
 {\mathbb L}^{\mbox{\tiny dTac}}_2&+{\mathbb L}^{\mbox{\tiny dTac}}_4 
  =     \int_{L+} \frac{du}{2\pi \I}
 e^{ 2u^2} \overbrace{\oint_{\Ga_\iy}\!\!\frac{dv} { 2\pi\I  }\!\frac{    e^{-v^2 - \theta_1 v  }}{v^{\tau_1}(v-u)} }^{\frac{\widetilde E^{\theta_1}_{\tau_1}(u)}{u^{\tau_1}}}\overbrace{\oint_{\Ga_0}\!\!\frac{dz}{2\pi\I} 
\frac{e^{- z^2 +  \theta_2 z  }}
{z^{ -\tau_2}(u\!-\!z)}}^{\frac{\widetilde F^{-\theta_2}_ {-\tau_2}(u)}{u^{-\tau_2}}}~~~~~(A)
\\+2&  \pi\sqrt{2}\int_{L_{ +}}\!\frac{e^{w^2}dw}{2\pi \I}\int_{L_{ +}} \!\frac{e^{ u^2}du}{2\pi \I} 
\widetilde K_r(  w,  u)\underbrace{\oint_{ \Ga_\iy } \frac{ e^{-v^2 - \theta_1 v  }dv} { 2\pi\I  v^{\tau_1} (v\!-\!u)}}_{\frac{\widetilde E^{\theta_1}_{\tau_1}(u)}{u^{\tau_1}} } \underbrace{\oint_{ {\Ga_0 }
 } 
\frac
{e^{ -z^2 +  \theta_2 z  }dz}{2\pi\I z^{-  \tau_2}(z\!-\!w)}}_{-\frac{\widetilde F^{-\theta_2}_ {-\tau_2}(w)}{w^{-\tau_2}}}  ~(B)
 \\=  & 
    \int_{L+}  
 \frac{e^{ 2u^2}du}{2\pi \I u^{\tau_1-\tau_2}}
\underbrace{\widetilde E^{ \theta_1}_{ \tau_1}(u)}_{ =-\widetilde F^{ \theta_1}_{ \tau_1}(u)+e^{ -u^2-\theta_1u}}\widetilde F^{-\theta_2}_{-\tau_2}(u)~~~~~~~(=\LR^+_{13}+\LR^+_{12})
\\&+2   \pi\sqrt{2}\int_{L_{ +}}\frac{e^{w^2}dw}{2\pi \I}\int_{L_{ +}} \frac{e^{ u^2}du}{2\pi \I} 
\widetilde K_r(  w,  u) \frac{\widetilde E^{\theta_1}_{\tau_1}(u)}{u^{\tau_1}}  \left( \frac{\widetilde E_{-\tau_2}^{-\theta_2}(w)}{w^{-\tau_2}}-\underbrace{\frac{e^{-w^2+\theta_2 w}}{w^{-\tau_2}}}_{(** )}\right)
 \\ =&\LR^+_{12}+\LR^+_{13}+\LR_2-(** )
.\el\ee
So, the term $(**)$ in (\ref{L2L4}), with the  replacement $w\to z$, is identical to $(** )$ in (\ref{L013}) .
 
 \noindent\underline{\em Step 3:}~ So, adding the two contributions (\ref{L013}) and (\ref{L2L4}) from steps 1 and 2  gives:
  %
 %
  %
  $$\bl
   \sum_{i=0}^4   {\mathbb L}^{\mbox{\tiny dTac}}_i & =(  {\mathbb L}^{\mbox{\tiny dTac}}_0+ {\mathbb L}^{\mbox{\tiny dTac}}_1+ {\mathbb L}^{\mbox{\tiny dTac}}_3)+( {\mathbb L}^{\mbox{\tiny dTac}}_2+ {\mathbb L}^{\mbox{\tiny dTac}}_4)
  \\ =& (\LR_{0}+\LR^+_{11})+(\LR^+_{12}+\LR^+_{13}+\LR_2)
  ,\el$$
  establishing formula (\ref{L-kernel'}) for $ {\mathbb L}^{\mbox{\tiny dTac}}_{r,0,0}$ in Lemma  \ref{lemma:Lkernel'}. \qed
  
  \vspace*{.5cm}
  
   \noindent {\em Proof of Proposition \ref{prop:L-kernel}}: The expression  (\ref{L-kernel}) for the kernel 
   $$ {\mathbb L}^{\mbox{\tiny dTac}}_{r,\rho,\beta}  (\tau_1, \xi_1\sqrt2;\tau_2, \xi _2\sqrt2)\sqrt2 d\xi\Bigr|_{\rho=\beta=0}$$
   is immediately obtained from setting $\theta_i=\xi_i\sqrt{2}$ and $u\to \frac{u}{\sqrt2}$ in formula (\ref{L-kernel'}) of Lemma \ref{lemma:Lkernel'}, and remembering the expressions $\widetilde F^{\sqrt2\xi}_{\tau}(\frac u{\sqrt2})=  F^{ \xi}_{\tau}(u)$,~   $\widetilde E^{\sqrt2\xi}_{\tau}(\frac u{\sqrt2})=   E^{ \xi}_{\tau}(u)$ and $\widetilde K_r(u,v)$ as in (\ref{Ktilde}). This establishes formula (\ref{L-kernel}) in Proposition \ref{prop:L-kernel}.\qed


 



\subsection{The ${\mathbb L}_2$-part of the kernel and Proof of Theorem  \ref{Th:prodop''}}

\noindent{\em Proof of Theorem  \ref{Th:prodop''}}. In view of the representation (\ref{L-kernel}) of the kernel 
$ {\mathbb L}^{\mbox{\tiny dTac}}_{r,\rho,\beta}  (\tau_1, \xi_1\sqrt2;\tau_2, \xi _2\sqrt2)\sqrt2 d\xi\Bigr|_{\rho=\beta=0} $ in Proposition \ref{prop:L-kernel}, it will suffice to prove the Proposition below.\qed

\begin{proposition} \label{prop:L2'}The ${\mathbb L}_2$-part of the kernel (\ref{L-kernel}) can be expressed in terms of the  $\FR^{(-\tau_1)}_{\xi_1}$ and $\FR^{( \tau_2)}_{-\xi_2}$ applied to the GUE-kernel $K_r(u,v)$: (for definitions, see (\ref{FR0}) and (\ref{KGUE}))
\be\label{L2''}\bl {\mathbb L}_2
 (\tau_1,\xi_1;\tau_2,\xi_2)  d\xi&= \int_{L+}  
  \frac{  e^{ \frac{u^2}2}du}{ \sqrt{2\pi} ~\I  u^{\tau_1}} \int_{L+} \frac{ e^{\frac{v^2}2}dv}{ \sqrt{2\pi}~ \I  v^{-\tau_2}}  K_r( -\I u,-\I v )  E^{ \xi_1}_{ \tau_1}(u)   E^{-\xi_2}_{-\tau_2}(v)d\xi
  \\&= 
(-\vr_1)^{\tau_1} (-\vr_2) ^{\tau_2}\FR^{-\tau_1}_{\vr_1 \xi_1}(u)\FR^{ \tau_2}_{-\vr_2 \xi_2}(v)K_r( \vr_1 u,-\vr_2 v) d\xi 
,\el\ee
which is  $\vr_1,\vr_2$-independent; nevertheless we will sometimes refer to 
${\mathbb L}_2={\mathbb L}_2^{\vr_1\vr_2}$.  
 %
 Given the involutions:\be\bl\label{invol}\IR_1:&~~\xi_1\leftrightarrow \xi_2,
  ~\tau_1\leftrightarrow -\tau_2, ~~\vr_1\leftrightarrow -\vr_2
  \\ \IR_2:&~~\xi_i\leftrightarrow -\xi_i,~~\vr_i \leftrightarrow -\vr_i
\mbox{, for }i=1,2,
 \el\ee
 we have $\IR_i({\mathbb L}_2)=(-1)^{\tau_1-\tau_2}{\mathbb L}_2$ for $i=1,2$.
\end{proposition}

\newpage

\vspace*{1cm}
  
 \proof Setting $u= \I x$ and $v= \I y$ in ${\mathbb L}_2$, given in (\ref{L-kernel}), changes $L_+$ into $\BR_{\mp}$, with an arbitrary choice of sign, since  $\frac{ E^{\xi}_\tau(w)}{w^{\tau }}$ never   has a pole at $0$, regardless of the sign of $\tau$. One uses the representation (\ref{KGUE}) of the GUE-kernel in terms of $\KR_r(f(x,z)/f(y,w))$. This leads to the following expressions using the representation of the $x$-integral  (\ref{IntcurlyF}) :
%
\be\label{L2''}\bl\frac{{\mathbb L}_2d\xi}{ (-1)^{\tau_1-\tau_2 }}&=
  \int_{\BR}  
  \frac{  e^{ -\frac{x^2}2}dx}{ \sqrt{2\pi} (- \I x)^{\tau_1}} \int_{\BR} \frac{ e^{-\frac{y^2}2}dy}{ \sqrt{2\pi }   (- \I y)^{-\tau_2}}  K_r( x,y )   E^{ \xi_1}_{ \tau_1}(\I x)   E^{-\xi_2}_{-\tau_2}(\I y)d\xi
\\&=
\KR_r^{z,w}\! \int_{\BR_{-\vr_1}} 
\!\frac{e^{ -\frac{x^2}2}  
 E^{ \xi_1}_{ \tau_1}(\I x)
   f(  x,  z)  dx}{\sqrt{ 2\pi}(-\I x)^{\tau_1}}
\!\int_{\BR_{-\vr_2}} \!
\frac{e^{ -\frac{y^2}2}  {   E^{-\xi_2}_{-\tau_2}(\I y)} dy}{\sqrt{2 \pi}(-\I y)^{-\tau_2}f(y,w)}d\xi
\\&= \vr_1^{\tau_1} \KR_r^{z,w} \FR^{-\tau_1}_{\vr_1 \xi_1}(u)[f( -\vr_1   u,  \I  z)]
%
%
\!\int_{\BR_{-\vr_2}} \!
\frac{e^{ -\frac{y^2}2}  {   E^{-\xi_2}_{-\tau_2}(\I y)} dy}{\sqrt{2 \pi}(-\I y)^{-\tau_2}f(y,w)}d\xi.
\el  \ee
  Next, we need the representation of the GUE-kernel (\ref{KGUE}) and its Fourier transform (\ref{GUEhat}) as the operator $\KR_r$ applied to ratios of $f$'s, always using $f(x,-y)=f(-x,y)$; this leads to the following expression :
 %
 \be\label{KR1} \bl
 \KR_r^{(z,w)}\frac{f(-\vr u,\I z)}{f(y,w)} =\widehat K_r(-\vr  u ,y)  &= \int_{\BR}\frac{dx}{\sqrt{2\pi}} e^{ -\vr  \I u x}\overbrace{K_r(x,y)}^{=K_r(y,x)}
 \\&=\KR_r^{(z,w)}\int_{\BR}\frac{dx}{\sqrt{2\pi}} e^{-\vr  \I u x}\frac{f(  y,  z)}{f(x,w)}
.\el\ee
 %
The left hand side of (\ref{KR1}) above is precisely the expression appearing in the last expression of (\ref{L2''}). So, using (\ref{KR1}), we have the first expression below, to which we apply  the identity (\ref{IntcurlyF}) again, giving the second equality below. Then, using the representation of $f(-\vr_2    v,  \I   z)$ as a Fourier transform of $f(y,z)$ (see (\ref{Fourierf})) gives the third identity below. Then pulling  $\KR_r^{z,w}$ all the way to the end gives us $K(x,y)$ in equality  $\stackrel{*}{=}$, leading to the double Fourier transform in $\stackrel{**}{=}$, using notation (\ref{FourierDef}). Finally, according to  formula (\ref{Fourierdouble}) in Lemma \ref{lemma:Fourier}, this gives us the last equality:  
$$ \bl&\hspace*{-.6cm}\frac{{\mathbb L}_2}{ (-1)^{\tau_1-\tau_2 }} d\xi=\frac{{\mathbb L}_2^{\vr_1\vr_2}}{ (-1)^{\tau_1-\tau_2 }} d\xi
\\  =&\vr_1^{\tau_1} 
 \KR_r^{z,w} \FR^{-\tau_1}_{\vr_1 \xi_1}(u)[\int_\BR\frac{dx}{\sqrt{2\pi}}e^{-\vr_1\I ux}f^{-1}(   x, w)]
\int_{\BR_{-\vr_2}} \!
\frac{e^{ -\frac{y^2}2}  {   E^{-\xi_2}_{-\tau_2}(\I y)}f(y,z) dy}{\sqrt{2 \pi}(-\I y)^{-\tau_2}}d\xi
 %
\\=&\vr_1^{\tau_1} \vr_2 ^{\tau_2}\KR_r^{z,w} \FR^{-\tau_1}_{\vr_1 \xi_1}(u)[\int_\BR\frac{dx}{\sqrt{2\pi}}e^{-\vr_1\I ux}f^{-1}(   x, w)]
\FR^{ \tau_2}_{-\vr_2 \xi_2}(v)[f(-\vr_2    v,  \I   z)]d\xi
\\=&\vr_1^{\tau_1} \vr_2 ^{\tau_2}\KR_r^{z,w} \FR^{-\tau_1}_{\vr_1 \xi_1}(u)[\int_\BR\frac{dx}{\sqrt{2\pi}}e^{-\vr_1\I ux}f^{-1}(  x, w)]
\FR^{ \tau_2}_{-\vr_2 \xi_2}(v)[\int_\BR\frac{dy}{\sqrt{2\pi}}e^{-\vr_2\I vy }f ( y, z)]d\xi
\\ \stackrel{*}{=}& \vr_1^{\tau_1} \vr_2 ^{\tau_2}\FR^{-\tau_1}_{\vr_1 \xi_1}(u)\FR^{ \tau_2}_{-\vr_2 \xi_2}(v) \int_\BR\frac{dx}{\sqrt{2\pi}}e^{-\vr_1\I ux}  
 \int_\BR\frac{dy}{\sqrt{2\pi}}e^{-\vr_2\I yv }  \underbrace{\KR_r^{z,w} \frac{f(y,z)}{f(x,w)}}_{=K(x,y)}d\xi
 \\\stackrel{**}=& \vr_1^{\tau_1} \vr_2 ^{\tau_2}\FR^{-\tau_1}_{\vr_1 \xi_1}(u)\FR^{ \tau_2}_{-\vr_2 \xi_2}(v)\widehat{\widehat K}(-\vr_1 u,-\vr_2 v)d\xi
 \\=&    \vr_1^{\tau_1} \vr_2 ^{\tau_2}\FR^{-\tau_1}_{\vr_1 \xi_1}(u)\FR^{ \tau_2}_{-\vr_2 \xi_2}(v) { K}( \vr_1 u,-\vr_2 v)d\xi.
\el$$
%
This establishes formula (\ref{L2'}) of Proposition \ref{prop:L2'}. \qed

\begin{corollary}\label{cor:LR22}
 The kernel ${\mathbb L}_{2}$, as in (\ref{L2'}) is a sum of four kernels 
 \be\label{L2sum}
 {\mathbb L}_{2} (\tau_1,\xi_1;\tau_2,\xi_2) d\xi=({\mathbb L}^{(\vr_1\vr_2)}_{21}+{\mathbb L}^{(\vr_1\vr_2)}_{22}+
{\mathbb L}^{(\vr_1\vr_2)}_{23}+
{\mathbb L}^{(\vr_1\vr_2)}_{24})d\xi,
\ee
 where the ${\mathbb L}^{(\vr_1\vr_2)}_{2i}$ depend on $\vr_i=\pm1$, although their sum is $\vr_i$-independent. The ${\mathbb L}^{(\vr_1\vr_2)}_{2i}$ are given by:
 $$\bl \hspace*{-.5cm}
 \frac{{\mathbb L}^{(\vr_1,\vr_2)}_{21}d\xi}{ (-\vr_1)^{\tau_1}(-\vr_2)^{\tau_2 }} \! =
  \KR_0^{v,w,\lb} \DR^{-\tau_1}_{ \vr_1\xi_1}(\mbox{\tiny$\bullet$}) \left[v^{-r}f_\lb (\mbox{\tiny$\bullet$},v)\right]~ \DR^{ \tau_2}_{ -\vr_2\xi_2}(\mbox{\tiny$\circ$} )\left[ w^r f_\lb ^{-1}( -\vr_1\vr_2( \mbox{\tiny$\circ$}) ,w) \right]
 d\xi \el$$
  \be\bl\label{LR22}\hspace*{-.5cm}\frac{{\mathbb L}^{(\vr_1,\vr_2)}_{22}d\xi}{ (-  \vr_1) ^{\tau_1} (- \vr_2) ^{\tau_2 }} 
 \! = 
  \KR_0^{v,w;\lb} \ER_{-\vr_2\xi_2}^{\tau_2}(\mbox{\tiny$\bullet$})\left[v^{-r}e^{\lb v}f(\I \vr_1\vr_2 (\mbox{\tiny$\bullet$}),\I v)\right] \DR_{\vr_1\xi_1}^{-\tau_1}(\mbox{\tiny$\circ$} )\left[ w^{ r} f_\lb^{-1}(\mbox{\tiny$\circ$} ,  w) \right]d\xi
 \el\ee
 $$\bl \hspace*{-.5cm}\frac{{\mathbb L}^{(\vr_1,\vr_2)}_{23}d\xi}{ (-\vr_1)^{\tau_1}(-\vr_2)^{\tau_2 }} 
\!  =\KR_0^{v,w;\lb} \ER_{\vr_1\xi_1}^{-\tau_1}(\mbox{\tiny$\bullet$})\!\left[v^{-r}e^{\lb v}f(\I \vr_1\vr_2 (\mbox{\tiny$\bullet$}),\I v)\right] \DR_{-\vr_2\xi_2}^{\tau_2}(\mbox{\tiny$\circ$} )\!\left[ w^{ r}  f^{-1}_\lb(\mbox{\tiny$\circ$} ,  w) \! \right]
 d\xi\el$$
 $$\bl \hspace*{-4.5cm}\frac{{\mathbb L}^{(\vr_1,\vr_2)}_{24}d\xi}{ (-\vr_1)^{\tau_1}(-\vr_2)^{\tau_2 }} 
 \!  = \ER_{\vr_1 \xi_1}^{  -\tau_1  }(v_1)  \ER^{ \tau_2 }_{-\vr_2\xi_2}( v_2 )  K_r(\vr_1\I  v_1,    \vr_2\I v_2)
 d\xi.\el$$
The involution $\IR_1$ in (\ref{invol}) acts as follows: 
$$\bl\IR_1({\mathbb L}^{(\vr_1 \vr_2)}_{21})&=(-1)^{\tau_1-\tau_2}{\mathbb L}^{(\vr_1 \vr_2)}_{21},~\IR_1({\mathbb L}^{(\vr_1 \vr_2)}_{23})=(-1)^{\tau_1-\tau_2}{\mathbb L}^{(\vr_1 \vr_2)}_{22},\\&\IR_1({\mathbb L}^{(\vr_1 \vr_2)}_{24})=(-1)^{\tau_1-\tau_2}{\mathbb L}^{(\vr_1 \vr_2)}_{24},
\el$$
and for the other involution $\IR_2$ in (\ref{invol}), we have for $i=1,..,4$:
\be\label{involJ2}\IR_2({\mathbb L}^{(\vr_1 \vr_2)}_{2i})=(-1)^{\tau_1-\tau_2}{\mathbb L}^{(\vr_1 \vr_2)}_{2i}
.\ee
  In the sector $\tau_1<0<\tau_2$, we have that ${\mathbb L}_2={\mathbb L}_{21}^{(\vr_1,\vr_2)}$.
\end{corollary}

\proof For the sake of brevity, this proof  uses momentarily the shorthand notation  $\widetilde\DR_i$ and $\widetilde\ER_i$:
 \be\bl\label{shortFR} \FR^{(-\tau_1)}_{\vr_1\xi_1}(u)&:=\DR^{(-\tau_1)}_{\vr_1\xi_1}(u )
+   \ER^{(-\tau_1)}_{\vr_1\xi_1}(v
 )
 \int_{\BR}\frac{d u }{\sqrt{2\pi}}
 e^{ u v  
  } =:(\widetilde\DR_1+\widetilde\ER_1)(u)
  \\  
    \FR^{(\tau_2)}_{-\vr_2\xi_2}(u)&= :\DR^{(\tau_2)}_{-\vr_2\xi_2}(u )
+   \ER^{(\tau_2)}_{-\vr_2\xi_2}(v 
 )
 \int_{\BR}\frac{d u }{\sqrt{2\pi}}
 e^{ u v  
  }=:(\widetilde\DR_2+\widetilde\ER_2)(u).\el\ee
So, working out the products, we find
 \be\label{LR'}\bl  
 {{\mathbb L}_{2}d\xi } = 
 &=   (-\vr_1)^{\tau_1}(-\vr_2)^{\tau_2}
 \FR^{(-\tau_1)}_{\vr_1\xi_1}(u_1)
  \FR^{(\tau_2)}_{-\vr_2\xi_2}(u_2)
 K_r( \vr_1u_1,- \vr_2 u_2) d\xi
 \\&=(-\vr_1)^{\tau_1}(-\vr_2)^{\tau_2} (\widetilde\DR_1+\widetilde\ER_1)(\widetilde\DR_2+\widetilde\ER_2)  K_r( \vr_1u_1,- \vr_2 u_2) d\xi
 \\&= (-\vr_1)^{\tau_1}(-\vr_2)^{\tau_2}(\widetilde\DR_1\widetilde\DR_2+\widetilde\DR_1\widetilde\ER_2+\widetilde\ER_1\widetilde\DR_2+\widetilde\ER_1\widetilde\ER_2) K_r( \vr_1u_1,- \vr_2 u_2) d\xi
 \\& =  :  ({\mathbb L}^{(\vr_1\vr_2)}_{21}+{\mathbb L}^{(\vr_1\vr_2)}_{22}+
{\mathbb L}^{(\vr_1\vr_2)}_{23}+
{\mathbb L}^{(\vr_1\vr_2)}_{24})d\xi,\el\ee
 which also yields the involution statement. 
 From expression (\ref{L2'}), using the definition (\ref{FR0}) of $\FR^\tau_\xi$, and using $K_r(x,y)=\KR_r^{v,w} \frac{f(x,v)}{f(y,w)}=\KR^{v,w;\lb}_0 \frac{v^{-r}f_\lb(x,v)}{w^{-r}f_\lb(y,w)}
  $ (see (\ref{KGUE})),
 we have for ${\mathbb L}_{21}$:
    $$\bl
 \frac{{\mathbb L}_{21}^{(\vr_1\vr_2)} }{ (-\vr_1)^{\tau_1}(-\vr_2)^{\tau_2 }}  &=
 \DR^{-\tau_1}_{\vr_1\xi_1}(u_1)
 \DR^{ \tau_2}_{-\vr_2\xi_2}(u_2)K_r(\vr_1u_1,-\vr_2 u_2) 
 \\& =\DR^{-\tau_1}_{\vr_1\xi_1}(u_1)
 \DR^{ \tau_2}_{-\vr_2\xi_2}(u_2)\KR_r^{vw}\frac{f (u_1,v)}{f (-\vr_1\vr_2 u_2,w)}
 \\& =\KR_0^{v,w;\lb}\DR^{-\tau_1}_{\vr_1\xi_1}(u_1)
 \DR^{ \tau_2}_{-\vr_2\xi_2}(u_2)\frac{v^{-r}f_\lb (u_1,v)}{w^{-r}f_\lb (-\vr_1\vr_2 u_2,w)}
, \el$$
which is formula ${\mathbb L}_{21}$ in (\ref{LR22}). Next we compute the ${\mathbb L}_{23}$-part, using (\ref{GUEhat}) and (\ref{Khatsym}):
 \be\label{LR23'}\bl
\frac{{\mathbb L}_{23}^{(\vr_1\vr_2)}  }{ (-\vr_1)^{\tau_1}(-\vr_2)^{\tau_2 }}  &=    \widetilde\ER_1(u_1)\widetilde\DR_2 (u_2)K_r( \vr_1u_1,- \vr_2 u_2) 
\\=&
 \ER_{\vr_1\xi _1}^{-\tau_1}(x )
 \int_{\BR}\frac{d u_1}{\sqrt{2\pi}}
 e^{ \I u_1(-\I   x )}
 \DR_{-\vr_2\xi_2}^{\tau_2}(u_2)
K_r( -\vr_1\vr_2u_2,   u_1) %
\\
=& \ER_{\vr_1\xi_1}^{-\tau_1}(x )
 \DR_{-\vr_2\xi_2}^{\tau_2}(u_2)
 \underbrace{\widehat K_r(- \vr_1 \vr_2 u_2, -\I x )}_{=\widehat K_r(\I \vr_1\vr_2 x ,u_2)}
\\\stackrel{ }{=}& \ER_{\vr_1\xi_1}^{-\tau_1}(x )
 \DR_{-\vr_2\xi_2}^{\tau_2}(u_2)
 \KR_0^{v,w;\lb}\frac{}{}\frac{v^{-r}e^{\lb v}f(\I \vr_1\vr_2 x ,\I v)}{w^{-r}e^{\lb w}f(u_2,  w)}
 \\=&
 \KR_0^{v,w;\lb} \ER_{\vr_1\xi_1}^{-\tau_1}(\mbox{\tiny$\bullet$})\left[v^{-r}e^{\lb v}f(\I \vr_1\vr_2 (\mbox{\tiny$\bullet$}),\I v)\right] \DR_{-\vr_2\xi_2}^{\tau_2}(\mbox{\tiny$\circ$} )\left[ w^{ r}( f^{-1}_\lb(\mbox{\tiny$\circ$} ,  w)) \right],\el\ee
 which is formula (\ref{LR22}). The formula (\ref{LR22}) for $\frac{{\mathbb L}_{22}d\xi}{ (-  \vr_1) ^{\tau_1} (- \vr_2) ^{\tau_2 }} $ is obtained immediately from applying the involution $\IR_1$ on $\frac{{\mathbb L}_{23}d\xi}{ (-  \vr_1) ^{\tau_1} (- \vr_2) ^{\tau_2 }} $.
 Finally, we compute 
 ${\mathbb L}_{24}^{(\vr_1\vr_2)} $ appearing in (\ref{LR'}), remembering the shorthand $ \widetilde\ER_i$ notation in (\ref{shortFR}):
$$\bl  &\hspace*{-.7cm}\frac{{\mathbb L}_{24}^{(\vr_1\vr_2)} }{ (-\vr_1)^{\tau_1}(-\vr_2)^{\tau_2 }}     =
 \widetilde\ER_1(u_1)\widetilde\ER_2 (u_2)K_r(\vr_1u_1,- \vr_2 u_2)
\\ =&
 \ER_{\vr_1 \xi_1}^{  -\tau_1  }(v_1)
 \int_{\BR}\frac{d u_1}{\sqrt{2\pi}}
 e^{-\I u_1(\I   v_1)}
  \ER^{ \tau_2 }_{-\vr_2\xi_2}( v_2 )  
 \int_\BR\frac{d u_2 }{\sqrt{2\pi}}
e^{-\I u_2( \I   v_2)}
 K_r( \vr_1u_1,- \vr_2 u_2)
\\ =&
 \ER_{\vr_1 \xi_1}^{  -\tau_1  }(v_1)  \ER^{ \tau_2 }_{-\vr_2\xi_2}( v_2 ) 
  \int_\BR\frac{d u_2 }{\sqrt{2\pi}}
e^{-\I u_2( \I   v_2)}
\underbrace{\int_{\BR}\frac{d u_1}{\sqrt{2\pi}}
 e^{\I u_1(-\I   v_1)}K_r(u_1,- \vr_1 \vr_2 u_2)}_{=\widehat K_r(-\I v_1,- \vr_1 \vr_2 u_2)=\widehat K_r(\I \vr_1 \vr_2 v_1,  u_2)}
 \el$$
\be\bl \label{L24}=& \ER_{\vr_1 \xi_1}^{  -\tau_1  }(v_1) \ \ER^{ \tau_2 }_{-\vr_2\xi_2}( v_2 ) 
\widehat {\widehat K_r}(\I \vr_1 \vr_2 v_1,  -\I v_2)
~~( \mbox{see (\ref{FourierDef}) and  (\ref{Fourierdouble})})
\\ =&
 \ER_{\vr_1 \xi_1}^{  -\tau_1  }(v_1)  \ER^{ \tau_2 }_{-\vr_2\xi_2}( v_2 )  K_r(\vr_1\I  v_1,    \vr_2\I v_2)
%
\\ =& 
 {(-1)^{\tau_1+\tau_2}}
\oint_{L_+}\frac{ dv_1e^{\frac{v_1^2}2 }}{\sqrt{2\pi}~ \I v_1^{\tau_1}}
 {     F_{\tau_1}^{ \vr_1\xi_1}( v_1  )}
\oint_{L_+}\frac{ dv_2e^{\frac{v_2^2}2 }}{\sqrt{2\pi}~ \I v_2^{-\tau_2}}
     {    F_{-\tau_2}^{- \vr_2\xi_2}( {v_2}{ })}
    K_r(\vr_1\I  v_1,    \vr_2\I v_2),
\el\ee
 using formula (\ref{Eoper''}) in the last equality, thus showing (\ref{LR22}). Notice that if $\tau_1<0$ (resp. $\tau_2>0$), we have $F_{\tau_1}=0$ (resp. $F_{-\tau_2}=0$). This ends the proof of Corollary \ref{cor:LR22}. \qed



 \section{Scaling and Asymptotics  of  the ${\mathbb L}_2$-part of the kernel
 }
 
In this section we focus on  the asymptotics of the ${\mathbb L}_2$-part of the kernel obtained in Theorem \ref{Th:prodop''} (see formulas (\ref{prodop''}) or (\ref{L2'})).  Corollary \ref{cor:LR22} tells us that ${\mathbb L}_2$, as in (\ref{L2sum}), can be written as a sum of four terms involving the operators $\DR_\xi^{ \tau} $ and  $ \ER^{  \tau }_{ \xi }$, as defined in (\ref{Doper}) and (\ref{Eoper''}). Therefore we shall deal with each of these terms in the next subsections.
 
 From (\ref{KGUE}) and (\ref{f}), one is reminded of the GUE-kernel, the operator $\KR^{(z,w;\lb)}$ and the function $f_\lb(x,z)$:
 \be\label{KGUE'}
  K_r(x,y)=\KR^{(z,w;\lb)}\frac{ f_\lb(x,z)}{f_\lb(y,w)},~\mbox{with }     f_\lb(x,z)=\sqrt2 e^{-z^2+(2x+\lb)z-\frac{x^2}{2}}.\ee

 \subsection{Asymptotics of the $\DR$-part}
 At first we introduce the two successive scales :
\be\label{scale1} \left\{{{z=\sqrt{r}z'}\atop{ w=\sqrt{r}w'}}\right\},~\lb=  \sqrt{ 2}\frac{\mu}{r^{1/6}}, ~\xi_i=\sqrt{2r}+\frac{\xi'_i}{\sqrt{ 2} r^{1/6}}
, x_{i}=1+\frac{y_{i}}{r^{2/3}},
 \ee
and then a second set of scales:
\be\bl \label{scale2}z'-z_0&=\frac{u}{\sqrt{2}r^{1/3}},~w'-z_0=\frac{v}{\sqrt{2}r^{1/3}},~~\mbox{with}~z_0=\frac1{\sqrt{2}} .
\el\ee
 %
%
Scalings (\ref{scale1}) and (\ref{scale2}) combined lead to  
\be\label{xi,z-as}\bl
&z  \sqrt{\frac{2}{r}}=1+\frac {u}{r^{1/3}}
,~w  \sqrt{\frac{2}{r}}=1+\frac {v}{r^{1/3}}, ~\frac{\lb(\xi_2-\xi_1)}2=\frac{\mu}{2r^{1/3}}(\xi'_2-\xi'_1)
\\
&\xi_1-2z=(-\sqrt 2 r^{1/6})(u-\frac{\xi'_1}{2r^{1/3}})\mbox{ and  }\xi_2-2w=(-\sqrt 2 r^{1/6})(v-\frac{\xi'_2}{2r^{1/3}}).
\el\ee
This scaling readily implies
:\be\label{dM}  {d\lb dzdwd\xi_i } =\frac12 {d\mu dudvd\xi'_i}{ } =:dM,~d\lb dw=d\mu dv
\mbox{ and }~~~d\lb dwdz=d\mu dvdu \frac{r^{1/6}}{\sqrt 2}.\ee  
Notice that, 
whenever $\tau<0$ and $\xi>0$, one has an alternative expression for the definition (\ref{Doper}) for $\DR_\xi^{ \tau}(u)$, upon setting $u=x\xi$, 
\be\label{Doper'} \DR_\xi^{ \tau}(u)[g(u)]=(-\xi)^{-\tau} \int_1^\iy dx \frac{(x-1)^{-\tau-1}}{(-\tau-1)!}g(x\xi).
\ee
Define $c_r$ and remember the Airy cubic $\AR_{u}(\xi)$ from (\ref{Airycub}):
\be\label{cA}
 c_r:=\left(\frac{r}{2e}\right)^{r/2}
  \mbox{  and }\AR_{u}(\xi):=\frac{u^3}3-\xi u
  .\ee
Given the function ${\frak f}(z')$ and its series about $z'=z_0=\frac1{\sqrt2}$,
\be\label{saddle}\frak f(z'):=z'^2-2\sqrt{2}z'+\log z'=-\tfrac12 (3+\log 2)+\tfrac{ (\sqrt2)^3}{3}(z'-z_0)^3+O((z'-z_0)^4),
\ee
we have, using the successive scalings (\ref{scale1}) and (\ref{scale2}), and the notation (\ref{cA}) for $c_r$ and $\AR_u$  ($x$ stands for $x_i$ and $y$ for $y_i$, $f_\lb$ is as in (\ref{KGUE'})) 
\be \label{Phi-scaled}\bl
\log ( \tfrac1{\sqrt2} &f_\lb(\xi_1x,z))  +\frac{\xi_1^2}2-r\log z \Bigr|_{\tiny\mbox{Scaling (\ref{scale1})}}
\\&= -z^2+(2x\xi_1+\lb)z-\frac{\xi_1^2(x^2-1)}{2}-r\log z\Bigr|_{\tiny\mbox{Airy scaling (\ref{scale1})}}
\\&=-r  \log\sqrt{r}
-r\frak f(z') +\lb z' \sqrt{r}+r^{1/3}\left(z'\sqrt{2}(2y+\xi_1)-2y \right)+O(r^{-1/3})
\\&=\frac r2 (3- \log \frac r2)+(\mu+\xi_1') r^{1/3}
-\left(\frac {{u}^3}{3}-(\xi'_1+\mu+2y)u\right)+O(r^{-1/3})
 \\&=r-\log c_r+(\mu+\xi_1') r^{1/3} %
 -\AR_u(\xi'_1+\mu) +2yu+O(r^{-1/3})
. \el\ee %
Using (\ref{xi,z-as}) and (\ref{HermP}), we now estimate the probabilistic Hermite polynomials $\Hp$ for $\tau_1,\tau_2\geq 0$ :
\be\bl\label{H-as}
 \Hp_{\tau_1}\!\left( {\xi_1\!-\!2z} \right) &=(\xi_1\!-\!2z)^{\tau_1}+\dots=\left(-\sqrt 2 r^{1/6} u\right)^{\tau_1}(1+O(r^{-1/3}))\\
\I^{\tau_2}\Hp_{\tau_2}\!\left(\frac{\xi_2\!-\!2w}{\I}\right)&=(\xi_2\!-\!2w)^{\tau_2}+\dots=\left(-\sqrt 2 r^{1/6} v\right)^{\tau_2}(1+O(r^{-1/3}))
.\el\ee

\begin{lemma}\label{lemma:asymptotD1}
The following holds
for any $\tau_1\in \BZ$, except 
  $\tau_1>0$ requires $\Re z<\sqrt {\tfrac r2}$ or equivalently $\Re z'<z_0=\frac1{\sqrt2}$ (i.e., $\Re u<0$  )
\be \label{asymptotD1}\bl 
  \tfrac1{\sqrt2}\DR_{\xi_1}^{ -\tau_1}(\mbox{\tiny$\bullet$}
   )&\Bigl[z^{-r} e^{\frac{\xi_1^2}2}    f_\lb(\mbox{\tiny$\bullet$}
    ,z)\Bigr]\Bigr|_{\tiny\mbox{Airy scaling (\ref{scale1})}}
\\&=\frac{e^r}{ c_r}e^{(\mu+\xi' _1)r^{1/3}}
 %
e^{-\AR_u(\mu+\xi'_1)
 }
\left( \sqrt{2}r^{1/6}u\right)^{ -\tau_1}  
(1+O(r^{-1/3})).
\el\ee
 %
For $\tau_2\in \BZ$, we have the corresponding formula, except    $\tau_2<0$ requires $\Re w>\sqrt {\tfrac r2}$ or equivalently $\Re w'>w_0=\frac1{\sqrt2}$ (i.e., $\Re v>0$),
\be\label{asymptotD2}\bl \sqrt2~
\DR_{   \xi_2}^{ \tau_2}(\mbox{\tiny$\bullet$}) \Bigl[(- 1)^{\tau_2} w^{ r} e^{-\frac{\xi_2^2}2}&   f_\lb^{-1}( \mbox{\tiny$\bullet$},w) \Bigr] \Bigr|_{\tiny\mbox{Scaling (\ref{scale1})}}
\\   &= \frac{c_r}{e^{r}} 
 e^{-(\mu+\xi_2') r^{1/3} }
e^{ \AR_v(\mu+\xi'_2)
 }
\left( \sqrt{2}r^{1/6}v\right)^{  \tau_2}  
(1+O(r^{-1/3}))
.\el\ee
In particular, we have:
\be\label{asymptotD2'}\bl  
 \sqrt2 (- 1)^{\tau_2} w^{ r}& \DR_{   \xi_2}^{ \tau_2}(\mbox{\tiny$\bullet$}) f^{-1}_\lb( \mbox{\tiny$\bullet$},w) \Bigr|_{\tiny\mbox{Airy scaling (\ref{scale1})}} 
 \\    &= c_r e^{- \mu  r^{1/3} }
e^{ \AR_v(\mu+\xi'_2)
 }
\left( \sqrt{2}r^{1/6}v\right)^{  \tau_2}  
(1+O(r^{-1/3}))
.\el\ee
\end{lemma}

\proof  
\underline{\em Case $\tau_1>0$}. Using the anti-derivative 
$\DR_{\xi_1}^{-\tau_1}$, as in (\ref{Doper'}), one obtains the first equality below. Then use (\ref{Phi-scaled}), one obtains the next expression, where $x$ and $y$ are related by $x-1=\frac{y}{r^{2/3}}$ (from (\ref{scale1})); then performing this change of variables, and using the scaling $\xi_i\to \xi'_i$ as in (\ref{scale1}), one finds the equality $(\stackrel{*}{=})$. The integral in that expression is finite and equals $(-2u) ^{-\tau_1}$, provided $\Re u<0$ \footnote{ For $\Re \al>0$:  $\dis\int_0^\iy dx \frac{x^{n-1}e^{- \al x}}{(n-1)!}=\frac{1}{\al^{n}} . $}; i.e., from (\ref{scale2}) this is  $\Re z'< z_0$. So, we have the formula below, where \underline{for $\tau_1>0$ it is required that  $\Re z'< z_0$, i.e., $\Re u<0$ }:
%
%
\be\label{z-as}\bl 
\tfrac{1}{\sqrt2}&z^{-r} e^{\frac{\xi_1^2}2}\DR_{\xi_1}^{-\tau_1} (\mbox{\tiny$\bullet$})   f_\lb( \mbox{\tiny$\bullet$} ,z) =\tfrac1{\sqrt2}(-\xi_1)^{\tau_1}  \int_{ 1}^\iy  
  dx \frac{(x-1)^{\tau_1-1}}{(\tau_1-1)!}
   z^{-r}e^{\frac{\xi_1^2}{2}} f_\lb(x\xi_1,z)  
 \\\stackrel{*}{=}&  (-1)^{\tau_1}
 \frac{e^r}{c_r}e^{ (\xi'_1+\mu)r^{1/3}}
e^{-\AR_u(\xi'_1+\mu)}\left( \frac{\sqrt{2r}+\frac{\xi'_1}{\sqrt{ 2} r^{1/6}} }{r^{2/3}} \right)^{\tau_1}\overbrace{\int_0^\iy dy \frac{y^{\tau_1-1}e^{2uy}}{(\tau_1-1)!}}^{ =(-2u)^{-\tau_1}}(1+O(r^{-1/3}))
\\=& (-1)^{\tau_1}\frac{e^r}{c_r}e^{ (\xi'_1+\mu)r^{1/3}}
e^{-\AR_u(\xi'_1+\mu)}
\left(-\sqrt{2}r^{1/6}u\right)^{-\tau_1}  
(1+O(r^{-1/3})).
\el \ee
\noindent \underline{\em Case $\tau_1\leq 0$}. Using (\ref{fder}), (\ref{Phi-scaled})(with $y=0$) and (\ref{H-as}), the following holds:
\be\label{z-as}\bl 
 \tfrac{1}{\sqrt2}z^{-r} &e^{\frac{\xi_1^2}2}\DR_{\xi_1}^{ -\tau_1}(\mbox{\tiny$\bullet$}) f_\lb(\mbox{\tiny$\bullet$},z)=\tfrac{1}{\sqrt2}z^{-r} e^{\frac{\xi_1^2}2}\left(\frac{\pl}{\pl \xi_1}\right)^{-\tau  _1}  f_\lb(\xi_1,z) \\&=\tfrac{1}{\sqrt2}(-1)^{\tau_1} z^{-r}e^{\frac{\xi_1^2}2}
   f_\lb(\xi_1,z)   \Hp_{\tau_1}\!\left( {\xi_1\!-\!2z} \right)
%
 %
 %
\\&= (-1)^{\tau_1}\frac{e^r}{c_r}e^{ (\mu +\xi'_2)r^{1/3}} 
e^{ -\AR_u(\xi'_1+\mu)}
\left(-\sqrt{2}r^{1/6}u\right)^{ -\tau_1}  
(1+O(r^{-1/3}))
.\el \ee
%
%
\noindent\underline{\em Case $\tau_2\geq 0$}. Using (\ref{fder}),(\ref{Phi-scaled}) and the estimate (\ref{H-as}) for the Hermite polynomials, we have, as in the calculation above:
\be\label{z-as}\bl 
 {\footnotesize\mbox{$\sqrt2$}}(-1)^{\tau_2} w^{ r}&e^{-\frac{\xi_2^2}2}\DR_{ \xi_2}^{ \tau_2}(u_2) f^{-1}_\lb(u_2,w) 
= {\sqrt2}w^{ r} e^{-\frac{\xi_2^2}2}\left(-\frac{\pl}{\pl \xi_2}\right)^{\tau_2}f^{-1}_\lb(\xi_2,w) 
\\&=  {\sqrt2}(-1)^{\tau_2}w^{ r} e^{-\frac{\xi_2^2}2}f^{-1}_\lb(\xi_2,w)\I^{\tau_2}\Hp_{\tau_2}\!\left(\frac{\xi_2\!-\!2w}{\I}\right)
%
\\&= (-1)^{\tau_2}
 \frac{c_r}{e^{r}} 
 e^{-(\mu+\xi_2') r^{1/3} } 
e^{ \AR_v( \mu+\xi'_2)
 }  \left(-\sqrt{2}r^{1/6}v\right)^{ \tau_ 2}(1+O(r^{-1/3})).
\el \ee

\noindent\underline{\em Case $\tau_2<0$}. For $\DR_{  \xi_2}^{ \tau_2}$, we use (\ref{Doper}) and then again the estimate (\ref{Phi-scaled});  remember the integral in  footnote 8, where here one needs $\Re v>0$, (and thus the proviso in the statement of the Lemma); here we also use the fact that $\xi_2=\sqrt{2r}+...>0$ in the third equality.

\be\label{z-as}\bl 
{\footnotesize\mbox{$\sqrt2$}}
w^{ r}&e^{-\frac{\xi_2^2}2}(-1)^{\tau_2}\DR_{  \xi_2}^{ \tau_2}(\mbox{\tiny$\bullet$}) f_\lb(\mbox{\tiny$\bullet$},w)^{-1}
  =  \sqrt2 \xi_2 ^{ -\tau_2}\int_1^\iy dx\frac{(x-1)^{-\tau_2-1}}{(-\tau_2-1)!}  \left(w^{- r} e^{\frac{\xi_2^2}2}  f_\lb(\xi_2x,w)\right)^{-1} 
 \\&=  \xi_2 ^{- \tau_2}\frac{c_r}{e^{r}} 
 e^{-(\mu+\xi_2') r^{1/3} } e^{\AR_v(\xi'_2+\mu)  }\int_0^\iy dy\frac{y^{-\tau_2-1}e^{-2yv}}{(-\tau_2-1)!}%
 \\&=  
 \frac{c_r}{e^{r}} 
 e^{-(\mu+\xi_2') r^{1/3} } 
e^{\AR_v(\xi'_2+\mu)  } 
\left( \frac{\sqrt{2r}+\frac{\xi'_2}{\sqrt{ 2} r^{1/6}} }{r^{2/3}} \right)^{-\tau_2}
 \left(2v\right)^{ \tau_ 2}(1+O(r^{-1/3}))
\\&=  
 \frac{c_r}{e^{r}} 
 e^{-(\mu+\xi_2') r^{1/3} } 
e^{ \AR_v  (\xi'_2+\mu)  } 
(\sqrt2   r^{1/6}v)^{\tau_2}
(1+O(r^{-1/3}))
.\el \ee
Finally, formula (\ref{asymptotD2'}) is obtained by noticing that the missing piece compared to (\ref{asymptotD2}) reads
$e^{-  \frac{\xi_2^2}2}=e^{-r}e^{-\xi_2'r^{1/3}}(1+O(r^{-1/3}))$.
%
This ends the proof of Lemma \ref{lemma:asymptotD1}.\qed



 \subsection{Asymptotics of the $\ER$-part  }

Remember from (\ref{Eoper''}) the operator $ \ER^{  -\tau }_{ \xi }$, with the replacement $w\to x$ in the $\Ga_0$-integration:
$$
 \ER^{  -\tau }_{ \xi }[g]  =\sqrt{2\pi}
\oint_{\Ga_0} \!\frac{dx ~ e^{- \frac{x ^2}{2}-  \xi  x }}{2\pi \I (-x )^{  \tau}}  
\int_{L+} \frac{dv ~e^{ \frac{  v ^2}2} }{2 \pi \I  (x \!-\!v )}g(v)
$$
and the constant $c_r$ as in (\ref{cA}). Then
given the scaling
\be\label{scaleE}\bl  ~z=\sqrt{\frac r2}z' ,~&\left\{{{v=\sqrt{2r}v'}\atop{x=\sqrt{2r}x'}}\right\},~\lb  =  \sqrt2 \frac{\mu}{r^{1/6}}, ~\xi_i=\sqrt{2r}+\frac{\xi'_i}{\sqrt{ 2} r^{1/6}}
\\&z'=\pm 1+\frac u{\sqrt r},
~~ x'=\tilde xr^{-1/3} 
,\el\ee
we have the following asymptotics for the $\ER^{  -\tau }_{ \xi }$-term, which nicely  combines with the integral over $\Ga_0$ in the GUE-kernel (\ref{KGUE}).

\begin{lemma}\label{lemma:ERf'}
Remembering $  f(-\I v,\I z)=\sqrt2 e^{\frac{v^2}2+2vz+z^2}
$, we have the following limit, for the scaling (\ref{scaleE}) :
\be\bl\label{ERf'}{\mathbb E}_r^{\tau_1,\xi_1,\lb} :=\left(2\oint_{\Ga_0}\frac{ dz}{2\pi \I}\right)&
 \ER _{ \xi_1}^{-\tau_1}(  \mbox{\tiny$\bullet$}
 )\left[\tfrac{1}{\sqrt2}z^{-r}e^{\lb z}  f(-\I     (\mbox{\tiny$\bullet$}))
  ,\I z)\right]\Bigr|_{{\tiny\mbox{scaling }}(\ref{scaleE})}
\\=& 
 \frac{ c_r^{-1} e^{ r^{1/3}\mu}
 }{(-\sqrt{2 }r^{1/6})^{ \tau_1-1} }\oint_{\Ga_0} \!   \frac{d\tilde x   ~ e^{     
 -    (\xi'_1+\mu) \tilde x  } }{2\pi \I   {\tilde x  }^{  \tau_1 }}(1+O(r^{-1/3}))
\\  =&
 \frac{ c_r^{-1 }e^{  \mu r^{1/3}}  }{( \sqrt{2 }r^{1/6})^{ \tau_1-1} } \frac{   (\xi'_1+\mu) ^{\tau_1-1}}{(\tau_1-1)!} (1+O(r^{-1/3}))
 . \el\ee

\end{lemma}

\proof
For later use, consider 
 the expansion of the function ${\frak g}(z')$ about the points $z'=\pm 1$ and the scaling for $z'\mp 1=u/\sqrt{r}$ as in (\ref{scaleE}):
\be\bl\label{saddleg}r{\frak g}(z') = r(\frac{ z'^2}{2}-\log z')
 &=r {\frak g}(\pm 1)+r\left((z'\mp1)^2+O( z'\mp1)^3)\right) 
 \\&=r {\frak g}(\pm 1)+u^2+O(\tfrac{u^3}{ r^{1/2}}) =
 -r\left\{ {0}\atop{ \I\pi}\right\}
 +\frac r2 +u^2+O(\tfrac{u^3}{ r^{1/2}}),
 .\el\ee
%
%
Notice that, along the circle of radius $1$ (i.e., for $z'=e^{\I\theta}$), we have that  $\Re (\frac{ z'^2}{2}-\log z')=\tfrac 12 \cos 2\theta$ has local maxima at $\theta=0$ and $\theta= \pi$, as in Fig. 7.

Since the operator $\ER_{ \xi_1}^{-\tau_1}(v)$ (setting the  $v$ for $\mbox{\tiny$\bullet$}$) contains $e^{\frac{v^2}{2}}$, we consider the log of the function on which $\ER_{ \xi_1}^{-\tau_1}$ acts, as in (99),
and remembering the constant $c_r$ as in (\ref{asymptotD2'}), the scaling (\ref{scaleE}) and the expansion (\ref{saddleg}), we find near the two saddle points $z'=\pm 1$:\newpage

\vspace*{10.5cm}
  
  \setlength{\unitlength}{0.017in}\hspace*{-4cm}\begin{tikzpicture}(0,160) \put(180,270){  \foreach \r/\c  in { 
        2/{}
        } 
      {
         \node[circle, draw, minimum size=2*\r cm,label=right:\c] {};
       }
  }
 \put(227,200){\vector(0,1){140}}
 \put(133,340){\vector(0,-1){140}}
 
 \put(180,314){$<$}
 
 \put(224,264){$\bullet\hspace*{1cm}z'=+1$}
 \put(78,264){$z'=-1\hspace*{1cm}\bullet$}


\end{tikzpicture}

\vspace*{-9.9cm}
Figure 7. For large $r$, the integrand $z^{-r}e^{\lb z}    f(-\I v,\I z)$ of the $z$-integral $\oint_{\Ga_0}\frac{2dz}{2\pi \I}$ for $z=\sqrt{\frac r2}z'$ has two saddle points $z'=\pm 1$ along the circle $\Ga_0$ of radius $1$, leading to two  $ \sqrt{r}(z'\mp 1)=:u$ integrations along the vertical lines $L_{\pm}$ in $\BC$ through $z'=\pm 1$.  

\medbreak

\be\label{Erf'}\bl
\log&\left(\tfrac1{\sqrt2}  z^{-r}e^{\lb z}    f(-\I v,\I z)\right)
 =\frac{v^2}2+2vz+z^2 -r\log z+\lb z
\\ =&-\frac r2 \log\frac r2 + rv'^2+2rv'z'+r\underbrace{\left(\frac{z'^2}{2}-\log z'\right)}_{{\frak g}(z')} +r^{1/3}\mu z'
\\ 
=&-\left\{ {0}\atop{\!\! \I\pi r \!\!}\right\} +\underbrace{\frac r2(1-  \log\frac r2 )}_{
 =:\log c_r^{-1}\mbox{\tiny as in (\ref{cA})}}
   + r(v'^2
\pm  2v')+\underbrace {u^2+2u(\sqrt r v'   +\frac{ \mu} {2r^{1/6}}) }_{\footnotesize\mbox{then completing the square}}\pm r^{1/3}\mu+O(r^{-1/2})
\\
=&-\left\{ {0}\atop{\! \I\pi r \!}\right\} +\log c_r^{-1}
\pm 2 rv'-   v' { \mu}r^{1/3} -\frac{\mu^2}{4r^{1/3}}  \\&~~~~~ +  \underbrace{(u +  \sqrt r v'   +\frac{ \mu} {2r^{1/6}})^2}_{=:Q(u)} \pm \mu r^{1/3} +O(r^{-1/2}).
\el\ee
So, we have: 
\be\label{ERf}\bl\frac1{\sqrt2}\ER_{ \xi_1}^{-\tau_1} (v
 ) \left[z^{-r}e^{\lb z}  f(-\I    v
  ,\I z)\right]& = c_r^{-1}e^{\mu r^{1/3} }d_r^{\pm}(\mu)
 \\&~~~\times \ER_{ \xi_1}^{-\tau_1}( v )
 \Bigr|_{v=\sqrt{2r}v'}
e^{   
\pm 2r v' 
 -\mu v' r^{1/3}-\frac{\mu^2}{4r^{1/3}} }
e^ {Q(u)
}, \el\ee
 for
  \be\label{dr}d^\pm_r(\mu):=\left\{{1~~~~~~~~~~~~~~~~ \mbox{\footnotesize for}+}\atop{\!(-1)^re^{-2\mu r^{1/3} }\!~~\mbox{\footnotesize for}-} \right. 
.\ee
Then, using the scaling $x\to x'$ and $\xi \to \xi'$, as in (\ref{scaleE}), we have
   \be\label{x-scaling}
  e^{-  \frac {{x }^2}{2}  -    \xi x  } \frac{dx}{{(-x) }^{   \tau }~} =  
~(\sqrt{2r})^{-\tau+1} e^{-  r({x'}^2+2 x' )   -   r^{1/3} \xi' x'   } \frac{dx' }{{(-x' ) }^{   \tau }}
.
\ee  
Also, using the scaling $z\to u$ and $v\to v'$ as in  (\ref{scaleE}), for which $dz=\frac{du}{\sqrt 2}$, and the scaling $x\to x'$ as in (\ref{x-scaling}), we define
  \be\label{saddle'2}
  rf^\pm(v'):=r(v'^2\pm 2v')=-r+(\sqrt{r}(v'\pm 1))^2
.  \ee
  As mentioned,  $\KR_0^{z,w;\lb}$ involves a $z$-integration $2\oint_{\Ga_0}\frac {dz}{2\pi \I}$ acting on the $\ER_{ \xi_1}^{-\tau_1}( v )[~] $-term and a $w$-integration acting on the other part, besides a $\lb$-integration.   
We have the $z$-integration, acting on the Gaussian $e^{Q(u)}$, using the scaling (\ref{scaleE})  $z\to z'\to u$  (see (\ref{Erf'}) for $Q(u)$). The $z'$-integration along the circle (chosen of radius $1$) localizes about the saddle point $z'=\pm 1$, turning the $z$-integration into a $u$-integration about the line $L_\pm$, as illustrated in Fig. 7. This expresses ${\mathbb E}_r^{\tau_1,\xi_1,\lb}$  (using the notation (\ref{dr}))  as the sum of  two ${\mathbb E}_r^{\pm}$, each  depending on a saddle point in Fig. 7:
\be\label{E-part}\bl
&c_re^{-\mu r^{1/3}} {\mathbb E}_r^{\pm}\\&:=
  d_r^{\pm}(\mu) \oint_{\Ga_0}\frac{2dz}{2\pi \I}\Bigr|_{{\tiny\mbox{scaling }}z\to u}
  \ER_{ \xi_1}^{-\tau_1}( v )
\left[e^{\pm 2rv'-\mu v' r^{1/3}-\frac{\mu^2}{4r^{1/3}} }
  e^ {Q(u)
   }\right]
\\&=2d_r^{\pm}(\mu)
\ER_{ \xi_1}^{-\tau_1}( v )\Bigr|_{v=v'\sqrt{2r}  }
e^{\pm 2rv'-\mu v' r^{1/3} -\frac{\mu^2}{4r^{1/3}}
 }
 \underbrace{ \tfrac 1{\sqrt 2} \oint_{\Ga_0\to L_\pm}\frac{du}{2\pi \I}e^ {Q(u)
  }}_{=\pm\tfrac 1{2\sqrt{2\pi}}}
\\&=\pm d_r^{\pm}(\mu) 
\oint_{\Ga_0} \!\frac{dx ~ e^{- \frac{x ^2}{2}-  \xi_1   x }}{2\pi \I  (- x) ^{  \tau_1 }}  
\int_{L+} \frac{dv  ~  e^{\frac{v^2}2} }{2 \pi \I  (x \!-\!v )}\Bigr|_{{x = \sqrt{2r}x'}\atop{v=\sqrt{2r}v'}}e^{\pm 2rv'-\mu v' r^{1/3} 
  }(1\!+\!O(r^{-1/3}))
\\&=
\frac{ \pm d_r^{\pm}(\mu)  
 }{(\sqrt{2r})^{ \tau_1-1} }\underbrace{\oint_{\Ga_0}
\frac{dx' e^{-  r f^+(x')
    -   r^{1/3} \xi'_1 x'   } }{{2\pi \I(-x' ) }^{   \tau_1 }}}_{*}
\oint_{\left\{{L_-+\Ga_0^+}\atop{L_+}\right\} } \!\!\!\frac{dv'e^{rf^{\pm}(v')
 -\mu v' r^{1/3} }}      {2 \pi \I  (x' \!-\!v')}
  (1\!+\!O(r^{-1/3}))
 ,\el\ee
  upon combining (\ref{x-scaling}) and (\ref{saddle'2}). Notice that for the {\em upper-sign} in the formulas above, we move the $L_+$ contour to the left of $\Ga_0$, giving $L_+=L_-+\Ga_0^+$, which contains the saddle point at $v'=-1$ , whereas for the {\em lower-sign}, we keep $L_+$, which contains the other saddle point $v'=1$. 
  
  So, we need to \underline{\em estimate the $x'$-integral $(*)$ about $\Ga_0$} in (\ref{E-part}), in particular the absolute value of the integrand, upon setting $x'=\epsilon e^{\I \theta}$, where $\epsilon=$ radius of $\Ga_0$ :
\be\label{estimx'}\bl
2\pi \epsilon^{\tau_1}\left| 
\frac{  e^{-  r f^+(x')
    -   r^{1/3} \xi'_1 x'   } }{{2\pi \I(-x' ) }^{   \tau_1 }}
\right|
&=  \left| e^{-r (x'^2+2x')-r^{1/3}\xi_1'x' } 
\right|
\\&=|e^{-r(\vr^2 e^{ 2\I\theta }+2\epsilon e^{  \I\theta  })-r^{1/3}\xi_1'\epsilon e^{\I\theta}}|\leq e^{ r(\epsilon^2  +2\epsilon  )+r^{1/3}|\xi_1'|\epsilon } .\el\ee
Next we \underline{\em estimate the $v'$-integrals about $L_\mp$}, using the change of scale $v'=\mp 1+\frac{\tilde v}{\sqrt r}$ or $\tilde v=\sqrt r (v'\pm1)$ (suggested above in (\ref{saddle'2})), with saddle point at $v'=\mp 1\in L_{\mp}$:
\be\label{estimv'}
\oint_ {L_{\mp}} \!\!\!\frac{dv'e^{rf^{\pm}(v')
 -\mu v' r^{1/3} }}      {2 \pi \I  (x' \!-\!v')}=e^{-r\pm \mu r^{1/3}}\oint_{L_{\mp}}\frac{ e^{\tilde v^2-\mu \tilde v r^{-1/6}}d\tilde v}{2\pi \I\sqrt{r}(\pm1+x'-r^{-1/2}\tilde v)}
(1+O(  r^{-1/2})).  \ee
So, the product of the $x'$-integrand (\ref{estimx'}) over $\Ga_0$ with the $v'$-integral (\ref{estimv'}) over $L_{\mp}$ has a zero limit for $r\to \iy$,   because the product of (\ref{estimv'}) and (\ref{estimx'}) behaves as $e^{-r}$ for a sufficiently small choice of $\epsilon$ in (\ref{estimx'}); say $\epsilon=r^{-1/2}$.

So, it remains to compute in ${\mathbb E}_r^{+} $, as in (\ref{E-part}), the $x'$-integral over $\Ga_0$, and the $v'$-integral over $\Ga_0^+$. By an obvious residue computation  at $v'=x'$, the last line of (\ref{E-part}) becomes the following, remembering that $d_r^{+}(\mu)=1$ (see (\ref{dr})) and using the scaling $x\to x'\to \tilde x$, namely $x'=\tilde xr^{-1/3}$ as in (\ref{scaleE}),
$$\bl
\sum_{\pm}{\mathbb E}_r^{\pm}= & \frac{c_r^{-1} e^{ \mu r^{1/3}} 
 }{(-\sqrt{2r})^{ \tau_1-1} } \oint_{\Ga_0} \!   \frac{dx'  ~ e^{     
 -   r^{1/3} (\xi'_1+\mu) x' } }{2\pi \I   {x' }^{  \tau_1 }}(1+O(r^{-1/3}))
\\  =&
 \frac{ c_r^{-1 }e^{  \mu r^{1/3}}  }{(-\sqrt{2 }r^{1/6})^{ \tau_1-1} }\oint_{\Ga_0} \!   \frac{d\tilde x   ~ e^{     
 -    (\xi'_1+\mu) \tilde x  } }{2\pi \I   {\tilde x  }^{  \tau_1 }}
 (1+O(r^{-1/3})) 
  \\  =&
 \frac{ c_r^{-1 }e^{  \mu r^{1/3}}  }{( \sqrt{2 }r^{1/6})^{ \tau_1-1} }\left(\frac{   (\xi'_1+\mu) ^{\tau_1-1}}{(\tau_1-1)!}\right)(1+O(r^{-1/3})),\el$$
 confirming the statement of Lemma \ref{lemma:ERf'}.\qed

\subsection{Asymptotics of the ${\mathbb L}_{2 } $-term 
 }

In this section we give the asymptotics of the ${\mathbb L}_{2 }(\tau_1,\xi_1;\tau_2,\xi_2) $-term, as in (\ref{L2sum}), which, unless $\tau_2<0<\tau_1 $, equals the sum of  
the 
terms ${\mathbb L}_{21}$, ${\mathbb L}_{22}$   and ${\mathbb L}_{23}$, as defined in (\ref{L2sum}) and (\ref{LR22}). Indeed  ${\mathbb L}_{24}\neq 0$ when $\tau_2<0<\tau_1 $. Remember from  Cor. 6.4 that ${\mathbb L}_{2 }= {\mathbb L}_{21 }$ for $\tau_1<0<\tau_2 $. For $A_\tau$ and ${\mathcal A}_u$ below, see (6) and (8).

 
 
 \begin{proposition}\label{prop:limL2}
For $\tau_1,~\tau_2\in \BZ$ arbitrary, except $\tau_2<0<\tau_1 $,  and $\vr=\pm1$, the following holds, for the $\vr$-dependent scaling (\ref{escale}), which amounts to the scale (82) times $\vr$:
 \be\label{limL2}\bl   {\mathbb L} _2^{\vr,-\vr} d\xi &:=\sum_{i=1}^3 {\mathbb L} _{2i}^{\vr,-\vr}d\xi :=\tfrac{d \xi' (1+O(r^{-1/3}))}{(-\vr\sqrt2 r^{1/6})^{\tau_1\!-\!\tau_2} } 
 \\& \times\left\{\bl   
 &+ \int_0^\iy d\mu\int_{\CR_L
  }
 e^{-\AR_u(\mu+\xi'_1)  }\frac{du}{2\pi \I u^{\tau_1}} 
  \int_{\CR_R
   }
e^{ \AR_v(\mu+\xi'_2) } 
 \frac{dv}{2\pi \I v^{-\tau_2}} 
\\&+
 \int_0^\iy d\mu  %
  \int_{\CR_L
    }
 e^{-\AR_v(\mu+\xi'_1) }\frac{du}{2\pi \I u^{ \tau_1}}\oint_{-\Ga_0}e^{-(\mu+\xi'_2 ) v}
  \frac{dv}{2\pi \I v^{-\tau_2}}
 ~  
\\&
 +
 \int_0^\iy d\mu \oint_{ \Ga_0}e^{ (\mu+\xi'_1 ) u}
  \frac{du}{2\pi \I u^{ \tau_1}}
 \int_{\CR_R
  }e^{\AR_v(\mu+\xi'_2)  }\frac{dv}{2\pi \I v^{-\tau_2}}   \el\!\right\} 
 \\&=\tfrac{d \xi' (1+O(r^{-1/3}))}{(-\vr\sqrt2 r^{1/6})^{\tau_1\!-\!\tau_2} }
(-1)^{\tau_2}\int_0^\iy
d\mu A_{\tau_1}(\mu+\xi'_1) A_{-\tau_2}(\mu+\xi'_2),  \el\ee
 with ${\mathbb L} _{22}^{\vr,-\vr}=0$ for $\tau_2\geq 0$ and ${\mathbb L} _{23}^{\vr,-\vr}=0$ for $\tau_1\leq 0$.

\end{proposition}




%
\proof
We now consider the asymptotics of the ${\mathbb L}_2$-part in  $( \sqrt2)^{\tau_2-\tau_1} {\mathbb L}^{\mbox{\tiny dTac}}_{r } 
     \sqrt2 d\xi$, as given in (\ref{prodop''}) or (\ref{L2sum}), with ${\mathbb L}_2$ given as a sum (\ref{L2sum}) of ${\mathbb L}^{\vr_1,\vr_2}_{2i}$, with  ${\mathbb L}_{2i}^{\vr_1,\vr_2}$ given by (\ref{LR22}). At first, we consider ${\mathbb L}_{21}^{\vr_1,\vr_2}$, as in (\ref{LR22}) for 
     $(\vr_1,\vr_2)=(\vr,-\vr)$. As will be explained below, it suffices to consider the case $\vr=1$. So, we have from (\ref{LR22}), upon inserting $e^{\frac{\xi_1^2-\xi_2^2}{2}-(\xi_1 -\xi_2)\sqrt{2r}}=1+O(r^{-1/3})$ on the second line and upon using the asymptotic expressions (\ref{asymptotD1}) and (\ref{asymptotD2}) of Lemma \ref{lemma:asymptotD1}: (for this choice of 
    $ \vr_1,~\vr_2$, we have $\vr_1 \vr_2=-1$)
 $$\bl
 & {\mathbb L}^{(\vr_1, \vr_2)}_{21}   d\xi \Bigr|_{\tiny\mbox{scaling (\ref{escale}) for  }  \vr_1=-\vr_2=1}
  \\&= 
 (-\vr_1)^{\tau_1} (-\vr_2)^{\tau_2} d\xi
 \KR_0^{z,w,\lb} \DR^{-\tau_1}_{ \xi_1}(\mbox{\tiny$\bullet$}) \left[z^{-r}f_\lb (\mbox{\tiny$\bullet$},z)\right]~ \DR^{ \tau_2}_{ \xi_2}(\mbox{\tiny$\circ$} )\left[
  w^r f_\lb ^{-1}(  \mbox{\tiny$\circ$} ,w) \right]
 \\& = 
 (-\vr_1)^{\tau_1} (-\vr_2)^{\tau_2} d\xi \KR_0^{z,w,\lb}
\left(\DR^{-\tau_1}_{\xi_1}(\mbox{\tiny$\bullet$})\bigl[z^{-r}e^{\frac{\xi_1^2}2-\sqrt{2r}\xi_1}  f_\lb(\mbox{\tiny$\bullet$},z)\bigr]\right)
\\&~~~~~~~\times\left(\DR^{ \tau_2}_{ \xi_2}(\mbox{\tiny$\circ$} )\bigl[w^{ r}e^{-\frac{\xi_2^2}2+\sqrt{2r}\xi_2}
 f_\lb^{-1}( \mbox{\tiny$\circ$} ,w)\bigr]\right)\Bigr|_{\tiny\mbox{Airy scaling (\ref{scale1})}}(1+O(r^{-1/3}))
\\&\stackrel{*}=
  (-\vr_1)^{\tau_1} (-\vr_2)^{\tau_2}\frac{d \xi'  }2 \widetilde\KR_0^{u,v,\mu}
\left(e^{ -\sqrt{2r}\xi_1} \frac{e^r}{ c_r}e^{(\mu+\xi' _1)r^{1/3}}
e^{-\AR_u(\mu+\xi'_1) }\left( \sqrt{2}r^{1/6}u\right)^{ -\tau_1}  
 \right)
 \\&~~~~~~~\times(-1)^{\tau_2}\left(e^{ \sqrt{2r}\xi_2}\frac{c_r}{e^{r}}   e^{-(\mu+\xi_2') r^{1/3} }
e^{ \AR_v(\mu+\xi'_2)} \left( \sqrt{2}r^{1/6}v\right)^{  \tau_2}  
\right)
(1+O(r^{-1/3}))  \el$$
 $$\bl
  &\hspace*{-.6cm}= (-\sqrt2 r^{1/6})^{\tau_2-\tau_1}
 d\xi' \int_0^\iy d\mu
 \\&~~~~~\times\int_{\CR_L
  }
u^{ -\tau_1} e^{-\AR_u(\mu+\xi'_1)  }\frac{du}{2\pi \I}
\int_{\CR_R
 } v^{\tau_2}
e^{ \AR_v(\mu+\xi'_2) } 
 \frac{dv}{2\pi \I}
(1+O(r^{-1/3})),
\el$$
upon using in the last equality that 
$e^{ \sqrt{2r}(\xi_2-\xi_1)}=e^{(\xi' _2-\xi' _1)r^{1/3}}$, and where in $\stackrel{*}=$,
$$\widetilde\KR_0^{u,v,\mu}:=
\int_0^\iy d\mu\int_{\CR_L}\frac{du}{2\pi \I}\int_{\CR_R}\frac{dv}{2\pi \I}
.$$
 Indeed, the circle $\Ga_0$ going with the $z$-integral in the GUE-operator $\KR_0^{z,w,\lb}$ gets deformed into a steepest descent curve  $\CR_L$, as described in Fig. 2,  passing through the saddle to the left of $0$. 
  Similarly, the complex line $L_+$ going with the $w$-integral in $\KR_0^{z,w,\lb}$ leads to $\CR_R$,  with a saddle to the right of $0$, as in  Fig. 2.

For the scaling (\ref{escale}), with $\vr=-1$, remember that according to the involution $\IR_2$ in (\ref{invol}),  we have:
 $$\bl\IR_2({\mathbb L}^{( \vr_1, \vr_2)}_{2i}(\tau_1, \xi_1;\tau_2, \xi_2)) &={\mathbb L}^{(-\vr_1,-\vr_2)}_{2i}(\tau_1,-\xi_1;\tau_2,-\xi_2) \\&=(-1)^{\tau_1-\tau_2}{\mathbb L}^{( \vr_1, \vr_2)}_{2i}(\tau_1, \xi_1;\tau_2, \xi_2) \el,$$ 
from which
$$\bl
 &
{\mathbb L}^{(\vr,-\vr)}_{21}(\tau_1,\xi_1;\tau_2,\xi_2) \Bigr|_{{\tiny\mbox{scaling (\ref{escale})}}\atop{ \vr=-1}}=(-1)^{\tau_1-\tau_2}{\mathbb L}^{(\vr,-\vr)}_{21}(\tau_1,\xi_1;\tau_2,\xi_2) \Bigr|_{{\tiny\mbox{scaling (\ref{escale})}}\atop{ \vr= 1}}
 ,\el$$
  leading to the same estimate as before, except for the multiplicative factor $(-1)^{\tau_1-\tau_2}$, thus proving the  first line (in the bracket) of ${\mathbb L}^{(\vr,-\vr)}_{2}$, as in (\ref{limL2}).
  
  \medbreak
  
  

%


We now turn to the asymptotics of the expression ${\mathbb L}^{(\vr_1, \vr_2)}_{23}$ as  in (\ref{L2sum})  and (\ref{LR22}), for $(\vr_1,\vr_2)=(\vr,-\vr)$. As before, it suffices to do the case $\vr=1$. From the estimates (\ref{ERf'}) in Lemma \ref{lemma:ERf'} for $\ER_{\xi_1}^{-\tau_1}$ and  (\ref{asymptotD2'}) of Lemma \ref{lemma:asymptotD1} for $\DR_{ \xi_2}^{\tau_2}$, we have :
%
$$\bl
&
   {\mathbb L}^{( \vr_1,  \vr_2 )}_{23}d\xi\Bigr|_{\tiny\mbox{scaling (\ref{escale}) for  } \vr_1=-\vr_2=1}
 \\&=(-\vr_1)^{\tau_1}(-\vr_2)^{\tau_2} 
 \KR_0^{z,w;\lb} \ER_{ \xi_1}^{-\tau_1}(\mbox{\tiny$\bullet$})\left[z^{-r}e^{\lb z}f(-\I  (\mbox{\tiny$\bullet$}),\I z)\right] \DR_{ \xi_2}^{\tau_2}(\mbox{\tiny$\circ$} )\left[ 
  w^{ r}  f_\lb^{-1}(\mbox{\tiny$\circ$} ,  w)  \right]d\xi
\\&= 
 (-\vr_1)^{\tau_1}(-\vr_2)^{\tau_2}\int_0^\iy d\lb
\left(2\oint_{\Ga_0}\frac{ dz}{2\pi \I}\ER_{ \xi_1}^{-\tau_1}(\mbox{\tiny$\bullet$})\left[z^{-r}e^{\lb z}f(-\I  (\mbox{\tiny$\bullet$}),\I z)\right]\right)
\\&~~~~~~~~~~ \times \oint_{L+}\frac{dw}{2\pi \I  }\left( \DR_{ \xi_2}^{\tau_2}(\mbox{\tiny$\circ$} )\left[ 
 w^{ r}  f_\lb^{-1}(\mbox{\tiny$\circ$} ,  w)  \right]\right)
 d\xi\el$$

 $$\bl \stackrel{**}{=}&
  \frac{(-\vr_1)^{\tau_1}(-\vr_2)^{\tau_2}d\xi '}{\sqrt2 r^{1/6}} \int_0^\iy d\mu\frac{ c_r^{-1} e^{\mu r^{1/3} }
 }{( \sqrt{2 }r^{1/6})^{ \tau_1-1} }
 \frac{   (\xi'_1+\mu) ^{\tau_1-1}}{(\tau_1-1)!}
 \\&\times (-1)^{\tau_2} \oint_{\CR_R}\frac{dv}{2\pi \I  }
 \frac{ c_r e^{- \mu  r^{1/3} }}{\left( \sqrt{2}r^{1/6} \right)^{ - \tau_2}}
 v^{\tau_2}e^{ \AR_v(\mu+\xi'_2)  } (1+O(r^{-1/3}))
 \\=& (-\sqrt2 r^{1/6})^{\tau_2-\tau_1}d\xi '
   \int_0^\iy d\mu \underbrace{\frac{   (\xi'_1+\mu) ^{\tau_1-1}}{(\tau_1-1)!}}_{*}
    \oint_{\CR_R}\frac{dv}{2\pi \I  }
 \frac{  }{ }
 v^{\tau_2}e^{ \AR_v(\mu+\xi'_2)  } (1+O(r^{-1/3}))
 ,\el$$
 using in equality $\stackrel{**}{=}$  the fact that $d\lb dw=d\mu dv$ from (\ref{dM}). The ratio (*) can then be written as: 
\be\label{residue}\frac{   (\xi' +\mu) ^{\tau -1}}{(\tau -1)!}=
\oint_{ \Ga_0} e^{ (\mu+\xi'  ) u}
  \frac{du}{2\pi \I u^{\tau }}
 \mbox{, for }\tau \geq1.\ee 
Setting $\vr=-1$, we have from the involution $\IR_2$, as before:
$$ {\mathbb L}^{( \vr,- \vr )}_{23}d\xi\Bigr|_{\tiny\mbox{scaling (\ref{escale})},~\vr=-1}=(-1)^{\tau_1-\tau_2}{\mathbb L}^{( \vr,- \vr )}_{23}d\xi\Bigr|_{\tiny\mbox{scaling (\ref{escale})},~\vr=1},$$
which establishes the third estimate for ${\mathbb L}^{( \vr,- \vr )}_{23}$ in the bracket (\ref{limL2}).
Also notice that Lemma \ref{lemma:asymptotD1} requires $\Re v>0$, which is so, since $v\in L_+$. 

\medbreak

 Finally, the estimate for ${\mathbb L}^{(\vr_1\vr_2)}_{22}$ in (\ref{limL2}), given  $(\vr_1,\vr_2) =(\vr, -\vr)$, can be obtained via the involution $\IR_1$, as in (\ref{invol}), acting  on ${\mathbb L}_{23}$ as in (\ref{involJ2}), namely 
$$\bl{\mathbb L}_{22}^{\vr_1\vr_2}(\tau_1,\xi_1;\tau_2,\xi_2) \Bigr|_{(\vr_1,\vr_2)=(\vr,-\vr)}&=
(-1)^{\tau_1-\tau_2}\IR_1 
{\mathbb L}_{23}^{ \vr,- \vr }( \tau_1,\xi_1; \tau_2,\xi_2) 
\\&=(-1)^{\tau_1-\tau_2}
{\mathbb L}_{23}^{ \vr ,-\vr }(-\tau_2,\xi_2;-\tau_1,\xi_1) 
.\el$$
    So, we obtain the same estimate (\ref{limL2}) for ${\mathbb L}_{23}^{\vr,-\vr}$, but in the new variables  $(-\tau_2,\xi'_2;-\tau_1,\xi'_1)$. In this new expression so obtained, 
we change $u\to -u,~v\to -v$, which  gives a multiplicative factor $(-1)^{\tau_1-\tau_2}$    and which changes the $(u,v)$-contours $\Ga_0\times\CR_R 
  \to$
$   
 \Ga_0\times -\CR_L
   $
     $   
 =-\Ga_0\times\CR_L
   $. %
  This establishes the second formula ${\mathbb L}_{22}^{\vr,-\vr}$ appearing in the curly bracket in (\ref{limL2}).   
  
  The first expression in (\ref{limL2}) can then readily be rewritten outside the range $\tau_2<0<\tau_1$ as follows, which then can be expressed in terms of the Airy-like functions $A_\tau(\xi)$ defined in (\ref{Atau}).
   $$
\bl &  {\mathbb L} _2^{\vr,-\vr}  d\xi  :=
\tfrac{d \xi' (1+O(r^{-1/3}))}{(-\vr\sqrt2 r^{1/6})^{\tau_1\!-\!\tau_2} } 
\int_0^\iy\! d\mu
\left[\left(\int_{\CR_L }
 e^{-\AR_u(\mu+\xi'_1)  }
  +\oint_{ \Ga_0}e^{ (\mu+\xi'_1 ) u}
  \right)\frac{du}{2\pi \I u^{ \tau_1}}\right] 
  \\&
 \hspace*{6cm}\times\left[\left(
 \int_{\CR_R} 
e^{ \AR_v(\mu+\xi'_2) } 
 +\oint_{-\Ga_0}e^{-(\mu+\xi'_2 ) v} \right) \frac{dv}{2\pi \I v^{-\tau_2}}\right]
\\&=\tfrac{d \xi' (1+O(r^{-1/3}))}{(-\vr\sqrt2 r^{1/6})^{\tau_1\!-\!\tau_2} }
(-1)^{\tau_2}\int_0^\iy
d\mu A_{\tau_1}(\mu+\xi'_1) A_{-\tau_2}(\mu+\xi'_2).\el$$
This ends the proof of Proposition \ref{prop:limL2}.\qed


\subsection{Asymptotics of the
Heaviside ${\mathbb L}_0$ combined with  the  ${\mathbb L}_1$-term}

  We will consider the scaling 
(\ref{scale1}) for each of the regions  $\tau_i\geq 0$ and $\tau_i\leq 0$; also  we let the $\xi_i\simeq \pm \sqrt{2r}$ for $r\to \iy$, in both directions; i.e., $\xi_i\simeq \sqrt{2r}$ and $\xi_i\simeq  -\sqrt{2r}$ for $r\to \iy$. That is to say, we look at the behavior near the upper-cut and near the lower-cut. 
So, $\xi_i$ is rescaled as follows:  
  \be\label{escale}\xi_i=\vr\bigl(\sqrt{2r}+\frac{\xi'_i}{\sqrt{ 2} r^{1/6}}\bigr)\mbox{, for }\vr=\pm1.\ee 
In this section we consider the asymptotics of the sum of the  terms ${\mathbb L}_0$ and ${\mathbb L}_1$, appearing in the expression (\ref{prodop''}) for the kernel   $(\sqrt2)^{\tau_2-\tau_1}{\mathbb L}^{\mbox{\tiny dTac}}_{r,0,0 } 
     \sqrt2 d\xi$, namely,  $$ ( {\mathbb L}_0+
{\mathbb L}_1)(\tau_1,\xi_1;\tau_2,\xi_2)=
( {\mathbb L}_0+
\sum_{i=1}^3{\mathbb L}_{1i})(\tau_1,\xi_1;\tau_2,\xi_2).$$        
 %
%
%

\begin{proposition} \label{prop:L01}The asymptotics of $ {\mathbb L}_0+{\mathbb L}_1$ reads as follows
 \be\label{L01}\bl
    (& {\mathbb L}_0  +{\mathbb L}_1) 
      d\xi \Bigr|_{\xi_i=\vr\bigl(\sqrt{2r}+\frac{\xi'_i}{\sqrt{ 2} r^{1/6}}\bigr)}
      =  ( -\vr{\sqrt2 r^{1/6}})^{\tau_2-\tau_1}
   \\ & \times  \left\{\bl&  
  - \BH^{\tau_1\!-\!\tau_2}  (\xi'_1\!-\!\xi'_2)+O(e^{-r}))
     \\&
   \!-\! (-    1)  ^{\tau_1\!-\!\tau_2 }\BH^{\tau_1-\tau_2}   (\xi'_2\!-\!\xi'_1) +O(e^{-r})
   \el\right\}
  d\xi'
  \mbox{, for}
  \left\{\bl \tau_1,\tau_2\geq 0\\
  \tau_1,\tau_2\leq 0. \el\right. 
  \\&=( -\vr{\sqrt2 r^{1/6}})^{\tau_2-\tau_1}\left(-\Id_{0\leq\tau_2< \tau_1}\Id_{\xi'_1\geq \xi'_2 }+ \Id_{ \tau_2< \tau_1\leq 0}\Id_{\xi'_2\geq \xi'_1 } \right)
\frac{(  \xi'_1\!-\!  \xi'_2)^{\tau_1-\tau_2-1}}{(\tau_1-\tau_2-1)!} +O(e^{-r})
  \el\ee
 
  \end{proposition}
  
  \medbreak
  
 \noindent To deal with the ${\mathbb L}_{1i}$  terms in (\ref{prodop''}), we first need the following Lemma:

\begin{lemma}\label{lemma:limLF} For $\xi,~\xi'\in \BR$, $\sg_1\geq 1$ and $\vr=\pm 1$ (otherwise $=0$)
\be\label{limLF}
 \oint_{ 
    L_{-\vr }} 
   \frac{e^{\frac{v^2}2+\xi' v}dv}{2\pi \I v^{\sg_1-\sg_2}}   F^{\xi }_{\sg_1 }(v) =O(e^{-\frac{\xi'^2}2})\mbox{, for }\xi'\to \vr\iy.
\ee

\end{lemma}

\proof Using (\ref{EF}) for $\widetilde F^{\theta_1}_{\tau_1}(u) $, and applying  (\ref{Phi}) to the integral over $L_+$, we have for $\sg_1\geq 1$ and $\theta,\theta'\in \BR$: 
$$
  \oint_{ 
    L_{\pm}} 
   \frac{e^{u^2+\theta' u}du}{2\pi \I u^{\sg_1-\sg_2}} F^{\theta  }_{\sg_1 }(u) 
   =\sum_{\ell=0}^{\sg_1-1}\frac{H_\ell(-\tfrac{\theta }2)}{\ell!} 
   \oint_{ 
    L_{\pm}} 
   \frac{e^{u^2+\theta' u}du}{2\pi \I u^{\sg_1-\sg_2-\ell}} 
$$
$$
=\frac{\pm 1}{\sqrt \pi}
\sum_{\ell=0}^{\sg_1-1}\frac{H_\ell(-\tfrac{\theta }2)}{\ell!} 
\left\{\bl
&\int_{\mp\frac{\theta'}{2}}^\iy \frac{( \theta' \pm 2 \ze)^{\sg_1-\sg_2-\ell-1}}{(\sg_1\!-\!\sg_2\!-\!\ell\!-\!1)!}
e^{- \ze  ^2}d\ze
\mbox{, for }\ell< \sg_1-\sg_2
\\&(\pm 2) ^{\sg_1-\sg_2-\ell-1}
e^{-(\frac{\theta'}2)^2 } H_{\ell-\sg_1+\sg_2}(\mp\tfrac {\theta'}2)\mbox{, for }\ell\geq \sg_1-\sg_2
\el\right\}
.$$
For the upper-sign (resp. lower-sign), the terms above tend to $0$ like $e^{-\theta'^2/4}$  for $\theta'\to -\iy$ (resp. $\theta'\to +\iy$).  This proves (\ref{limLF}) and thus ends the proof of Lemma \ref{lemma:limLF}.\qed

\medbreak

  \noindent{\em Proof of Proposition \ref{prop:L01}}:  A first immediate observation is that ${\mathbb L}^+_{13}\neq 0$, only if $\tau_2<0<\tau_1$ since $  F^{\xi}_{\tau}(v)=0$ whenever $\tau<0$; see (\ref{EFtilde}). 
%
%
%
%
For the same reason, we have:  
\be\label{Lplusmin}{\mathbb L}_{1i}^+={\mathbb L}_{1i}^-
 =0\mbox{, for  $\dis\left\{\ {{i=1,~ {\tau_1<0}}\atop{i=2,~{\tau_2>0}}} \right\}$}
 .\ee
 Also it is  immediate from straight substitution of the $\xi$-scaling (\ref{escale}) in the Heaviside part and  from Lemma \ref{lemma:limLF} for $\xi,\xi'$ blowing up at the same rate that, for $\vr=\pm 1$:
 \be\label{L1i'}\bl & {\mathbb L}_{11}^\vr\Bigr|_{\xi=-\vr\sqrt{2r}}={\mathbb L}_{11}^+\Bigr|_{\xi=-\sqrt{2r}} ={\mathbb L}_{11}^-\Bigr|_{\xi=\sqrt{2r}}=
  O(e^{-r})\mbox{, for  $\tau_1>0$}
  \\
 &  {\mathbb L}_{12}^\vr\Bigr|_{\xi= \vr\sqrt{2r}}={\mathbb L}_{12}^+\Bigr|_{\xi= \sqrt{2r}} ={\mathbb L}_{12}^-\Bigr|_{\xi=-\sqrt{2r}}=
  O(e^{-r})\mbox{, for  $\tau_2<0$}
 .\el\ee
 However, for ${\mathbb L}_{11}^\vr\Bigr|_{\xi= \vr\sqrt{2r}}$ and ${\mathbb L}_{12}^\vr\Bigr|_{\xi= -\vr\sqrt{2r}}$, it is (much) more convenient to express ${\mathbb L}_{1i}^\vr$ in terms of 
  ${\mathbb L}_{1i}^{-\vr}$, as follows.
 Namely, from the kernel (\ref{prodop''}) or (\ref{L-kernel}) and the form (\ref{EFtilde}) of $  F^{\xi_1}_{\tau_1}(v)$, one checks the expression below, upon noticing that $L_+-L_-=\Ga_0^+$. Then one first performs the $v$ integration about $v=w$ and then the $w$-integration about $w=0$ for each of the cases $i=1,~\tau_1>0$ and $i=2,~\tau_2<0$:
 \be\label{L1i''}\bl&{\mathbb L}_{1i}^+
 -{\mathbb L}_{1i}^- \stackrel{\left\{\!\!{{i=1,\tau_1>0}\atop{i=2,\tau_2<0}}\! \!\right\}}{=}
 \oint_{\Ga_0^+}\tfrac{e^{\frac{v^2}{2}}dv}{2\pi \I  }
 \left\{ { {\frac{e^{\xi_2v}}{v^{-\tau_2}}   \frac{  F^{\xi_1}_{\tau_1}(v)}{v^{\tau_1}}} \atop{ {  }   \frac{ e^{-\xi_1v}}{v^{\tau_1}}\frac{ F^{-\xi_2}_{-\tau_2}(v)}{v^{-\tau_2}}}}\right\}
 =\oint_{\Ga_0^+}\tfrac{e^{\frac{v^2}{2}}dv}{2\pi \I  }
\left\{ { {\frac{e^{\xi_2v}}{v^{-\tau_2}}   \oint_{ \Ga_0 }\frac{dw} { 2\pi\I }\frac{e^{ -\frac{w^2}2  -  \xi_1    w  }}{w^{\tau_1 }(v-w )} } \atop{ {  }   \frac{ e^{-\xi_1v}}{v^{\tau_1}}\oint_{ \Ga_0 }\frac{dw} { 2\pi\I }\frac{e^{ -\frac{w^2}2  +  \xi_2    w  }}{w^{-\tau_2 }(v-w )} }}\right\}
 \\&=
   \frac{(\xi_2-\xi_1)^{\tau_1-\tau_2-1}}{(\tau_1-\tau_2\!-\!1)!}\Id_{\tau_1-\tau_2\geq 1}
    =
   \underbrace{\BH^{\tau_1-\tau_2}(\xi_2-\xi_1)}_{=:-{\mathbb L}_0^-}+\underbrace{(-1)^{\tau_1-\tau_2+1}\BH^{\tau_1-\tau_2}(\xi_1-\xi_2) }_{=:  {\mathbb L}^+_0}
   , \el\ee
  where we define two  Heaviside pieces ${\mathbb L}^\pm_0$. The last equality is straightforward from the definition of the Heaviside function.
   From (\ref{L1i'}) and (\ref{L1i''}), it immediately follows that \underline{\em for $\tau_1,\tau_2\geq 0$}:
 $$\begin{array}{llll}{\mathbb L}_{0}^- +{\mathbb L}^+_{11}\Bigr|_{\xi=\sqrt{2r}}&= {\mathbb L}_{0}^++
 {\mathbb L}^-_{11}\Bigr|_{\xi=\sqrt{2r}}
   &=  {\mathbb L}_{0}^+\Bigr|_{\xi= \sqrt{2r}}+O(e^{-r})
 \\{\mathbb L}_{0}^-  +{\mathbb L}^+_{11}\Bigr|_{\xi=-\sqrt{2r}}& &= {\mathbb L}_{0}^-\Bigr|_{\xi=-\sqrt{2r}}+
  O(e^{-r})\mbox{,~  for $\tau_1,\tau_2\geq 0$},
  \end{array} $$
  and so, from the above and  (\ref{Lplusmin}), we have, for $\tau_1,\tau_2\geq 0$ (Note ${\mathbb L}_{0}^-={\mathbb L}_{0}$ of Prop. 6.1)
  \be\label{estim+}\bl( {\mathbb L}_{0}^-  +{\mathbb L}^+_{11}+{\mathbb L}^+_{12})&d\xi\Bigr|_{\xi_i=\vr\bigl(\sqrt{2r}+\frac{\xi'_i}{\sqrt{ 2} r^{1/6}}\bigr)}=
  {\mathbb L}_{0}^\vr d\xi\Bigr|_{\xi_i=\vr\bigl(\sqrt{2r}+\frac{\xi'_i}{\sqrt{ 2} r^{1/6}}\bigr)}+O(e^{-r})
  \\&=-\left(\frac{-\vr}{\sqrt2 r^{1/6}}\right)^{\tau_1-\tau_2}{\mathbb H}^{\tau_1-\tau_2}(\xi'_1-\xi'_2)d\xi'+O(e^{-r}).
 \el \ee
  Again, from (\ref{L1i'}) and (\ref{L1i''}), it follows that \underline{\em for $\tau_1,\tau_2\leq 0$}
  $$\begin{array}{llll}
  {\mathbb L}_{0}^- +{\mathbb L}^+_{12}\Bigr|_{\xi= \sqrt{2r}}&&= {\mathbb L}_{0}^-\Bigr|_{\xi= \sqrt{2r}}+
  O(e^{-r})
  \\ 
   {\mathbb L}_{0}^- +{\mathbb L}^+_{12}\Bigr|_{\xi=-\sqrt{2r}}&=  {\mathbb L}_{0}^+ +
 {\mathbb L}^-_{12}\Bigr|_{\xi=-\sqrt{2r}}
&=  {\mathbb L}_{0}^+\Bigr|_{\xi=-\sqrt{2r}} +O(e^{-r})
\mbox{,~~~~~~~~~ for $\tau_1,\tau_2\leq 0$} \end{array}$$
 and so, from the above and again (\ref{Lplusmin}), we have  for $\tau_1\!,\!\tau_2\leq 0$:
  \be\label{estim-}\bl( {\mathbb L}_{0}^- +{\mathbb L}^+_{11}  +{\mathbb L}^+_{12}&)d\xi   \Bigr|_{\xi_i=\vr\bigl(\sqrt{2r}+\frac{\xi'_i}{\sqrt{ 2} r^{1/6}}\bigr)} ={\mathbb L}_{0}^{-\vr}d\xi\Bigr|_{\xi_i=\vr\bigl(\sqrt{2r}+\frac{\xi'_i}{\sqrt{ 2} r^{1/6}}\bigr)}\!+\!O(e^{-r})\mbox{}\\&=-\left(\frac{ \vr}{\sqrt2 r^{1/6}}\right)^{\tau_1-\tau_2}{\mathbb H}^{\tau_1-\tau_2}(\xi'_2-\xi'_1)d\xi'+O(e^{-r})
  .\el\ee
  Combining the estimates (\ref{estim+}) and (\ref{estim-}) 
amounts to formula (\ref{L01}) for $ {\mathbb L}_0+{\mathbb L}_1$. Finally notice that for $\tau_1,\tau_2\leq 0$, we have
$$(-1)^{\tau_1-\tau_2-1} {\mathbb H}^{\tau_1-\tau_2}(\xi'_2-\xi'_1)=\Id_{ \tau_2< \tau_1\leq 0}\Id_{\xi'_2\geq \xi'_1 }  
\frac{(  \xi'_1\!-\!  \xi'_2)^{\tau_1-\tau_2-1}}{(\tau_1-\tau_2-1)!} 
,$$ and so the second expression for (\ref{L01}) follows. This ends the proof of Proposition  \ref{prop:L01}.\qed

\section{Proof of the main Theorem \ref{MainTheo}}

We now prove the main Theorem \ref{MainTheo} of the paper:

\medbreak

\noindent{\em Proof of Theorem \ref{MainTheo}}: Remember the expression (\ref{prodop''}) for ${\mathbb L}^{\mbox{\tiny dTac}}_{r,\rho,\beta} $ (for $\rho=\beta=0$) in terms of the  ${\mathbb L}_i$ and notice that ${\mathbb L}_{13 }= {\mathbb L}_{24 }=0$ outside the sector $\tau_2<0<\tau_1$.  Then using the estimates (\ref{L01}) (Proposition \ref{prop:L01})  for $ {\mathbb L}_0+{\mathbb L}_1 $ and (\ref{limL2}) (Proposition \ref{prop:limL2}) for ${\mathbb L}_2={\mathbb L}_2^{\vr,-\vr}$, we have: 
$$\bl
       ( \sqrt 2)^{\tau_2-\tau_1} &  {\mathbb L}^{\mbox{\tiny dTac}}_{r,\rho,\beta}  (\tau_1, \xi_1\sqrt2;\tau_2, \xi _2\sqrt2)\sqrt2 d\xi\Bigr|_{\rho=\beta=0} 
   =({\mathbb L}_0+ {\mathbb L}_1+{\mathbb L}_2)d\xi
   \\ =&\left( {\mathbb L}_0+({\mathbb L}_{11}+{\mathbb L}_{12})+({\mathbb L}_{21}+
   {\mathbb L}_{22}+{\mathbb L}_{23})\right)d\xi
  \\ =&( -\vr{\sqrt2 r^{1/6}})^{\tau_2-\tau_1}d\xi'
\\ \times&\left\{\bl&\left(-\Id_{0\leq\tau_2< \tau_1}\Id_{\xi'_1\geq \xi'_2 }+ \Id_{ \tau_2< \tau_1\leq 0}\Id_{\xi'_2\geq \xi'_1 } \right)
\frac{(  \xi'_1\!-\!  \xi'_2)^{\tau_1-\tau_2-1}}{(\tau_1-\tau_2-1)!}  +O(e^{-r})\\&+
 (-1)^{\tau_2}\int_0^\iy
d\mu A_{\tau_1}(\mu+\xi'_1) A_{-\tau_2}(\mu+\xi'_2)  (1+O(r^{-1/3}))
\el\right\}.
\el$$
 This 
leads immediately to the limit statement (\ref{limLtac-Ledge}) of the main Theorem \ref{MainTheo}, thus yielding the ${\mathbb L}^{\mbox{\tiny T-cuspAiry}}$-statistics.\qed

\section{The term ${\mathbb L}_{13}^+ +{\mathbb L}_{24} $ is asymptotically oscillatory in the range $\tau_2<0<\tau_1$}

Using expression (\ref{L-kernel}) for ${\mathbb L}_{13}^+$, with $e^{\frac{v^2}{2}}$ replaced by the left hand side of (\ref{eigenf}), we have (we indeed have $v_1^{\tau_2}F_{-\tau_2}^{- \xi_2}( {v_{1}} )$ in the $v_2$-integral)
\begin{equation}
  {\mathbb L}_{13}^+=
   - \oint_{L_+} \frac{e^{\frac{v_1^2}2}dv_1}{\sqrt{2\pi} \I v_1^ {\tau_1}} 
    { F_{\tau_1}^{ {\xi_1}}( {v_{1}}{ })}
    \oint_{L_-} \frac{e^{\frac{v_2^2}2}dv_2}{\sqrt{2\pi} \I v_1 ^{-\tau_2}}  {F_{-\tau_2}^{- \xi_2}( {v_{1}} )} 
    K_r(\I v_1, \I v_2)
\end{equation}
and expression (\ref{L24})  for ${\mathbb L}_{24}$, written  for $\vr_1=-\vr_2=1$, $v_2\to -v_2$(which changes $L_+\to L_-$),
$$
{\mathbb L}_{24}=
\oint_{L_+}\frac{ dv_1e^{\frac{v_1^2}2 }}{\sqrt{2\pi}  \I v_1^{\tau_1}}
 {     F_{\tau_1}^{  \xi_1}( v_1  )}
\oint_{L_-}\frac{ dv_2e^{\frac{v_2^2}2 }}{\sqrt{2\pi}  \I v_2^{-\tau_2}}
     {    F_{-\tau_2}^{- \xi_2}(  {v_2} )}
    K_r( \I  v_1,     \I v_2)
   . $$
  Then we have, setting $y_{j}= \I v_j, \ell= \tau_1 - 1 - m, k= -\tau_2 - 1 - \ell$  and using (\ref{tildeF}), (\ref{KGUE}), and (\ref{Herm})
    \be\bl{\mathbb L}_{13}^++{\mathbb L}_{24}&=
       \oint_{L_+}\frac{   e^{v_1^2}dv_1}{\I\sqrt{2\pi}   {v_1^ {\tau_1}} }\sum_{\ell=0}^{\tau_1 -1}\frac{\Hp _\ell  (-\xi_{1}) }{\ell!}  \oint_{L_-} \frac{e^{v_2^2}dv_2}{{\mathrm i}\sqrt{2\pi} }    \sum_{k=0}^{-\tau_2 -1}\frac{\Hp _k (\xi_2)}{k!} 
\\& 
\times    \left( \frac{v_1^\ell}{v_2 ^{- \tau_2 - k}} - \frac{v_1^\ell}{v_1^{\tau_1-k}} \right) 
    \left( \frac{H_r(\I v_1) H_{r-1}(\I v_2) - H_{r-1}(\I v_1) H_r(\I v_2)}{2^{2r-1}  \I c^2_{r-1} (v_1 - v_2)} \right)
%
 %
 %
  \\&   {\stackrel{*}{=}} -2(-1)^{\tau_1}   \sum_{m=0}^{\tau_1 - 1} \frac{\Hp_{\tau_1-1-m} (\xi_1)}{(\tau_1 -1- m)!\I^m} 
    \sum_{\ell=0}^{-\tau_2 -1 } \frac{\Hp_{-\tau_2-1-\ell} (\xi_2)}{(-\tau_2 -1 - \ell)! (-\I)^\ell}
  \\&
   \times \sum_{s=0}^{\ell} 
    \left( I^+_{2 + m+ s} (r) I^-_{1+\ell-s} (r-1) - I^+_{2+m-s}(r-1) I^-_{1+\ell-s} (r)\right)
    \left (1 + O(\frac{1}{r})\right),
\el\ee
where 
$$
    I^{\pm}_{k} (\alpha)= 
     \left( \frac{\pi \alpha}{2} \right)^{\frac{1}{4}}
   \oint_\mathbb{R^{\pm}}  \frac{dye^{-y^2} 
    }{2\pi \I y^k}  \frac{H_{\alpha}(y)}{\sqrt{2^{\alpha} \alpha!}}.
$$

\begin{proposition} \label{prop:L13+L24}We have:
$$
 {\mathbb L}_{13}^++{\mathbb L}_{24}
\simeq (-1)^{\tau_1} \frac{ ( \sqrt{2r})^{\tau_1 - \tau_2 -1} O_{sc}^{(r)}(\tau_1, -\tau_2)}{\pi^2}
.$$
where $O_{sc}^{(r)}$ can be expressed in terms of a periodic function $I_{\ell}(r) = I_{\ell}(r+16)$:
$$\bl
O_{sc}^{(r)}(\tau_1, -\tau_2)& = \sum_{0 \leq m \leq \tau_1 -1} \sum_{0 \leq \ell \leq -\tau_2 -1} 
\tfrac{\left( \frac{\I}{\sqrt{2}} \right)^{m+\ell} (-1)^m}
{(\tau_1 -1 -m)! (-\tau_2 -1 -\ell)!} 
\\&
\times\sum_{s=0}^{\ell} 
\left({{I}_{2+m+s}(r) { I}_{1+\ell-s}(r-1)  
- {I}_{2+m+s}(r-1) {I}_{1+\ell-s}(r)}\right)
,\el$$
where
\be\label{I(r)}\bl
I_{1+2\alpha}(r)  = a(\alpha)  \sin \frac{r \pi}{8} - b(\alpha) \sin \frac{r \pi}{2}
,~ I_{2+2\alpha}(r)  = c(\alpha) \cos \frac{r \pi}{8} - d(\alpha) \cos \frac{r \pi}{2},
\el\ee
with
$$\bl
a(\alpha) &= 2^{-\alpha} \oint_{0}^{\pi} \frac {d\theta}{2} \, e^{-2\I\alpha \theta} \sin \left( \frac{e^{\I\alpha \theta}}{2} \right),
\quad
b(\alpha) = 2^{\alpha} \oint_{2}^{\infty} \frac{\sin u}{u^{1+2\alpha}} du
\\ 
c(\alpha)& = \frac{\I 2^{-\alpha}}{\sqrt{2}} \oint_{0}^{\pi}\frac{ d\theta}{2} \, e^{-(2\alpha+1)\I \theta} \cos \left( \frac{e^{\I\alpha \theta}}{2} \right),
\quad
d(\alpha) = 2^{\alpha} \sqrt{2} \oint_{2}^{\infty} \frac{\cos u}{u^{2+2\alpha}} du
.\el$$
\end{proposition}

\textbf{Remark:}   \( a(\alpha) \), \( c(\alpha) \) are  explicitly computable.

\medbreak

\noindent{\em Proof of Proposition \ref{prop:L13+L24}}: Remember that
$
\xi \simeq \sqrt{2r}, $ and $ \Hp_\ell  (x) = x^ \ell{+ \dots}
$
Then we must apply the following Lemma to (120):

\textbf{Lemma:} For \( \ell = 0,1, \dots \),
\[ I^{\pm}_{1+\ell}\simeq
-\frac{\sqrt{r}^\ell}{\I \pi} I_{1+\ell}(r) \left( 1 + O(\frac{1}{r}) \right)
.\]

\textbf{Sketch of Proof of the Lemma:} The difficulty in computing the integrals \( I_k^{\pm}(r) \) for \( r \) large is the pole at \( r = 0 \), for \( k \geq 1 \). The integrals \( \mathbb{R^{\pm}} \) are along \( \mathbb{R} \) except for a small semi-circle above or below the origin for \( k \geq 1 \).

Thus, we can use the Szeg\"o asymptotics \cite{Szego} on $ \mathbb{R} $ except near the origin, where we must use the strong asymptotics of Deift et al. \cite{Deift} which is valid near the origin in \( \mathbb{C} \). The Szeg\"o asymptotics for the Hermite polynomials beyond the Airy region yield such strong decay that their contribution to \( I_k^{\pm}(r) \) is \( O(e^{-r}) \). The Szeg\"o asymptotics in the sinusoidal region require more care, with $\vr>0$, sufficiently small, but for \( r ^\vr  \leq |y| \leq \sqrt{2r} \), it yields a contribution to \( I_{\ell}^{\pm}(r) \) which is \( O(e^{-c r^\vr}) \). The contribution in the Airy region to \( I_{\ell}^{\pm}(r) \) is \( O(e^{-r}) \). Thus, the contribution of these three regions can be neglected. The contribution to \( I_{\ell}^{\pm} (r) \) in the sinusoidal region, restricted to $(\frac r2)^{-1/2}  \leq |y| = r^\vr $, yields the \( \sin \frac{r\pi}{2}, \cos \frac{r\pi}{2} \) contributions in (\ref{I(r)}), while the contributions along the \textbf{semi-circle} of radius $(\frac r2)^{-1/2}$   in the integral \( I_{\ell}^{\pm}(r) \) yield, by the asymptotics by Deift et al.\cite{Deift}, the \( \sin \frac{r\pi}{8} \) and \( \cos \frac{r\pi}{8} \) contributions to (\ref{I(r)}). These two contributions are the difficult part of the calculation and require care.

  \section{Epilogue: Transversal Cusp-Airy versus Cusp-Airy and the $\{\rho\}$-strip versus the $\{\sg\}$-strip}
      
 In this section, we discuss informally when the Cusp-Airy, versus  the  transversal Cusp-Airy statistics, will take place for   the model discussed in this paper. To do so, we  consider Petrov's \cite{Petrov2} seminal paper on hexagonal models with cuts along the upper-edge and no cut below. In the context of this paper, we apply this to a hexagon with one cut above, and no cut below,   since it seems plausible to believe that the lower-cut will have little impact on the gross behavior near the upper-cut.      So, for the one-cut model in Fig. 8a and 8b, with a cut of size $d$, at distances $n _1$ and $n _2$ from the corners  (as in Figs. 5 and 4), with sides $b$ and $c$ on the left and $c'$ and $b'$ on the right, we prove the following statement:
 
 \begin{proposition}\label{prop:cusp-cut}
     Referring to Figs. 8(a) and 8(b), the cusp of the arctic curve is tangent to (i) the extension of the right hand segment of the cut or (ii) the extension of its left hand segment if \be\label{cusp-cut} (i)~n_2b-n_1c'>dc' ~~ \mbox{ or }~~~(ii)~~n_2b-n_1c'<-db .  \ee
 %
 For the arctic curve to belong to the polygonal domain, we must have\footnote{Such a condition must be implicit in Petrov's paper \cite{Petrov2}.}   $n_2b-n_1c' \notin d[-b,c']$.   \end{proposition}

   For the scaling (\ref{geomscale}) of our paper, used to go from the  discrete-discrete kernel $ {\mathbb L}^{\mbox{\tiny blue}}$ to the  discrete-continuous kernel ${\mathbb L}^{\mbox{\tiny dTac}} _{r,\rho,\beta}  $, we have that, for $ \rho=0$ and $\beta_i=\ga_i=0$:
     $$n_i=\tfrac{\ga+1}{\ga-1} d+(-1)^{i }r, \mbox{ for $ i=1,2$ and $1<\ga<3$}.$$
     To connect with Petrov's scaling, all variables must be divided by $d$ and so, the inequality (\ref{cusp-cut}(i)) is satisfied if 
     $$\frac{n_1+d}{n _2}=\frac{2\ga-\frac rd(\ga-1)}{\ga+1+\frac rd(\ga-1)}<0<\frac{b}{c'},$$
     which is satisfied as long as $\frac{2\ga}{\ga-1}<\frac rd$. 
\newline  {\em Heuristic statement}:  So, if $\frac rd >\frac{2\ga}{\ga-1}$ is satisfied, then this puts us in situation {\em (i)} above and so the cusp-Airy statistics would take  place along the extension of the right hand segment of the cut and the T-cusp-Airy fluctuations along the left hand segment of the cut. This is also confirmed by the simulation of Fig. 4.

    It remains an open problem to obtain the cusp-Airy process via the analogue of the   $ {\mathbb L}^{\mbox{\tiny blue}}$-process; namely one would have to consider the $ {\mathbb L}^{\mbox{\tiny yellow}}$-point-process  along parallel lines to the $\{\sigma\}$-strip, as defined in (\ref{sigma}). 
    
     \newpage
      
      \vspace*{7cm}
      
          \hspace*{3.9cm} \setlength{\unitlength}{0.017in}\begin{picture}(0,60)
\put(100.5,  33.9){\makebox(0,0) {\rotatebox{0}{\includegraphics[width=267mm,height=225mm]
{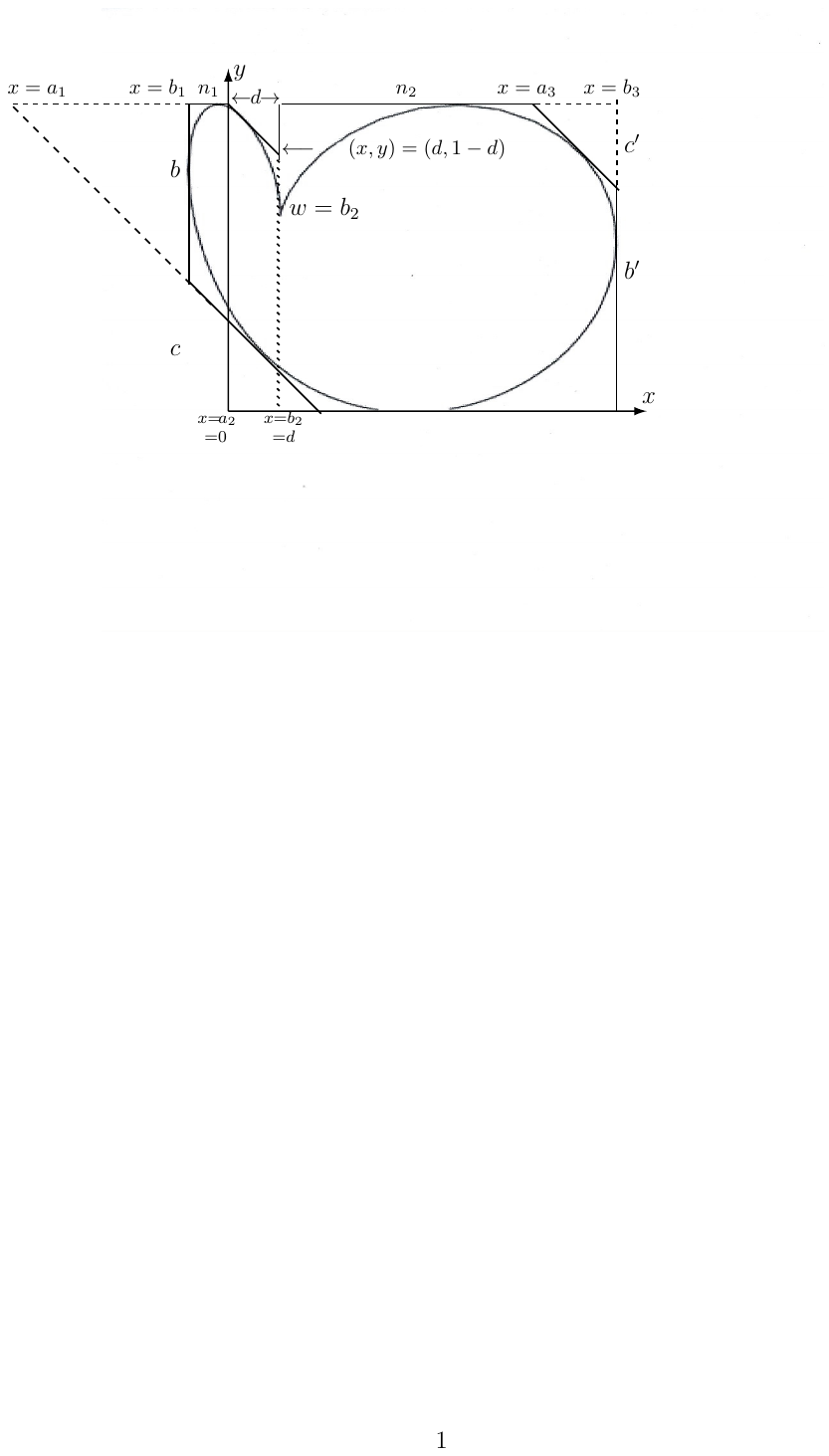}}}}
\end{picture}

\vspace{-6cm}

Fig. 8(a). The arctic curve for a non-convex shape ${\bf P}$, with  one cut for  
$n_1=2,~n_2=8, ~d=\frac1 3, ~b=\frac1 4, ~c=1-b=\frac 3 4, ~b'=b+d=\frac7 {12},~ c'=1-b'=\frac5{ 12}$.

 \vspace{9.5cm}

     \hspace*{3cm} \setlength{\unitlength}{0.017in}\begin{picture}(0,60)
\put(106.5,  33.9){\makebox(0,0) {\rotatebox{0}{\includegraphics[width=327mm,height=225mm]
{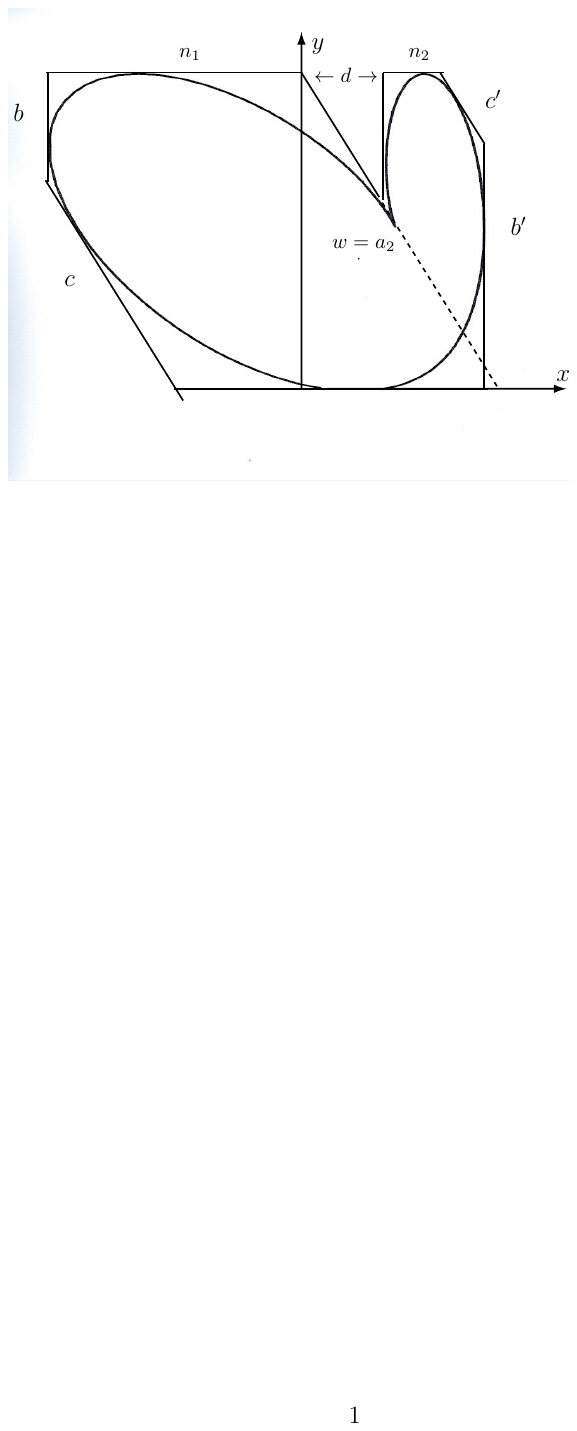}}}}
\end{picture}

\vspace{-6.6cm}

Fig. 8(b). The arctic curve for a non-convex shape ${\bf P}$, with  one cut for  
$n_1=8,~n_2=2, ~d=\frac1 3, ~b=\frac1 4, ~c=1-b=\frac 3 4, ~b'=b+d=\frac7 {12},~ c'=1-b'=\frac5{ 12}$. The shape of ${\bf P}$ in the figure is out of scale compared to the Maple-generated arctic curve (\ref{xwyw}).

\bigbreak

  \noindent  This could be derived from the known inverse Kasteleyn matrix. The same combinatorial arguments would imply the same {\em doubly interlacing patterns} as in Fig. 6  for the yellow tiles (instead of blue tiles for the $\{\rho\}$-strip). This puts us in the situation of the double interlacing used for the cusp-Airy statistics by Duse, Johansson and  Metcalfe \cite{DJM}.




\bigbreak

      

\bigbreak

    \noindent{\em Proof of Proposition \ref{prop:cusp-cut}}:     Consider the $(x,y)$ coordinates of Figs. 8, with the $y$-axis passing through the left-hand corner of the upper-cut $[a_2,b_2]$ in $x$-coordinates.  The two triangles  to the left and to the right of Fig. 8a are viewed as ``cuts" $[a_1,b_1]$ and $[a_3,b_3]$. Petrov\cite{Petrov2} computes the arctic curve for a polygon with (one or several) cuts along the upper-boundary.  Incidentally, it is still an open problem to compute the arctic curve for the model with several cuts, one above and one below.
      
      For $r$ large, we rescale the figure so that $b+c=b'+c'=1$. So, in the $(x,y)$ coordinates, we have, corresponding to the three cuts, satisfying $\sum_1^3(b_i-a_i)=1$:
      \be\label{aibi}a_1=  -n_1-b,~b_1=-n_1; ~~~a_2=0,~b_2=d; ~~~a_3=n_2+d,~   b_3=n_2+d+c'
     , \ee 
     with $c=1-b$,~ $b'=1-c'$ and $b+d=1-c'=b'$. So, following Petrov\cite{Petrov2}, given the product and the sum:
     \be\label{PS}P:=P(w):=\prod_{i=1}^3
     \frac{w-b_i}{w-a_i}\mbox{ and }
     S:=S(w):=\sum_{i=1}^3\left(\frac1{w-b_i}-\frac1{w-a_i}\right)
    , \ee
  we have the following parametric curve $(x(w),y(w))$ in $w$, together with its tangent $dy/dx$:
  \be\label{xwyw}x(w)=w+\frac{P-1}{S }(w),~~y(w)=1-\frac{(P-1)^2}{PS}(w),\mbox{ with }\frac{dy}{dx}=\frac{1-P}{ P}(w).\ee
  
  \noindent {\em $(i)$ Vertical cusp}. Here we refer to Fig. 8a. This takes place at the points $w=b_i$, for which  $P(b_i)=0$ and so $\frac{dy}{dx}(b_i)=\iy$. This means that  the arctic curve is tangent to the right hand side of the cut (or an extension of it).
    We therefore have a cusp when $y(b_2)$ is below the lower tip $(x(b_2),y(b_2))=(b_2,b+c-d)=(b_2,1-d)$ of the cut  along the vertical direction $x(b_2)=b_2$; namely for
    $$(x(b_2),y(b_2))=(b_2,y(b_2)),\mbox{ for }y(b_2)< 1-b_2=1-d .$$
   Then,   substituting the parameters $a_i,b_i$ from (\ref{aibi}) into $(P,S)$ (as given by (\ref{PS}))  and then computing $y(w)$ in terms of $(P,S)$ from   (\ref{xwyw}) and using the identities (above (\ref{PS})), $b'+c'=1$ and $b+d=1-c'$, one finds the following condition, since $n_i,d,c'>0$, confirming (\ref{cusp-cut}(i)):
  \be\label{ineq}1-b_2-y(b_2)= \frac{d\left(n_2b-(n_1+d)c'\right)}{(n_1+d)(n_2+c')}>0~~~\Longrightarrow~~~ \frac{n_1+d}{n _2}<\frac{b}{c'}.\ee
   %

  
   \noindent {\em $(ii)$ Oblique cusp}. Here we refer to Fig. 8b. This is  at the points $w=a_i$, where $P(a_i)=\iy$ implying $\frac{dy}{dx}(a_i)=-1$ from (\ref{xwyw}). This means that  the arctic curve is tangent to the left hand side of the cut (or an extension of it). 
    So, the cusp takes place when the cuspidal point $ (x(a_2),y(a_2)) $ belongs
    
    \newpage
    
    \vspace*{-.8cm}
    
    \noindent to the extension of the  line $(-1,1)$,  strictly below the lower tip of the cut; 
     namely for
    $$(x(a_2),y(a_2))=(d,y(a_2)) ,\mbox{ such that } x(a_2)>a_2+d \mbox{ and }y(a_2)< 1-b_2=1-d .$$
 
 \noindent  Then,   substituting as before the parameters $a_i,b_i$ from (\ref{aibi}) into $(P,S)$ (as given by (\ref{PS}))  and then computing $(x(w),y(w))$ in terms of $(P,S)$ from   (\ref{xwyw}) and using the same identities as above (\ref{PS})), one finds the following condition, since $n_i,d,b>0$,
  \be\label{ineq}1-b_2-y(a_2)= x(a_2)-a_2-d=\frac{d\left(n_1c'-(n_2 +d)b\right)}{(n_1+b)(n_2+d)}>0~~~\Longrightarrow~~~ \frac{b}{c'}<\frac{n _1 }{n_2+d} ,\ee
which is (\ref{cusp-cut}(ii)), thus  ending the proof of Proposition \ref{prop:cusp-cut}. \qed%


 \section{\bf The Evidence for  Conjecture 2.4 for $\rho\geq 0$}
 Recalling:
 
 \noindent {\bf Conjecture 11.0}
%
{\em Given the strip $\{\rho\}$ of width $\rho\geq 0$, with $\tau_1\leq 0$ and $\tau_2\geq \rho$, 
the following limit holds:
\be\label{mainlimit}\bl\lim_{r\to \iy}
(-r^{1/6})^{\tau_1-\tau_2}&
 {\mathbb L}_{r,\rho,\beta}^{\mbox{\tiny dTac}}  ( \tau_1,\sqrt 2   \xi_1 ;\tau_2, \sqrt2   \xi_2) \sqrt{2} d\xi \Bigr|
_{_{{\rho\geq 0,~~\beta=0}\atop{\xi_i=  \sqrt{2r}+\frac{\xi'_i}{\sqrt{ 2} r^{1/6}} }}}
\\&= (-1)^{\tau_2-\rho}   
\int_0^\iy d\mu A_{\tau_1-\rho}(\mu+\xi'_1)A_{-(\tau_2-\rho)}(\mu+\xi'_2)
.\el\ee}

\vspace{.3cm}

In the sector $\tau_1\leq 0$ and $\tau_2\geq \rho$, and $\beta=0$, the kernel $ {\mathbb L}_{r,\rho,\beta}^{\mbox{\tiny dTac}}$ 
 reduces to the one term ${\mathbb L}_{3}^{\mbox{\tiny dTac}}$ in (\ref{Final0}); the latter can readily   be expressed  as $\xi_1,~\xi_2-$partial  derivatives of a double Fourier transform\footnote{In this section, $\widehat{\widehat g}(\xi_1,\xi_2) 
 =\int_{\BR}\frac{dx ~e^{ -\I \xi_1 x}}{\sqrt{2\pi}}
\int_{\BR}\frac{dy ~e^{ \I \xi_2 y}}{\sqrt{2\pi}}g(x,y)  $, which is the same as  (\ref{FourierDef}) upon replacing $\xi_1\to -\xi_1$.} , upon setting $\ka=r-\rho$: 
  \be\label{L3'}\bl  {\mathbb L }_{r,\rho,\beta} ^{\mbox{\tiny dTac}} ( & \tau_1,\sqrt2\xi_1 ;\tau_2, \sqrt2\xi_2  )\sqrt2d\xi ={\mathbb L}_{3}^{\mbox{\tiny dTac}}\sqrt2d\xi= 
  (\sqrt2)^{\tau_1-\tau_2 }\left(-\frac {\pl}{\pl  \xi_1} \right)^{-\tau_1} \left( \frac {\pl}{\pl \xi_2 } \right)^{ \tau_2-\rho}\\&\times\int_{\BR}\frac{dx ~e^{ -\I \xi_1 x}}{\sqrt{2\pi}}
\int_{\BR}\frac{dy ~e^{ \I \xi_2 y}}{\sqrt{2\pi}}
   \underbrace{\left( e^{-\frac{x^2+y^2}2}
     \frac{r\Theta_{r-1}^+(\frac{\I x}{\sqrt2},\frac{\I y}{\sqrt2}}{4\pi \Theta_r(0,0)}\right)}_{=: \frac{\I (-1)^{\rho}}{c_{\ka-1}^2}\Phi(x,y)} d\xi 
 %
 \\&=(\sqrt2)^{\tau_1-\tau_2 }\left(-\frac {\pl}{\pl  \xi_1} \right)^{-\tau_1} \left( \frac {\pl}{\pl \xi_2 } \right)^{ \tau_2-\rho}
 \left(\frac{ \I (-1)^\rho}{ c_{\ka-1}^2}
 \widehat{\widehat\Phi}(\xi_1,\xi_2)\right)d\xi  
 .\el\ee

\subsection{Some  (rigorous) statements on $\ell$-spaced Christoffel-Darboux formulas and their double Fourier transform}
 
  \begin{lemma} Remembering $\ka:= r-\rho$, the ratio $\Phi(x,y)$  of the functions $\Theta$ in the equation (\ref{L3'}) above equals the ratio of two determinants:
  \be\bl \label{Phi(x,y)} 
  \Phi(x,y)=\frac{e^{-\frac{ x^2 + y^2}2}}{x-y}
 \frac{\det \HR(x,y)}{\det \HR_\rho}, \el\ee
 where $ \HR(x,y)$ and $\HR_{\rho}$ are two square matrices of size $\rho+2$ and $\rho$, with $\HR_{\rho}$ being a submatrix of $\HR(x,y)$, containing (one-sided) Gaussian moments $\mu_k=\int_0^\iy \frac { e^{-u ^2} }{\sqrt{\pi}}  {u ^{ k }}  du$, and containing two rows of   
monic Hermite polynomials $\pi_k(x) $ and $\pi_k(y)$ (see (\ref{Herm})):
 \be\label{HR}\bl
 \HR (x,y):=
 \left(
\begin{array}{cccc}
& \left(\frac{\mu_{ {\ka-1 +\al+\beta}    }}{\mu_{\ka-1} }\right)_{{0\leq \al\leq \rho-1}\atop{0\leq \beta\leq \rho+1}}
\\  \\ \hline
\\
&    \left(\bl&\frac{\pi_{\ka-1+\beta}(x)}{(-\I)^{\beta}}
\\&
\frac{\pi_{\ka-1+\beta}(y)}{(-\I)^{\beta}}
\el\right)_{0\leq \beta\leq \rho+1 } 
\end{array}\right) 
\mbox{, and  }
  \HR_{\rho}: = \left( \frac{\mu_{ {\ka-1 +\al+\beta}    }}{\mu_{\ka-1} }  \right)_{{0\leq \al\leq \rho-1}\atop{1\leq \beta\leq \rho}}  .\el\ee

 \end{lemma}
 
 \proof The  identity   (\ref{Phi(x,y)}) uses a formula in \cite{BDS} on the expressions (\ref{Theta}) for $\Theta^+_{r-1}$ and $\Theta _{r }$ in terms of   the Cauchy transform of Hermite polynomials (and its derivatives)
 $
h_k(\vr)=\frac1{2\pi \I}\int_{ _{\!\! -\!-\!\!{}^{\dis\cap}\!\!-\!- \hspace{-4.92mm} {o} }}
\frac{\pi_k(x+\vr)}{x}e^{- {(x+\vr)^2} }dx$, with $\I \vr>0$, upon deforming the pole $w_{ \al}^{-\rho}\leftrightarrow \prod_{\al=1}^{ \rho} (w_\al-\vr_j)^{-1}$ and then letting $\vr_j \to  0$, and using various identities. 
\qed

In order to compute   $\det \HR(x,y)$, one  expands the determinant along its last two rows which involves all $2\times 2$ determinants; i.e. $\ell$-spaced  Christoffel-Darboux  formulas for Hermite polynomials as in (\ref{Wnabla}), of which one needs to take the double Fourier transform as in  (\ref{FourierW}). 

To do so, we will need a $s$-generalization of the GUE-kernel (\ref{KGUE}) for integer $s\geq 0$ and $\ell\in \BZ$, with $f$ as in (\ref{f}): (for $s=0$ this is the GUE-kernel)
  \be\label{sKGUE}\bl
K^{ \mbox{\tiny \!\! GUE(s)} }(n+1,x ;n+\ell,y)& =2\oint_{\Ga_0}\frac{dz}{2\pi \I}
\oint_{L_+}\frac{dw}{2\pi \I (w-z)^{s+1}}\frac{w^{n+\ell}}{z^{n+1}}
\frac{f(x,z )
 }{f(y,w )
 } 
,\el\ee
whose {\em ``discrete Fourier transform''} reads:  ( notation $[~]_{[0,\ell-1]}$ denotes the $\ell-1$-truncated series in $\lb$; see (\ref{tildeF})):
\be\label{F,Phi}\bl 
F_{n+1,n+\ell}&(\xi_1,\xi_2)d\xi_2 :=
\frac 12 \sum_{s=0}^{\ell-1}
\left(\frac{\xi_2-\xi_1}{2} \right)^s
K^{ \mbox{\tiny \!\! GUE(s)} }(n+1,\xi_1 ;n +\ell,\xi_2)d\xi_2
\\&=
\int_0^\iy d\lb \left[ e^{\frac{\lb(\xi_2-\xi_1)}2}\right]_{[0,\ell-1 ]} 
\oint_{\Ga_0}\frac{ dz }{2\pi \I}\oint_{L}\frac{dw}{2\pi \I }
\frac{w^{n+\ell }f(\xi_1,z )e^{\lb z}}{z^{n+1}f( \xi_2,w)e^{ \lb w}}
.\el\ee

\begin{lemma}
The following {\em $\ell$-spaced  Christoffel-Darboux formula}  for Hermite polynomials (\ref{Herm}), with $\ell\geq 1$,   
 relates to the $s$-GUE-kernel, as follows, upon setting $\nabla:=-\frac12\left(\frac{\pl}{\pl x}+\frac{\pl}{\pl y}\right)$:  
%
\be\bl\label{Wnabla}
W_{n+\ell ,n } (x,y) &:=\frac1{c_n^2}
\frac{\left(\pi_{n+\ell }(x) \pi_{n }(y)-\pi_{n+\ell }(y) \pi_{n }(x)\right)}{x-y}
        e^{-\frac{x^2+y^2}2} 
 \\&= \frac12\sum_{s=0}^{\ell-1} \nabla^s\Bigl( K^{  \tiny \mbox{GUE(s)} }(n+1,x ;n\! +\!\ell,y)+ K^{  \tiny \mbox{GUE(s)} }(n+1,y ;n \!+ \!\ell,x)\Bigr)
.\el\ee
Its double Fourier transform is expressed in terms of the discrete Fourier transform $F_{n+1,n+\ell}$ of $K^{  \tiny \mbox{GUE(s)} }$, defined   in (\ref{F,Phi}):
\be\label{FourierW} 
\bl
 \widehat{\widehat W}_{n+\ell ,n}(\xi_1,\xi_2) 
 &=
\I^{\ell-1} F_{n+1,n +\ell}(\xi_1,\xi_2)+(-\I)^{\ell-1 }
F_{n+1,n +\ell}(\xi_2,\xi_1).\el\ee
\end{lemma}
\proof Identity (\ref{Wnabla}) for $W_{}$  hinges on the following inductive step for the GUE-kernel $K(x,y)$:
\be\label{Winduc}
W_{n+\ell ,n }(x,y)=\tfrac12\left( K (n\!+\!1,x ;n\!\!+\!\!\ell,y)+ K (n\!+\!1,y ;n\!\!+\!\!\ell,x)\right)
+\nabla
  \sum_{j=0}^{n  }W_{j+\ell-1,j}(x,y),
 \ee
 followed by using the same identity again for the $W_{j+\ell-1,j}$ and summing.\qed
\bigbreak 

Define 
  ${\mathbb E}_\rho(f)$ as the GUE spectral probability distribution on functions $f$ on the positive half-line: 
\be\label{Erho}{\mathbb E}_\rho(f) := \frac{1}{Z_\rho}\left(\prod_{\al=0}^{\rho-1}\int_0^\iy {e^{-u_\al ^2} du_\al} \right)
\Dt^2(u)
 f(u)
.\ee
We now state the following proposition for ${\mathbb L}^{\mbox{\tiny dTac}}  $,  remembering $\kappa:=r-\rho$ :

\begin{lemma} \label{Th3.1}The double Fourier transform of $\Phi(x,y)$ as in (\ref{L3'}) reads as follows, in terms of the discrete Fourier transform $F_{n+1,n+\ell}$, as in (\ref{F,Phi}):
\be\label{Phihathat}\bl\frac{\I(-1)^\rho}{c^2_{\ka-1}}  \widehat{\widehat\Phi}(\xi_1, \xi_2)
 &=
 \sum_{0\leq\ell< k\leq \rho+1}  {\mathcal E}_{\ell,k}
 \left(\prod_{\al=k}^\rho \tfrac{r-\al}2 \!\!\right)
\\&\times\left[(-1)^{k-1}F_{r-k+1,r-\ell}(\xi_1,\xi_2)+(-1)^\ell F_{r-k+1,r-\ell}(\xi_2,\xi_1)\right]\el\ee
where
\be\label{ER}{\mathcal E}_{\ell,k}=\frac{{\mathbb E}_\rho((\sg_\ell \sg_{k-1} -\sg_{\ell-1}\sg_{k } ) \sg_\rho^{\ka-1})}{{\mathbb E}_\rho\sg_\rho^\ka  }=\sum_{\mu=0}^\iy
\left(\frac r2\right)^{\frac{\ell+k-\rho-1-\mu}2}
   \sum_{{\al,\beta\geq 0}\atop{0\leq \al+\beta\leq \mu}} \vr^{(\rho)}_{\ell,k;\al,\beta}
  \Ga^\mu_{\al,\beta}  \ee
 is a ratio  of expectations  (\ref{Erho}), in terms of elementary symmetric functions\footnote{$ \prod_{i=0}^{\rho-1}(z-u_i)=\sum_{j=0}^\rho (-1)^j z^{\rho-j}\sg_j(u_0,\dots,u_{\rho-1})$
} $\sg_j(u_0,\dots,u_{\rho-1})$. This ratio admits an expansion in powers of (large)  $\sqrt r$.  Its coefficients are further  expressed in terms of $\rho$-independent expressions $\Ga^\mu_{\al,\beta} =\Ga^\mu_{\beta,\alpha} $, which are rational functions in the expectations (137) and such that $\Ga^0_{0,0}=1$.
The
 coefficients  $\vr^{(\rho)} _{\ell,k;\al,\beta}$ in (139) are defined for integer  $\al,\beta\geq 0$ as follows: 
\be\label{epsilon-rho}\vr^{(\rho)}_{\ell,k;\al,
 \beta}:=C^{ \al}_\ell  C^{ \beta}_{k-1}-C^{ \al}_k C^{ \beta}_{\ell-1}\mbox{, for }  C^{ \al}_\ell := \left({\rho-\al}\atop{\ell-\al}\right)\mbox{, for ${0\leq \ell<k\leq\rho}$}.
  \ee

 
  \end{lemma}
  
  \proof
  One  computes the $\det \HR(x,y)$, given by (\ref{HR}), by expanding the determinant along its last two rows which involves all $2\times 2$ determinants; i.e. $\ell$-spaced  Christoffel-Darboux  formulas   (\ref{Wnabla}), of which one  takes the double Fourier transform (\ref{FourierW}). This establishes formula (\ref{Phihathat}).
 
  In order to deal with the $\ER_{\ell,k}$-term for $r\to \iy$, we notice that $\ER_{\ell,k}$ involves a crucial  integrand   $e^{-u^2}u^r F(u)du =
e^{\frac 12 r\log r+rf(u')}F(u'\sqrt r)du'\sqrt r$, with $f(u'):=\log u'-u'^2$, upon setting $u=u'\sqrt r$ and then $u'=\frac{1}{\sqrt 2}+\frac{u''}{\sqrt r}$ about the saddle $u'=\frac{1}{\sqrt 2}$; that means, all together, $u=\sqrt{\tfrac r2}+u''$. So, we have 
\be\sg_\ell(u)\Bigr|_{u=\sqrt{\tfrac r2}+u''}=\sum_{\al=0}^{\ell}
(r/ 2)^{\frac{\ell-\al}2}C^\al_\ell \sg_\al(u''), \mbox{ where  $C^\al_\ell$ is as in (\ref{epsilon-rho})} \label{sigma-ell},\ee
which then leads  to the identity (\ref{ER}) for  ${\mathcal E}_{\ell,k}$. This ends the proof of Lemma  \ref{Th3.1} .\qed


  \newpage
  

%



\subsection{The limit of the kernel ${\mathbb L}^{\mbox{\tiny dTac}}  $ by saddle analysis} 

In view of kernel ${\mathbb L }_{r,\rho,\beta} ^{\mbox{\tiny dTac}} (   \tau_1,\sqrt2\xi_1 ;\tau_2, \sqrt2\xi_2  )$ in (\ref{L3'}), we will need to do the asymptotics on the partials in (\ref{L3'}), acting on $\widehat{\widehat \Phi}$ in (\ref{Phihathat}); namely, 
\begin{proposition}\label{cor:F+F}For fixed $0\leq \ell< k\leq \rho+1$, the following holds for the derivatives of the $F$-expression (\ref{F,Phi})  
  : (here $[n]$ denotes the parity of $n$)\footnote{where $\int dM :=\tfrac12 d\xi'_2\int_{0}^\iy d\mu\int_{\CR_L }\int_{\CR_R } {dudv} $ and $e^{\sum_1^\iy t^i z^i} =\sum_0^\iy p_n(t)z^n$ are the elementary Schur polynomials.}
\be\label{F+F}\bl
 &(-\sqrt{2} r^{1/6})^{  \tau_1-\tau_2+\rho }   \left(-\frac{\pl}{\pl \xi_1}\right)^{-\tau_1}\left( \frac{\pl}{\pl \xi_2}\right)^{\tau_2-\rho}
\left(\sqrt{\frac r2}\right)^{\ell-(k-1)}
 \\&~~~~~~\times\Bigl[(-1)^{k+1}F_{r-k+1,r-\ell}(\xi_1,\xi_2)+(-1)^\ell F_{r-k+1,r-\ell}(\xi_2,\xi_1)\Bigr]\Bigr|_{\tiny\mbox{Airy scaling (\ref{scale1}), (\ref{scale2})}}
\\ = &
 \pl_{\xi_1'}^{-\tau_1}(-\pl_{\xi_2'})^{\tau_2-\rho}\int  \frac{dM}{(2\pi \I)^2}
\sum_ {{0\leq m,n,a,b<\iy}\atop{a+b=m}} \frac{\vr_{\ell,k}^{a,b;[ n]}u^a(-v)^b}{r^{\frac{n+m}3}}
\\&\times\!\!\left[p_n-(-1)^{k\!-\!\ell}\!\!\!\sum_{0< {k\!-\!\ell}\leq \beta\leq n-1}\!\!
\left({ \beta -1}\atop{k\!-\!\ell\!-\!1}\right)\frac{\left(\frac{\mu(\xi'_1-\xi'_2)}{2}\right)^{ \beta  }}{  \beta   !}p_{n-\beta }  \Id_{n\geq 2}\right](t^\mu)  
  \frac{e^{-\AR_u(\xi'_1+\mu)}}{e^{-\AR_v(\xi'_2+\mu)}},
\el\ee
in terms of elementary Schur polynomials $p_n(t)$, and where, for integer $0\leq a,b\leq  k-1$,
\be\label{eps_ab}\bl\vr_{\ell,k }^{a,b;\pm  } 
  &:=   {
    \mbox{$  (-1)^{k+1}\left({k-1}\atop{a}\right)\left({\ell-1+b}\atop{b}\right) \pm(-1)^{\ell } \left({k-1}\atop{b}\right)\left({\ell-1+a}\atop{a}\right)$}}
   . \el\ee

\end{proposition}

\proof  We now perform a saddle analysis for $r\to \iy$ on  the integrand of the double $(z,w)$-integral in the function $F_{n+1,n+\ell} (\xi_1,\xi_2)$ as in (\ref{F,Phi}), for $n=r-k$.
Given the Airy scaling $(z,w,\lb,\xi_i)\to
(u,v,\mu,\xi_i')$, as in (\ref{scale1}) and ({\ref{xi,z-as}), we perform a saddle analysis, by setting $z=z'\sqrt r$ and subsequently $z'-z'_0=\frac{u}{\sqrt 2 u^{1/3}}$, as in (\ref{scale2}), with saddle point $z_0'=\frac{1}{\sqrt 2}$, but, as is unusual in saddle point analysis,   keeping all the higher order terms of the Taylor series; this gives the following exact formula in terms of the elementary Schur polynomials 
 $p_n(t^{(\mu)})$ for $t^{(\mu)}=(t_1^{(\mu)},t^{(\mu)},\dots)$, given  in (\ref{tn}), and setting  $\mu=0$, with $\AR_u(\xi)=\frac{u^3}3-u\xi$, as in (\ref{Airycub}):
\be\label{phi/phi}\bl
 \frac{f( \xi_1,z)e^{\lb z -r\log z}}{f( \xi_2,w)e^{\lb w -r\log w}}\Bigr|_{\tiny{\mbox{scaling}}\atop\mbox{{above}}} 
 &=e^{^{ \sum_{n=1}^\iy t^{}_n  r^{-n/3}  }}
~\frac{e^{-\AR_u(\xi'_1+\mu)}}{e^{- \AR_v(\xi'_2+\mu)}}
=\sum_{n=0}^\iy p_n(t^{(\mu)})r^{-\frac n3 } \frac{e^{-\AR_u(\xi'_1+\mu)}}{e^{- \AR_v(\xi'_2+\mu)}}
,\el\ee
where
\be \label{tn}t^{(\mu)}_n=
 \frac{u^{n+3}-v^{n+3}
  }{(-1)^{n+3}(n+3)}-\frac{(\xi'_1+\mu)^2-(\xi'_2+\mu)^2}{4}\dt_{n,1}, \mbox{ and }t_n=t_n^{(\mu)} |_{\mu=0}.\ee 
  Also notice that  (\ref{eps_ab}) comes from  scaling and expanding the ratio $w^{-\ell}/z^{-k+1}$ appearing in the function $F_{r-k+1,r-\ell}$ and its symmetrization (\ref{Phihathat}):
 $$ \tfrac{\left(1+\frac{v}{r^{1/3}}\right)^{-\ell}}{\left(1+\frac{u}{r^{1/3}}\right)^{-k+1}}=\sum_{0\leq m\leq \rho}\sum_{{a+b=m}\atop{a,b\geq 0}}r^{-m/3}  \left({k-1}\atop{a}\right)\left({\ell-1+b}\atop{b}\right)+O(r^{-\frac{\rho+1}{3}})
 . $$
  This sketches the proof of Proposition \ref{cor:F+F}.\qed
  
  Before proceeding, we define  other   combinatorial expressions:
\be \label{eta-theta} \eta^a_\theta(k ):=\eta_\theta (k;\rho), \mbox{such that }\prod_{j=k}^\rho(1-\frac jr)=\sum_{\theta=0}^\rho \frac{\eta_\theta (k;\rho)}{r^{\theta}}
.\ee
Next, given the integer $a$, we will need to express $\eta_\theta(k;\rho)=\eta^a_\theta(k;\rho) $ in terms of a basis, given by  the ratios of binomial coefficients below; so, we have\footnote{in terms of ascending factorials $k^{(n)}=k(k+1)\dots (k+n-1)$}:
\be \label{eta-theta'} \eta^a_\theta(k;\rho)=\sum_{j=0}^{2\theta } c^{\prime(a)}_j(\theta)\tfrac{\left({k -1}\atop {a+j}\right)}{\left({k -1}\atop {a }\right)} 
  \mbox{, with } c^{\prime(a)}_{2\theta}(\theta)
   =\frac{(1+a^2)^{(2\theta)}}{\theta! 2^\theta}.  \ee

Setting the expression (\ref{F+F}) into the expression after the first equality in the kernel (\ref{Ltausigma}) below, leads to the proposition below. Notice  the  {\em presence of the $r^{\rho/3}$-term} next to   the integral $\int dM$ (after $\stackrel{*}{=}$ below), which blows up for $r\to \iy$, is the reason for keeping all the higher order terms in the saddle analysis of (\ref{phi/phi}); i.e., one has to keep $\rho$ terms. 

  \begin{proposition} The following holds for $r\to \iy$:
   \be\label{Ltausigma}\bl
 &(-r^{1/6}  )^{\tau_1-\tau_2}    
  {\mathbb L}^{\mbox{\tiny dTac}}  ( \tau_1, \sqrt2 \xi_1 ;\tau_2,  \sqrt2 \xi_2)    d\xi_2 
 \Bigr|_{\tiny\mbox{Airy scaling}}
\\ =&(-\sqrt2 r^{1/6})^{\tau_1-\tau_2 }\left(-\frac {\pl}{\pl  \xi_1} \right)^{-\tau_1} \left( \frac {\pl}{\pl \xi_2 } \right)^{ \tau_2-\rho}
 \sum_{0\leq\ell< k\leq \rho+1}  {\mathcal E}_{\ell,k}
 \left(\prod_{\al=k}^\rho \tfrac{r-\al}2 \!\!\right)
\\&\times\left[(-1)^{k-1}F_{r-k+1,r-\ell}(\xi_1,\xi_2)+(-1)^\ell F_{r-k+1,r-\ell}(\xi_2,\xi_1)\right]d\xi  \Bigr|_{\tiny\mbox{Airy scaling}}
%
\\\stackrel{*}{=}& 
 \tfrac{1}{ 2^{\rho+1}}
  \left( \frac {\pl}{\pl  \xi'_1} \right)^{-\tau_1} \left( -\frac {\pl}{\pl \xi'_2 } \right)^{ \tau_2-\rho}
 \int dM  
 \sum_{\tiny\begin{array}{ccc}&0\leq n<\iy\\
& 0\leq m\leq \rho
\\&0\leq \nu<\iy\\
&0\leq\theta\leq \rho
 \end{array}}
   r^{\frac{\rho -  n-m-3\theta }{3}   } 
 (\frac r2)^{-\frac{\nu}2} 
 \sum_{a+b=m} u^a(-v)^b  
 \\&\!\!\!\!\times  \!\! \sum_{{ }\atop{{0\leq \al+\beta \leq \nu}\atop{\al,\beta\geq 0}}}\left[\bl
& 
\!     
(A^{a,b;n}_{\al,\beta;\theta}+B^{a,b;n}_{\al,\beta;\theta})p_n(t^{(\mu)})%
  +C_{\al \beta;\theta}^{a,b;n}
 \el\right] \Ga_{\al \beta }^{\nu}\frac{e^{-\AR_u(\xi'_1+\mu)}}{e^{-\AR_v(\xi'_2+\mu)}}d\xi'
+O(1/r)\\=& 
 ({\mathbb K}_{\tau_1,\tau_2}^{(0)}+{\mathbb K}_{\tau_1,\tau_2}^{(1)}
+{\mathbb K}_{\tau_1,\tau_2}^{(2)})(\xi'_1,\xi'_2)d\xi'+O(1/r)%
,\el\ee
where the coefficients of $\Ga_{\al,\beta}^{(\nu)}$, for fixed $\al,\beta$ such that $0\leq \al+\beta\leq \nu$ are given below; remember  $\eta_\theta(k)$ defined in (\ref{eta-theta}) and (\ref{eta-theta'}), the quantities $\vr^{(\rho)}_{\ell,k;\al ,\beta }$ and $
  \vr_{\ell,k}^{a,b,\pm}$ as in (140) and (143), and the combinatorial expressions $C^\al_\ell$  defined in (\ref{epsilon-rho}) \footnote{The factorials within the bracket  $\left({ a \atop b}\right)=0$, if $b>a$ or $b<0$. 
}; so we have for $A^{a,b;n}_{\al,\beta;\theta} $, $B_{\al \beta;\theta}^{a,b;n}$, and  $C_{\al \beta;\theta}^{a,b;n}$ :
\be\label{Dtabn}\bl&A^{a,b;n}_{\al,\beta;\theta} :=(-1)^\rho\sum_{0\leq \ell,k\leq \rho+1}
  \vr^{(\rho)}_{\ell,k;\al ,\beta }
  \vr_{\ell,k}^{a,b,[   n]}\eta_\theta(k )
  \\& =\sum_{j=0}^{2\theta }c^{\prime (a)}_j(\theta)
   \sum_{\ell=0}^{\rho+1} \left({\ell\!+\!b-\!1\!}\atop{b}\right)\left[\bl& C^{\al}_{\ell}
   \left({{\beta} \atop {\rho-a-j}}\right) _{* }
 + C^{\beta}_{\ell-1} 
  \left({\al-1 }\atop{\rho\!-\!a\!-\! j\!-\!1}\right)_{** }
\!\!\!\el\right]
\\&
~~~+[n]\sum_{j=0}^{2\theta }  c^{\prime(b)}_j(\theta)  
\sum_{\ell=0}^{\rho+1}\left({\ell-1 }\atop{b+j}\right)\left[\bl&  C^{\al}_{\ell} 
 \left({{a+\beta}\atop{\rho}} \right)_{*** }
 + C^{\beta}_{\ell-1}
 \left({{a+\al-1}\atop{\rho-1}} \right)_{****}\!\!\!
\el\right]
,\el\ee 
 
$$
B_{\al \beta;\theta}^{a,b;n}:=(-1)^\rho
\sum_{ {0\leq \ell<k\leq \rho+1}\atop{ 
  }} 
  \!\!\! \vr^{(\rho)}_{\ell,k;\al ,\beta } 
 \left(\vr_{\ell,k}^{a,b,[  n]} \eta_\theta(k )  +   \vr_{k,\ell }^{a,b,[  n]}\eta_\theta(\ell )\right)
, $$
and
$$\bl C_{\al \beta;\theta}^{a,b;n}&:=2(-1)^\rho
\sum_{  0\leq \ell<k\leq \rho+1}
\vr^{(\rho)}_{\ell,k;\al ,\beta }
\vr_{\ell,k }^{a,b,[  n]}\eta_\theta(k )
\\&
\times\sum_{0\leq \ga\leq n-k+\ell\leq n-1}
\left({{n-\ga-1}\atop{k-\ell-1}} \right)
\frac{ \left(\frac{\mu(\xi'_1-\xi'_2)}{2}\right)^{ n-\ga  } }{(n-\ga)!}
p_\ga(t^{\mu}).\el$$
 
\end{proposition}

 \subsection{The Airy ideal}
 
 Given the Airy function, satisfying the Airy differential equation, 
 \be Ai (x)=\int_{ \CR_L}\frac{du }{ 2\pi \I }
 e^{e^{-\frac{u^3}{3} +xu}} 
 =\int_{ \CR_R
  }\frac{ dv}{ 2\pi \I } 
 e^{  \frac{v^3}{3}-xv }\mbox{, with }\pl_x^2Ai(x)=xAi(x)
,\label{Ai}\ee
we define the Airy ideal $\IR_A$
 as the following set of polynomials, with the understanding that $x,y$ always is to the left of $\pl_x,\pl_y$; 
\be\label{Airy ideal}
\IR_A:=\left\{\bl
&Q(x,u;y,-v), \mbox{ polynomials in $x,u,y,v$,} \\
&\mbox{such that }Q(x,\pl_x;y,\pl_y)\oint_{\CR_L}\frac{du}{2\pi\I}\oint_{\CR_R}\frac{dv}{2\pi \I} \tfrac{e^{-\frac{u^3}{3} +xu}}{ e^{ -\frac{v^3}{3}+yv}}=0
 \el\right\}.
 \ee
 Evidently, this implies that 
 $$
   \oint_{\CR_L}\frac{du}{2\pi\I}\oint_{\CR_R}\frac{dv}{2\pi \I} Q(x,u;y,-v)\tfrac{e^{-\frac{u^3}{3} +xu}}{ e^{ -\frac{v^3}{3}+yv}}=0.$$
%


We have at once that, from the Airy ode (\ref{Ai}), the following functions  $u^2-x\mbox{ ,~  }v^2-y,~u^4 -x^2 -2u\mbox{ and }v^4 -y^2 +2v\in\IR_A $. That $u^2-x\in \IR_A$ is exactly the Airy ode (\ref{Ai}).

We also have (by induction)  $u^n-\left(p^{(n)}(x)+q^{(n)}(x)u\right)\in   \IR_A$, and the same for $x\to y$ and $u\to -v$ for $n\in\BZ_{\geq 0}$, 
where the polynomials $p^{(n)}(x)$ and $q^{(n)}(x)$  are defined recursively for $n\geq 0$: 
\newline\hspace*{.6cm}$\left(\begin{array}{ccc} p^{(n+1)}\\
 q^{(n+1)}
\end{array}\right)
=\left(\begin{array}{ccc}
\pl_x&x\\ 1&\pl_x
\end{array}\right)
 \left(\begin{array}{ccc} p^{(n )}\\
 q^{(n )}
\end{array}\right)$ with $ p^{(0)}(x)=1,~q^{(0)}(x)=0 $.
\newline Hence any  polynomial in $P(u,v,x,y)$ can be reduced to an expression of the following type, with polynomial coefficients in $x,y$:
\be\label{Airy-reduction} P(u,x,v,y)= a(x,y)uv+b (x,y)u+c(x,y)v+d(x,y), \mbox{~mod }\IR_A.
\ee

 


 \subsection{Evidence for Conjecture 11.0 and 2.4}
 For $r\to \iy$, the negative powers of $r$ in (\ref{Ltausigma}) tend to $0$; so, they do not matter. Therefore, only   consider  the terms for which 
$\frac{\rho -  n-m-3\theta }{3}-\frac{\nu }2\geq 0,\mbox{ with }~~0\leq n,~~0\leq  m=a+b,~~0\leq \al +\beta \leq \nu$. \newline We now give a ({\em rigorous proof}) for the limit of the first kernel in (\ref{Ltausigma}): 
\newline \be\bl \label{K0}\hspace*{-.5cm}\bullet  ~~~ \lim_{r\to \iy} {\mathbb K}_{\tau_1,\tau_2}^{(0)}(\xi_1',\xi'_2  )d\xi'_2=
 d\xi'_2  
\int_0^\iy d\mu 
  A_{\tau_1-\rho}(\mu+\xi'_1)A_{-(\tau_2-\rho)}(\mu+\xi'_2)
. \el\ee
 %
Indeed, we have that the expression (\ref{Dtabn}) above satisfies:
\be\label{Aalphabeta}\bl A^{a,b;n}_{\al,\beta;\theta}&=0, \mbox{ whenever }
 \frac{\rho -  n-m-3\theta }{3}-\frac{\nu }2> 0,~~0\leq n,~~0\leq  m=a+b,~~0\leq \al +\beta \leq \nu 
 \\&= 2^{\rho+2}\mbox{, for }\rho=m=a \mbox{ and }\nu =\theta =b=n=\al=\beta=0 \mbox{ and so } [n]=1.\el \ee
 Indeed, if $A^{a,b;n}_{\al,\beta;\theta}> 0$, then at least one of the four (starred) binomial coefficients in the  bracket of (\ref{Dtabn}) must be non-zero; leaving four possibilities: 
 $$\bl& (*):\rho-\beta\leq a+j\leq \rho\mbox{ some } j\leq 2\theta,~~  (**):\rho-\al\leq a+j\leq \rho\mbox{ some } j\leq 2\theta
 \\&
  (***):\rho-\beta\leq a ,~~~~~~~~~~~~~~~~~~~~~~ (****):\rho-\al\leq a\leq \rho. \el$$Assuming first $\theta\geq 1$ and then $\theta=0$, the four cases above 
   would lead to 
$$\bl
&0\leq 3\al+\beta\leq -2 b 
&\mbox{or}~~0\leq \al+3\beta\leq -2 b 
, \el$$
which, since $b\geq 0$, forces $\al=\beta=b=0$ and then $ \nu=0$ and $m=a$; then $\Ga^\nu_{\al,\beta}\big|_{\al=\beta=\nu=0}=1$. Hence, in (\ref{Dtabn}),  we must have $\theta=0$. Then, one checks that each of the sums in (\ref{Dtabn}) equals $\pm 2^\rho (-1)^\rho$, using 
 $\sum_{\ell=0}^{\rho }C^\al_\ell=\sum_{\ell=1}^{\rho+1}C^\beta_{\ell-1}= \sum_{\ell=0}^\rho \left({\rho}\atop{\ell}\right)  =2^\rho$ for $\al=\beta=0$. This leads to the proof of identity (\ref{Aalphabeta}), which put in  formula (\ref{Ltausigma}), leads to  the limiting kernel (\ref{K0}). 
\newline $\bullet$ ~{\em Evidence} for~$\lim_{r\to \iy} {\mathbb K}_{\tau_1,\tau_2}^{(1)}=0$ in (\ref{Ltausigma}). 
 %
 For fixed $0\leq \al,\beta\leq \frac23\rho $ and fixed $0\leq \om\leq \rho-\frac32(\al+\beta)$, we claim that the following expression (which only depends on $\rho$ through the  $B^{a,b;n}_{\al,\beta;\theta}$)\footnote{Recalling from (\ref{tn}), we have $t^{(\mu)}=(t^{(\mu)}_n)_{n\geq 0}$, with:
  $$t^{(\mu)}_n=
 \frac{u^{n+3}-v^{n+3}
  }{(-1)^{n+3}(n+3)}-\frac{x^2-y^2}{4}\dt_{n,1},\mbox{ setting } \xi'_{_{({1\atop2})}} +\mu=:\left(x\atop y\right).  $$}:
  \be\label{Balphabeta}
  \sum_{\footnotesize\begin{array}{ccc}& \\
& 0\leq n,a, b,\theta  \leq \rho
\\ 
&      n+a+b+3\theta=\om
\\&
 \end{array}}~
 p_n
 \left(t^{(\mu)}
 \right)
   u^a(-v)^{ b}B^{a,b;n}_{\al,\beta;\theta}  
 \in \IR_A  
  .\ee
%
Expression (\ref{Balphabeta}) is a polynomial in $u,v,x,y$ with coefficients, depending themselves on the  purely combinatorial expressions   
 $B^{a,b;n}_{\al,\beta;\theta} $. Upon using the Airy reduction (\ref{Airy-reduction}) on this expression, one shows that the  respective coefficients all vanish; i.e.,$a(x,y)=b(x,y)=c(x,y)=d(x,y)=0$ in (\ref{Airy-reduction}).    By the very definition of the Airy ideal, the integral $dM$ (as in footnote 13) of (\ref{Balphabeta}) vanishes.  We have no conceptual way of checking  the statement (\ref{Balphabeta})!   However it  has been checked for $\rho=1$ and $\rho=2$, for which $B^{a,b;n}_{\al,\beta;\theta} $ can be computed explicitly.   
 \newline $\bullet$ ~{\em Evidence} for~$\lim_{r\to \iy} {\mathbb K}_{\tau_1,\tau_2}^{(2)}=0$ in (\ref{Ltausigma}). As for ${\mathbb K}_{\tau_1,\tau_2}^{(1)}$ above, we claim that for fixed $0\leq \al,\beta\leq \frac23\rho $ and fixed $0\leq \om\leq \rho-\frac32(\al+\beta)$, 
 this expression ${\mathbb K} _{\tau_1,\tau_2}^{(2)}$ is a {\em sum of two parts}:  a \underline{first part} in the Airy ideal and a \underline{second part}, a perfect derivative in $\mu$ (see footnote 12), where $\BC[u,v;\mu]$ is the set of polynomials (over $\BC$) in $u,v;\mu$:  
 $$ \Bigl(\!\!\!\!\!\sum_{\footnotesize\begin{array}{ccc}& \\
& 0\leq n,a, b,\theta  \leq \rho
\\ 
&      n+a+b+3\theta=\om
\\&
 \end{array}} u^a (-v)^b C_{\al \beta;\theta}^{a,b;n}\Bigr)
    \frac{e^{-\AR_u(x)}}{e^{-\AR_v(y)}}\in  \left(\IR_A  +\frac{\pl}{\pl \mu} \mu\BC[u,v;\mu]\right) \frac{e^{-\AR_u(\xi'_1+\mu)}}{e^{-\AR_v(\xi'_2+\mu)}}.
  $$
  Then, by $dM$ integration both parts vanish. To summarize, the only term left will be $\lim_{r\to \iy} {\mathbb K}_{\tau_1,\tau_2}^{(0)}(\xi_1',\xi'_2  )$ which  evidently leads to Conjecture 2.4.\qed

\vspace*{.5cm}

\noindent {\footnotesize{\em Mathematics Subject Classification}. Primary:60G60, 60G65, 35Q53; secondary: 60G10, 35Q58. {\em Keywords and Phrases}: Lozenge tilings, cusp-Airy kernels. 

\bigbreak

  {\footnotesize {*Mark Adler: Department of Mathematics, Brandeis University,
  Waltham, Mass 02454, USA. E-mail: adler@brandeis.edu.
  The support of a Simon's foundation grant \#278931
    is gratefully acknowledged.}}
    
      {\footnotesize{$\dagger$ Pierre
 van Moerbeke: {UCLouvain, 1348 Louvain-la-Neuve,  Belgium \& Brandeis University. E-mail: pierre.vanmoerbeke@uclouvain.be. The support of a Simon's foundation grant \#280945   is
  gratefully acknowledged.
}}
\end{document}